\documentclass[12pt]{article}
\usepackage{jheppub}
\usepackage{multicol}
\pdfoutput=1

\usepackage{epsfig}
\epsfclipon
\usepackage{multicol}
\usepackage{epsfig,bm}
\usepackage{amssymb,amsmath}

\newcommand{\roughly}[1]{\mathrel{\raise.3ex\hbox{$#1$\kern-0.85em
\lower1ex\hbox{$\sim$}}}}

\newcommand{\lsim}{\roughly<}
\newcommand{\gsim}{\roughly>}

\def\be{\begin{equation}}
\def\beq\begin{equation}
\def\ee{\end{equation}}
\def\bea{\begin{eqnarray}}
\def\eea{\end{eqnarray}}

\def\pref#1{(\ref{#1})}

\def\EFT{{\scriptscriptstyle EFT}}

\def\beq{\begin{equation}}
\def\eeq{\end{equation}}
\def\beqa{\begin{eqnarray}}
\def\eeqa{\end{eqnarray}}

\def\cB{{\cal B}}

\def\cM{{\cal M}}

\newcommand{\bmat}{\left(\begin{array}}
\newcommand{\emat}{\end{array}\right)}

\def\yzero{\smash{\hbox{$y\kern-4pt\raise1pt\hbox{${}^\circ$}$}}}

\def\-{\hphantom{-}}

\def\s2{\frac{1}{2}}

\def\IF{\relax{\rm I\kern-.18em F}}
\def\II{\relax{\rm I\kern-.18em I}}
\def\IP{\relax{\rm I\kern-.18em P}}
\def\IC{\relax{\rm I\kern-.48em C}}
\def\IR{\relax{\rm I\kern-.18em R}}
\def\IK{\relax{\rm I\kern-.20em K}}
\def\IM{\relax{\rm I\kern-.25em M}}

\def\cA{{\cal A}}

\def\Dsl{\,\raise.15ex\hbox{/}\mkern-13.5mu D} 
\def \one{\relax{\rm 1\kern-.26em I}}

\def\exd{{\rm d}}

\def\bfk{{\bf k}}

\def\bfp{{\bf p}}
\def\bfr{{\bf r}}

\def\bfv{{\bf v}}
\def\bfx{{\bf x}}
\def\bfy{{\bf y}}

\def\ssE{{\scriptscriptstyle E}}
\def\ssI{{\scriptscriptstyle I}}
\def\ssT{{\scriptscriptstyle T}}
\def\ssN{{\scriptscriptstyle N}}

\def\drho{{\delta \rho}}
\def\dpr{{\delta p}}
\def\dss{{\delta s}}
\def\dphi{{\delta \phi}}
\def\dbfv{{\delta {\bf v}}}

\def\nott#1{\setbox0=\hbox{$#1$}                
   \dimen0=\wd0                                 
   \setbox1=\hbox{/} \dimen1=\wd1               
   \ifdim\dimen0>\dimen1                        
      \rlap{\hbox to \dimen0{\hfil/\hfil}}      
      #1                                        
   \else                                        
      \rlap{\hbox to \dimen1{\hfil$#1$\hfil}}   
      /                                         
   \fi}                                         %

\def\nn{\nonumber}

\newcommand{\cO}{{\cal O}}
\newcommand{\cE}{{\cal E}}

\newcommand{\cL}{{\cal L}}

\newcommand{\cW}{{\cal W}}

\newcommand{\cV}{{\cal V}}
\newcommand{\cT}{{\cal T}}

\newcounter{oldcounter}
\addtocounter{equation}{1}
\def\pref#1{(\ref{#1})}

\begin{document}

\title{Intro to Effective Field Theories and Inflation}

\author{C.P. Burgess\\
\qquad{\it Department of Physics \& Astronomy, McMaster University\\
 \qquad and  Perimeter Institute for Theoretical Physics 
}}

\date{}

\abstract{These notes present an introduction to inflationary cosmology with an emphasis on some of the ways effective field theories are used in its analysis. Based on lectures prepared for the Les Houches Summer School {\em Effective Field Theory in Particle Physics and Cosmology}, July 2017. }

\preprint{}
\maketitle




These lectures are meant to provide a brief overview of two topics: the standard (Hot Big Bang, or $\Lambda$CDM) model of cosmology and the inflationary universe that presently provides our best understanding of the standard cosmology's peculiar initial conditions. There are several goals to this presentation: the first of which is to provide a particle-physics audience with some of the tools required by later lecturers in this school. After all, cosmology has become a mainstream topic within particle physics, largely because cosmology provides several of the main pieces of observational evidence for the incompleteness of the Standard Model of particle physics. 

A second goal of these lectures is to touch on the important role played in cosmology by many of the same methods of effective field theory (EFT) used elsewhere in physics. This second goal is particularly important for the cosmology of the very early universe (such as inflationary or `bouncing' models) for which a central claim is that quantum fluctuations provide an explanation of the properties of primordial fluctuations presently found writ large across the sky. If true, this claim would imply not only that quantum-gravity effects\footnote{Here `quantum-gravity effects' means quantum phenomena associated with the gravitational field, rather than (the much more difficult) foundational issues about the nature of spacetime in a strongly quantum regime.} are observable; but that their imprint has already been observed cosmologically. Such claims sharpen the need to clarify what parameters control the size of quantum effects in gravity, and along the way more generally to identify the domain of validity of semi-classical methods in cosmology.

In practice the cosmology part of these lectures is divided into two parts: homogeneous, isotropic cosmologies and the fluctuations about them. The first part provides a very brief description of the classic homogeneous and isotropic cosmological models usually encountered in introductory cosmology courses. One goal of this section is to highlight  the great success these models have describing the universe as we find it around us. A second goal is to describe the peculiar initial conditions that are required by this observational success. This section then highlights how these puzzling initial conditions suggest the universe once underwent an earlier epoch of accelerated expansion. It closes by describing several simple and representative single-field inflationary models that have been proposed to provide this earlier accelerated epoch. 

The second part of the cosmology part of these lectures repeats the same picture, but now for fluctuations about both standard and inflationary cosmologies. This section starts by describing the very successful picture of structure formation within the standard $\Lambda$CDM model, in which both fluctuations in the cosmic microwave background (CMB) and the distribution of galaxies are attributed to the amplification by gravity of a simple primordial spectrum of small fluctuations. Again the success of standard cosmology proves to rely on a specific choice for the initial spectrum of primordial fluctuations, and again the required initial spectrum can be understood as being produced by quantum fluctuations if there were an earlier epoch of accelerated expansion. Accelerated expansion plays double duty: potentially both explaining the initial conditions of the background homogeneous universe and of the primordial spectrum of fluctuations within it. 

Because of the important role played by gravitating quantum fluctuations, EFT methods are central to assessing the domain of validity of the entire picture. Consequently the third, non-cosmology, section of these notes summarizes several of the ways they do so, and how their application can differ in cosmology from those encountered elsewhere in particle physics. This starts by extending standard power-counting arguments to identify the small parameters that control the underlying semiclassical expansion implicitly used in essentially all cosmological models. In my opinion it is the quality of this control over the semiclassical expansion that at present favours inflationary models over their alternatives (such as bouncing cosmologies).\footnote{Of course, although this might explain the current preference amongst cosmologists for inflationary models, it does not mean that Nature prefers them. Rather, EFT arguments just help set the standard to which formulations of alternative proposals should also aspire to achieve equal credence.} Other EFT topics discussed include several issues of principle to do with how to define and use EFTs in explicitly time-dependent situations, and how to quantify the robustness of inflationary predictions to any peculiarities of unknown higher-energy physics. 

Parts of these lectures draw on some of my earlier review articles \cite{Stringy, GREFT, GhostBusters}. Meant as a personal viewpoint about the field rather than a survey of the literature, these lecture notes include references that are not comprehensive (and I apologize in advance to the many friends whose work I inevitably have forgotten to include).

\section{Cosmology: Background}

This section summarizes the standard discussion of background cosmology, both for the $\Lambda$CDM model and its inflationary precursors.

\begin{figure}[h]
\begin{center}
\includegraphics[width=160mm,height=50mm]{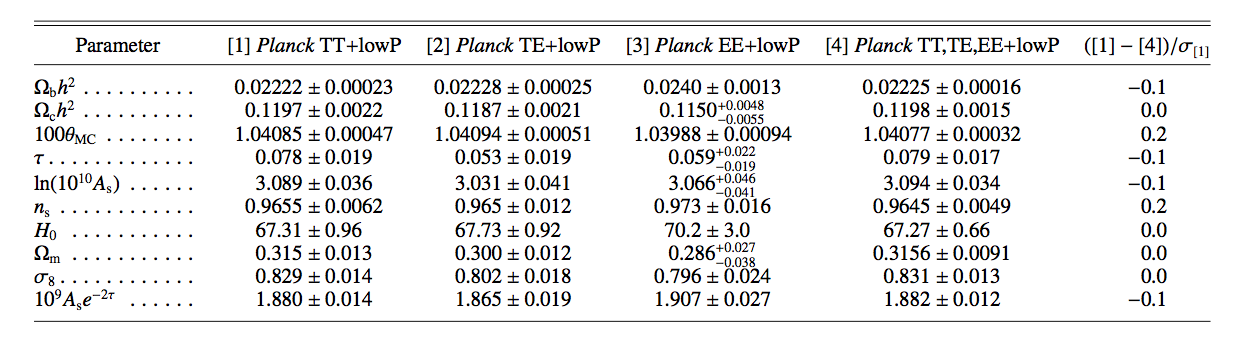} 
\caption{Summary of cosmological parameters (taken from \cite{Ade:2015xua}), as obtained by fits to various combination of data sets (all of which include the most detailed measurements of the properties of the CMB using the Planck satellite). The parameters described in the text are $H_0$ (present-day Hubble expansion rate, equal to $h$ in units of 100 km/sec/Mpc), $\Omega_b$ (baryon abundance), $\Omega_c$ (Dark Matter abundance) and $\Omega_m = \Omega_b + \Omega_c$ (nonrelativistic matter abundance), which describe the background cosmology, as well as $A_s$ and $n_s$ that describe properties of primordial fluctuations about this background. $\Omega_i := \rho_i/\rho_{\rm crit}$ is defined as the energy density in units of the critical density $\rho_{\rm crit} := 3H_0^2/(8\pi G)$, where $G$ is Newton's gravitational constant. Not discussed are $\theta_{{\scriptscriptstyle MC}}$ (the angular size of the sound horizon at last scattering), $\tau$ (the optical depth, which measures the amount of ionization in the later universe) and $\sigma_8$ (a measure of the amount of gravitational clustering).} \label{Fig:LCDMparams} 
\end{center}
\end{figure}

\subsection{Standard $\Lambda$CDM cosmology}

The starting point is the standard cosmology of the expanding universe revealed to us by astronomical observations. At present an impressively large collection of observations is described very well by the $\Lambda$CDM model, in terms of the handful of parameters listed in Fig.~\ref{Fig:LCDMparams}. The following sections aim to describe the model, and what some of these parameters mean.

\subsubsection{FRW geometries}

\begin{figure}[h]
\begin{center}
\includegraphics[width=80mm,height=60mm]{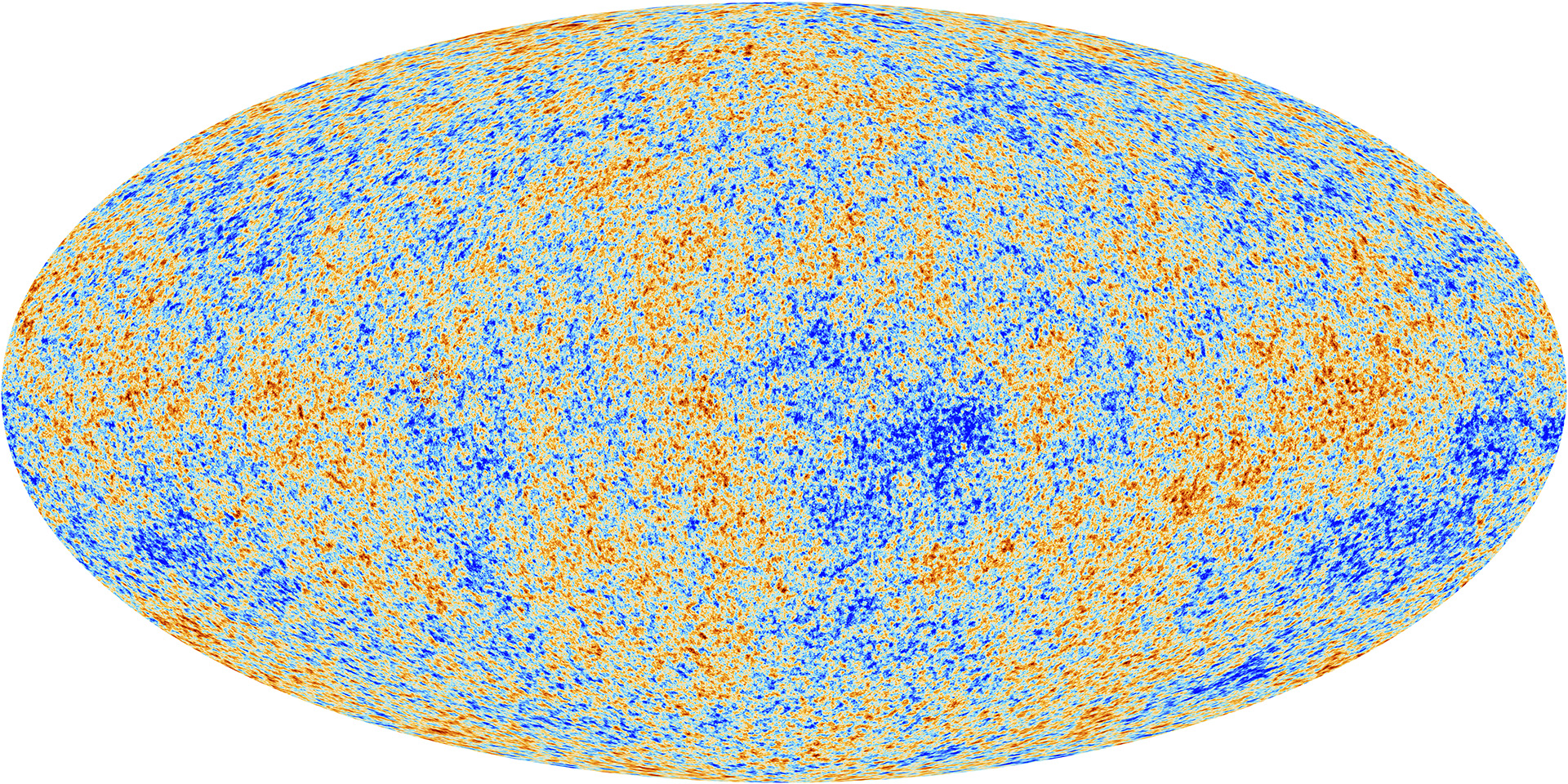} 
\caption{Temperature fluctuations in the CMB as a function of direction in the full sky (in galactic coordinates) as measured by the Planck collaboration \cite{Adam:2015rua}. The figure  subtracts foregrounds due to our galaxy and an order $\delta T/T \sim 10^{-3}$ dipole due to the Earth's motion through the CMB. The fluctuations that remain have a maximum amplitude of order $\delta T/T \sim 10^{-5}$. (Also, notice Stephen Hawking's initials just to the left of centre.) } \label{Fig:CMBmugshot} 
\end{center}
\end{figure}

Cosmology became a science once Einstein's discovery of General Relativity (GR) related the observed distribution of stress-energy to the measurable geometry of space-time. This implies the geometry of the universe as a whole can be tied to the overall distribution of matter at the largest scales. It used to be an article of faith that this geometry should be assumed to be homogeneous and isotropic (the `cosmological principle'), but these days it is pretty much an experimental fact that the stress-energy of the universe is homogeneous and isotropic on the largest scales visible. One piece of evidence to this effect is the very small --- one part in $10^5$ --- temperature fluctuations of the CMB, as shown in Fig.~\ref{Fig:CMBmugshot} (and more about which later).

On such large scales the geometry of space-time should therefore be homogeneous and isotropic, and the most general such a geometry in 3+1 dimensions is described by the Friedmann-Robertson-Walker (FRW) metric. The line-element for this metric can be written as\footnote{For those rusty on what a metric means and perhaps needing a refresher course on GR, an undergraduate-level introduction using the same conventions as used here can be found in {\em General Relativity: The Notes} at {\tt http/www.physics.mcmaster.ca/\~{}cburgess/Notes/GRNotes.pdf}.}
\bea \label{FRWMetric}
    \exd s^2 = g_{\mu\nu} \, \exd x^\mu \exd x^\nu &=& - \exd t^2 + a^2(t) \, \left[
    \frac{\exd r^2}{1 - \kappa
    r^2/R_0^2} + r^2 \, \exd\theta^2 + r^2 \sin^2\theta \,
    \exd \phi^2 \right] \\
    &=& - \exd t^2 + a^2(t) \, \left[
    \exd \ell^2 + r^2(\ell) \, \exd\theta^2 +
    r^2(\ell) \sin^2\theta \, \exd \phi^2 \right]
    \,, \nonumber
\eea
where $R_0$ is a constant with dimension length and $\kappa$ can take one of the following three values: $\kappa = 1,0,-1$. The coordinate $\ell$ is related to $r$ by $\exd \ell = \exd r/(1 - \kappa r^2/R_0^2)^{1/2}$, and so
\be  \label{rvsell}
    r(\ell) = \left\{ \begin{matrix}
    R_0 \, \sin (\ell/R_0) & \hbox{if} \quad \kappa = +1 \\
    \ell & \hbox{if}\quad \kappa = 0 \\
    R_0 \, \sinh (\ell/R_0) & \hbox{if}\quad \kappa = -1
    \end{matrix}
    \right. \,.
\ee

The quantity $a(t) R_0$ represents the radius of curvature of the spatial slices at fixed $t$, which are 3-spheres when $\kappa = +1$; 3-hyperbolae for $\kappa = -1$ and are flat for $\kappa = 0$. It is conventional to scale $R_0$ out of the metric by re-scaling the coordinates $\ell \to R_0\, \ell$ and $r \to R_0\, r$ while at the same time rescaling $a(t) \to a(t)/R_0$. This redefinition makes $r$ and $\ell$ dimensionless while giving $a(t)$ units of length, and it is often useful to choose cosmological units for which $a(t_0) = 1$ for some $t_0$ (such as at present). The case $\kappa = 0$ turns out to be of particular interest because all current evidence (coming, for instance, from the measured properties of the CMB) indicates that the spatial slices in the universe are consistent with being flat. 

Trajectories along which only $t$ varies are time-like geodesics of this metric and represent the motion of a natural set of static `co-moving' observers. The co-moving distance, $\Delta \ell$, between two such observers at a fixed time $t$ is related to their physical distance --- as measured by the metric \pref{FRWMetric} --- by
\be
    D(\Delta\ell,t) = \Delta \ell \, a(t) \,,
\ee
so the `scale-factor' $a(t)$ describes the common time-evolution of spatial scales. So long as $a(t)$ is monotonic one can use $t$ or $a$ interchangeably as measures of the passage of time. 

The trajectories of photons play a special role in cosmology since until very recently they brought us all of our information about the universe at large. Since they move at the speed of light their trajectories satisfy $\exd s^2 = 0$ and so
\be
    g_{\mu\nu} \left( \frac{\exd x^\nu}{\exd s} \right)
    \left( \frac{\exd x^\nu}{\exd s} \right) = 0 \,,
\ee
which for radial motion specializes to $\exd t/\exd s = \pm a(t) (\exd \ell/\exd s)$. Choosing coordinates that place us at the origin means all photons sent to us move along a radial trajectory. 

A photon arriving at $t=0$ from a galaxy situated at fixed co-moving position $\ell = L$ must have departed at time $t = -T$ where
\be \label{LvsTforPhotons}
    L = \int_0^T \frac{\exd t}{a(t)} \,.
\ee
Since the universe expands by an amount $a_0/a$ in this time (where $a_0 = a(0)$ is the present-day scale factor and $a = a(-T)$ is its value when the light was emitted), the redshift, $z$, of the light is given by $z := (\lambda_{\rm obs} - \lambda_{\rm em})/\lambda_{\rm em}$, with $\lambda_{\rm obs}/\lambda_{\rm em} = a_0/a$. Consequently $z$ and $a$ are related by
\be \label{FRWRedshiftFormula}
    1 + z =  \frac{a_0}{a} \,. 
\ee
This very usefully ties the universal expansion to the more easily measured redshift of distant objects.\footnote{In practice the redshift of any particular object depends also on its `peculiar' motion relative to co-moving observers, but because peculiar-motion effects are smaller than the cosmic redshift for all but relatively nearby galaxies they are ignored in what follows.}

For later purposes, it is worth introducing another useful time coordinate when discussing the evolution of light rays in FRW geometries. Defining `conformal time', $\tau$, by
\be
  \tau = \int \frac{\exd t}{a(t)} \,,
\ee
allows the metric \pref{FRWMetric} to be written
\be \label{ConformalCoords}
 \exd s^2 = a^2(\tau) \, \left[ - \exd \tau^2 +
    \exd \ell^2 + r^2(\ell) \, \exd\theta^2 +
    r^2(\ell) \sin^2\theta \, \exd \phi^2 \right] \,.
\ee
The utility of this coordinate system is that the scale-factor $a(\tau)$ completely drops out of the evolution of photons, which simplifies the identification of many of the causal properties of the spacetime ({\em i.e.} identifying which events can communicate with each other by exchanging photons).

\subsubsection{Implications of Einstein's equations}

So far so good, but the story so far is largely just descriptive. The FRW metric, with $a(t)$ specified, says much about how particles move over cosmological distances. But we also need to know how to relate $a(t)$ to the universe's stress-energy content. This connection is made using Einstein's equations,\footnote{These notes use the metric signature $(-+++)$ as well as units with $\hbar = c = k_B = 1$, and conform to the widely used MTW curvature conventions \cite{MTW} (which differ from Weinberg's conventions \cite{GnC} -- that are often used in my own papers -- only by an overall sign for the Riemann curvature tensor, ${R^\mu}_{\nu\lambda\rho}$). The world divides into two camps regarding the metric signature, with most relativists and string theorists using $(-+++)$ and many particle phenomenologists using $(+---)$. As students just forming your own habits now, you should choose one and stick to it. When doing so keep in mind that the $(+---)$ metric becomes the $(----)$ metric in Euclidean signature (such as arises when Wick rotating or for applications at finite temperature), leading to many headaches keeping track of signs because all vector norms become negative: $V^2 = g_{mn} V^m V^n < 0$. Your notation should be your friend, not your adversary.}
\be
  R_{\mu\nu} - \frac12 \, R \, g_{\mu\nu} = 8\pi G \, T_{\mu\nu} \,,
\ee
where $G$ is Newton's constant of universal gravitation, $R_{\mu\nu} = {R^\alpha}_{\mu\alpha\nu}$ is the geometry's Ricci tensor (where ${R^\alpha}_{\mu\beta\nu}$ is its Riemann tensor) and $R = g^{\mu\nu} R_{\mu\nu}$. 

The twin requirements of homogeneity and isotropy dictate that the most general form for the universe's stress-energy tensor, $T_{\mu\nu}$, is that of a perfect fluid,
\be
 T_{\mu\nu} = p \, g_{\mu\nu} + (p + \rho) \, U_\mu U_\nu \,,
\ee
where $p$ and $\rho$ are respectively the fluid's pressure and energy density, while $U^\mu \partial_\mu = \partial_t$ (or, equivalently, $U_\mu \, \exd x^\mu = -\exd t$) is the 4-velocity of the co-moving observers. 

Specialized to the metric \pref{FRWMetric} the Einstein equations boil down to the following two independent equations:
\be \label{FriedmannEqn}
  H^2 + \frac{\kappa}{a^2} = \frac{8\pi G}3 \; \rho = \frac{\rho}{3M_p^2} \qquad \hbox{(Friedmann equation)}
\ee
and
\be \label{EnergyConservationEqn}
  \dot \rho + 3 H (p + \rho) = 0 \qquad\qquad \hbox{(energy conservation)} 
\ee
where $G$ is Newton's gravitational constant and over-dots denote differentiation with respect to $t$ and the Hubble function is defined by $H = \dot a/a$. The last equality in eq.~\pref{FriedmannEqn} also defines the `reduced' Planck mass: $M_p = (8\pi G)^{-1/2} \simeq 10^{18}$ GeV. Differentiating \pref{FriedmannEqn} and using \pref{EnergyConservationEqn} gives a useful formula for the cosmic acceleration
\be \label{AccEq}
 \frac{\ddot a}{a} = -\; \frac{1}{6M_p^2} (\rho + 3 p ) \,.
\ee

Mathematically speaking, finding the evolution of the universe as a function of time requires the integration of
eqs.~\pref{FriedmannEqn} and \pref{EnergyConservationEqn}, but in themselves these two equations are inadequate to determine the
evolution of the three unknown functions, $a(t)$, $\rho(t)$ and $p(t)$. Another condition is required in order to make the problem well-posed. The missing condition is furnished by the equation of state for the matter in question, which for the present purposes we take to be an expression for the pressure as a function of energy density, $p = p(\rho)$. In particular, the equations of state of interest in $\Lambda$CDM cosmology have the general form
\be \label{EqnofState}
    p = w \, \rho \,,
\ee
where $w$ is a $t$-independent constant. 

The first step in solving for $a(t)$ is to determine how $p$ and $\rho$ depend on $a$, since this is dictated by energy conservation. Using eq.~\pref{EqnofState} in \pref{EnergyConservationEqn} allows it to be integrated to obtain
\be \label{rhovsa}
    \rho = \rho_0 \left( \frac{a_0}{a} \right)^{\sigma}
    \qquad \hbox{with} \quad \sigma = 3(1+w) \,.
\ee
Eq.~\pref{EqnofState} implies the pressure also shares this same dependence on $a$. Similarly using eq.~\pref{rhovsa} to eliminate $\rho$ from \pref{FriedmannEqn} leads to the following differential equation for $a(t)$:
\be
    \dot{a}^2 + \kappa = \frac{8 \pi G \rho_0 a_0^2}{3} \left( \frac{a_0}{a}
    \right)^{\sigma-2} \,.
\ee
When $\kappa = 0$ (and $w \ne -1$) this equation is easily integrated to give
\be
    a(t) = a_0 \left( \frac{t}{t_0} \right)^\alpha \qquad
    \hbox{with} \quad \alpha = \frac{2}{\sigma} = \frac{2}{3(1+w)}
    \,.
\ee

\subsubsection{Equations of state}

In the $\Lambda$CDM model of cosmology the total energy density is regarded as the sum of several components, each of which separately satisfies one of the following three basic equations of state.

\medskip\noindent {\bf Nonrelativistic matter}

\medskip\noindent
An ideal gas of non-relativistic particles in thermal equilibrium has a pressure and energy density given by
\be
    p = n \, T \qquad \hbox{and} \qquad
    \rho = n \, m + \frac{n \, T}{\gamma - 1} \,,
\ee
where $n$ is the number of particles per unit volume, $m$ is the particle's rest mass and $\gamma = c_p/c_v$ is its ratio of specific heats, with $\gamma = 5/3$ for a gas of monatomic atoms. For non-relativistic particles the total number of particles is usually also conserved,\footnote{The total {\em difference} between the number of nonrelativistic particles and their antiparticles can be constrained to be nonzero if they carry a conserved charge (such as baryon number, for protons and neutrons). In the absence of such a charge the density of such particles becomes quite small if they remain in thermal equilibrium since their abundance becomes Boltzmann suppressed, $n \propto e^{-m/T}$, at temperatures $T < m$. This suppression happens because the annihilation of particles and antiparticles is not compensated by their pair-production due to there being insufficient thermal energy.} which implies that
\be \label{NConservationEqn}
    \frac{\exd}{\exd t} \Bigl[ n \, a^3 \Bigr] = 0 \,.
\ee

Since $m \gg T$ (or else the atoms would be relativistic) the equation of state for this gas may be taken to be
\be
   \frac{ p }{\rho} \sim \frac{T}{m} \ll 1 \qquad \hbox{and so} \qquad
    w \simeq 0 \,.
\ee
Since $w \simeq 0$ energy conservation implies $\sigma = 3(1+w) \simeq 3$ and so $\rho \, a^3$ is a constant. This is appropriate for nonrelativistic matter for which the energy density is dominated by the particle rest-masses, $\rho \simeq n \, m$, because in this case energy conservation is equivalent to conservation of particle number which according to \pref{NConservationEqn} implies $n \propto a^{-3}$. 

Finally, whenever the total energy density is dominated by non-relativistic matter we know $w = 0$ also implies $\alpha = 2/\sigma = 2/3$ and so if $\kappa = 0$ then the universal scale factor expands like $a \propto t^{2/3}$. 

\medskip\noindent {\bf Radiation}

\medskip\noindent
Thermal equilibrium dictates that a gas of relativistic particles (like photons) must have an energy density and pressure given by
\be
    \rho = a_B \, T^4 \qquad \hbox{and} \qquad
    p = \frac13 \, a_B \, T^4 \,,
\ee
where $a_B = \pi^2/15 \simeq 0.6580$ is the Stefan-Boltzmann constant (in units where $k_B = c = \hbar = 1$) and $T$ is the temperature. Together, these ensure that $\rho$ and $p$ satisfy the equation of state
\be \label{Radw}
    p = \frac13 \, \rho \qquad \hbox{and so} \qquad w = \frac13 \,.
\ee
Eq.~\pref{Radw} also applies to any other particle whose temperature dominates its rest mass, and so in particular applies to neutrinos for most of the universe's history.

Since $w = 1/3$ it follows that $\sigma = 3(1+w) = 4$ and so $\rho \propto a^{-4}$. This has a simple physical interpretation for a gas of noninteracting photons, since for these the total number of photons is fixed and so $n_\gamma \propto a^{-3}$. But each photon energy is inversely proportional to its wavelength and so also redshifts like $1/a$ as the universe expands, leading to $\rho_\gamma \propto a^{-4}$.

Whenever radiation dominates the total energy density then $w = 1/3$ implies $\alpha = 2/\sigma = 1/2$, and so if $\kappa = 0$ then $a(t) \propto t^{1/2}$. 

\medskip\noindent {\bf The vacuum}

\medskip\noindent
If the vacuum is Lorentz invariant --- as the success of special relativity seems to indicate --- then its stress-energy must satisfy $T_{\mu\nu}  \propto g_{\mu\nu}$. This implies the vacuum pressure must satisfy the only possible Lorentz-invariant equation of state:
\be \label{VacEOS}
    p = - \rho \qquad \hbox{and so} \qquad w = -1 \,.
\ee
Furthermore, for $T_{\mu\nu} = - \rho \, g_{\mu\nu}$ stress-energy conservation, $\nabla^\mu T_{\mu\nu} = 0$, implies $\rho$ must be spacetime-independent (in agreement with \pref{rhovsa} for $w = -1$). This kind of constant energy density is often called, for historical reasons, a cosmological constant. 

Although counter-intuitive, constant energy density can be consistent with energy conservation in an expanding universe. This is because \pref{EnergyConservationEqn} implies the total energy satisfies $\exd(\rho \, a^3)/\exd t = - p \, \exd(a^3)/\exd t$. Consequently the equation of state \pref{VacEOS} ensures the pressure does precisely the amount of work required to produce the change in total energy required by having constant energy density.

When the vacuum dominates the energy density then $\alpha = 2/\sigma \to \infty$, which shows that the power-law solutions, $a \propto t^\alpha$, are not appropriate. Returning directly to the Friedmann equation, eq.~\pref{FriedmannEqn}, shows (when $\kappa = 0$) that $H = \dot{a} / a$ is constant and so the solutions are exponentials: $a \propto \exp[ \pm H (t-t_0)]$. Notice that \pref{VacEOS} implies $\rho + 3 p$ is negative if $\rho$ is positive. This furnishes an explicit example of an equation of state for which the universal acceleration, $\ddot{a}/a = - \frac43 \,\pi G(\rho + 3p)$, can be positive.

\subsubsection{Universal energy content}

At present there is direct observational evidence that the universe contains at least four independent types of matter, whose properties are now briefly summarized. 

\medskip\noindent{\bf Radiation}

\medskip\noindent
The universe is known to be awash with photons, and is also believed to contain similar numbers of neutrinos (that until very recently\footnote{Although neutrino masses play an important role in some things (like the formation of galaxies and other structure), I lump them here with radiation because for most of what follows the fact that they very recently likely became nonrelativistic does not matter.} could also be considered to be radiation). 

\medskip\noindent{\it Cosmic Photons:} 

\smallskip\noindent
The most numerous type of photons found at present in the universe are the photons in the cosmic microwave background (CMB). These are distributed thermally in energy with a temperature that is measured today to be $T_{\gamma 0} = 2.725$ K. The present number density of these CMB photons is determined by their temperature to be
\be
    n_{\gamma 0} = 4.11 \times 10^8 \; \hbox{m}^{-3} \,,
\ee
which turns out to be much higher than the average number density of ordinary atoms. Their present energy density (also determined by their temperature) is
\be
    \rho_{\gamma 0} = 0.261 \; \hbox{MeV m}^{-3} \qquad
    \hbox{or} \qquad \Omega_{\gamma 0} = 5.0
    \times 10^{-5} \,,
\ee
where $\Omega_{\gamma 0}:= \rho_{\gamma 0}/\rho_{c0}$ defines the fraction of the total energy density (also the `critical' density, $\rho_{c0} \simeq 5200$ MeV m${}^{-3} \simeq 10^{-29}$ g cm${}^{-3}$) currently residing in CMB photons.

\medskip\noindent {\it Relict Neutrinos:} 

\smallskip\noindent
It is believed on theoretical grounds that there are also as many cosmic relict neutrinos as there are CMB photons running around the universe, although these neutrinos have never been detected. They are expected to have been relativistic until relatively recently in cosmic history, and to be thermally distributed. The neutrinos are expected to have a slightly lower temperature, $T_{\nu 0} = 1.9$ K, and are fermions and so have a slightly different energy-density/temperature relation than do neutrinos.

Their contribution to the present-day cosmological energy budget is not negligible, and if they were massless would be predicted to be
\be
    \rho_{\nu 0} = 0.18 \; \hbox{MeV m}^{-3} \qquad
    \hbox{or} \qquad \Omega_{\nu 0} = 3.4
    \times 10^{-5} \,,
\ee
leading to a total radiation density, $\Omega_{R0} = \Omega_{\gamma 0} + \Omega_{\nu 0}$, of size
\be
    \rho_{R 0} = 0.44 \; \hbox{MeV m}^{-3} \qquad
    \hbox{or} \qquad \Omega_{r 0} = 8.4 \times 10^{-5} \,.
\ee 

\noindent{\bf Baryons}

\medskip\noindent
The main constituents of matter we see around us are atoms, made up of protons, neutrons and electrons, and these are predominantly non-relativistic at the present epoch. Although some of this material is now in gaseous form much of it is contained inside larger objects, like planets or stars. But the earlier universe was more homogeneous and at these times atoms and nuclei would have all been uniformly spread around as part of the hot primordial soup. (At least, this is the working hypothesis of the very successful Hot Big Bang picture.)

The absence of anti-particles in the present-day universe indicates that the primordial soup had an over-abundance of baryons ({\em i.e.}~protons and neutrons) relative to anti-baryons. The same is true of electrons, whose total abundance is also very likely precisely equal to that of protons in order to ensure that the universe carries no net electric charge.

Since protons and neutrons are about 1840 times more massive than electrons, the energy density in ordinary non-relativistic particles is likely to be well approximated by the total energy in baryons. It turns out it is possible to determine the total number of baryons in the universe (regardless of whether or not they are presently visible), in several independent ways. 

\begin{figure}[h]
\begin{center}
\includegraphics[width=100mm,height=80mm]{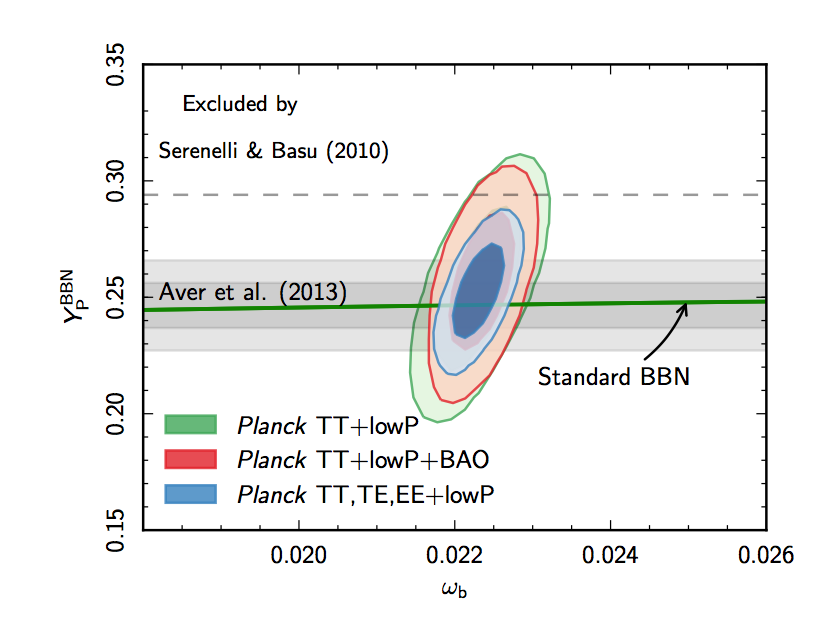} 
\caption{The universal baryon abundance as inferred from CMB measurements and from Big-Bang nucleosynthesis (BBN) calculations, taken from \cite{Ade:2015xua}. $Y_p$ denotes the primordial He/H abundance while $\omega_b = \Omega_b h^2$ represents the universal baryon fraction (with $h$ as in Fig. \ref{Fig:LCDMparams}).} \label{Fig:CMBvsBBN} 
\end{center}
\end{figure}

One way to determine the baryon density uses measurements of the properties of the CMB, whose understanding depends on things like the speed of sound or on reaction rates  -- and so also on the density -- for the Hydrogen (and some Helium) gas from which the CMB photons last scattered \cite{Ade:2015xua}. Another way uses the success of the predictions for the abundances of light elements as nuclei formed during the very early universe, which depends on nuclear reaction rates -- again proportional to the total nucleon density. A determination of the baryon abundance as inferred from the primordial He/H abundance ratio (measured from the CMB and from nuclear calculations of primordial element abundances -- or Big Bang Nucleosynthesis) is given in Figure \ref{Fig:CMBvsBBN}.

These two kinds of inferences are consistent with each other and indicate the total energy density in baryons is\footnote{$\Omega_{B0}$ is what is denoted $\Omega_b$ in the table in Figure \ref{Fig:LCDMparams}.}
\be
    \rho_{B 0} = 210 \; \hbox{MeV m}^{-3} \qquad
    \hbox{or} \qquad \Omega_{B 0} = 0.04 \,.
\ee
For purposes of comparison, this is about ten times larger than the amount of {\sl luminous} matter, found using the luminosity density for galaxies, $n L = 2 \times 10^8$ $L_\odot$ Mpc${}^{-3}$, together with the best estimates of the average mass-to-luminosity ratio of for galactic matter: $M/L \simeq 4 M_\odot/L_\odot$.

It should be emphasized that although there is more energy in baryons than in CMB photons, the {\sl number density} of baryons is much smaller, since
\be
    n_{B 0} = \frac{210 \; \hbox{MeV m}^{-3}}{940 \; \hbox{MeV}} =
    0.22 \; \hbox{m}^{-3} = 5 \times 10^{-10} \, n_{\gamma 0} \,.
\ee

\medskip\noindent{\bf Dark Matter}

\medskip\noindent
There are several lines of evidence pointing to the large-scale presence of another form of non-relativistic matter besides baryons, carrying much more energy than do the baryons. Part of the evidence for this so-called {\sl Dark Matter} comes from a variety of independent ways of measuring of the total amount of gravitating mass in galaxies and in clusters of galaxies. 

The differential rotation rate of numerous galaxies as a function of their radius indicates there is considerably more gravitating mass present than would be inferred by counting the luminous matter which can be seen. Furthermore, the motion of Hydrogen gas clouds and other things orbiting these galaxies indicates this mass is distributed well outside of the radius of the visible stars. 

Similarly, the total mass contained within clusters of galaxies appears to be much more than is found when adding up what is visible. This is equally true when galaxy-cluster masses are estimated using the motions of their constituent galaxies, or from the temperature of their hot inter-galactic gas or from the amount of gravitational lensing they produce when they are in the foreground of even more distant objects. 

Whatever it is, this matter should be non-relativistic since it takes part in the gravitational collapse which gives rise to galaxies and their clusters. (Relativistic matter tends not to cluster in this way, as is seen in later sections.)

All of these estimates appear to be consistent with one another, and with several independent ways of measuring energy density in cosmology (more about which below). They indicate a non-relativistic matter density of order 
\be
    \rho_{DM 0} = 1350 \; \hbox{MeV m}^{-3} \qquad
    \hbox{or} \qquad \Omega_{DM 0} = 0.26 \,.
\ee
The errors in this inference of the size of $\Omega_{DM 0}$ can be seen in Fig.~\ref{Fig:LCDMparams} (where $\Omega_{DM0}$ is denoted $\Omega_c$). Provided this has the same equation of state, $p \approx 0$, as have the baryons (as is assumed in the $\Lambda$CDM model), this leads to a total energy density in non-relativistic matter, $\Omega_{M 0} = \Omega_{B 0} + \Omega_{DM 0}$, which is of order
\be
    \rho_{M 0} = 1600 \; \hbox{MeV m}^{-3} \qquad
    \hbox{or} \qquad \Omega_{m 0} = 0.30 \,.
\ee
In Fig.~\ref{Fig:LCDMparams} the quantity $\Omega_{m0}$ is denoted $\Omega_m$.

\medskip\noindent{\bf Dark Energy}

\medskip\noindent
Finally, there are also at least two lines of evidence pointing to a second form of unknown matter in the universe, independent of the Dark Matter. One line is based on the recent observations that the universal expansion is accelerating, and so requires the universe must now be dominated by a form of matter for which $\rho + 3p < 0$. Whatever this is, it cannot be Dark Matter since the evidence for Dark Matter shows it to gravitate similarly to nonrelativistic matter. 

The second line of argument is based on the observational evidence about the spatial geometry of the universe, which favours the universe being spatially flat, $\kappa = 0$. (The evidence for spatial flatness comes from measurements of the angular fluctuations in the temperature of the CMB, since the light we receive from the CMB knows about the geometry of the intervening space through which it passed to get here.) These two lines of evidence are consistent with one another (within sizeable errors) and point to a {\sl Dark Energy} density that is of order
\be
    \rho_{DE 0} = 3600 \; \hbox{MeV m}^{-3} \qquad
    \hbox{or} \qquad \Omega_{DE 0} = 0.70 \,.
\ee

\begin{figure}[h]
\begin{center}
\includegraphics[width=80mm,height=60mm]{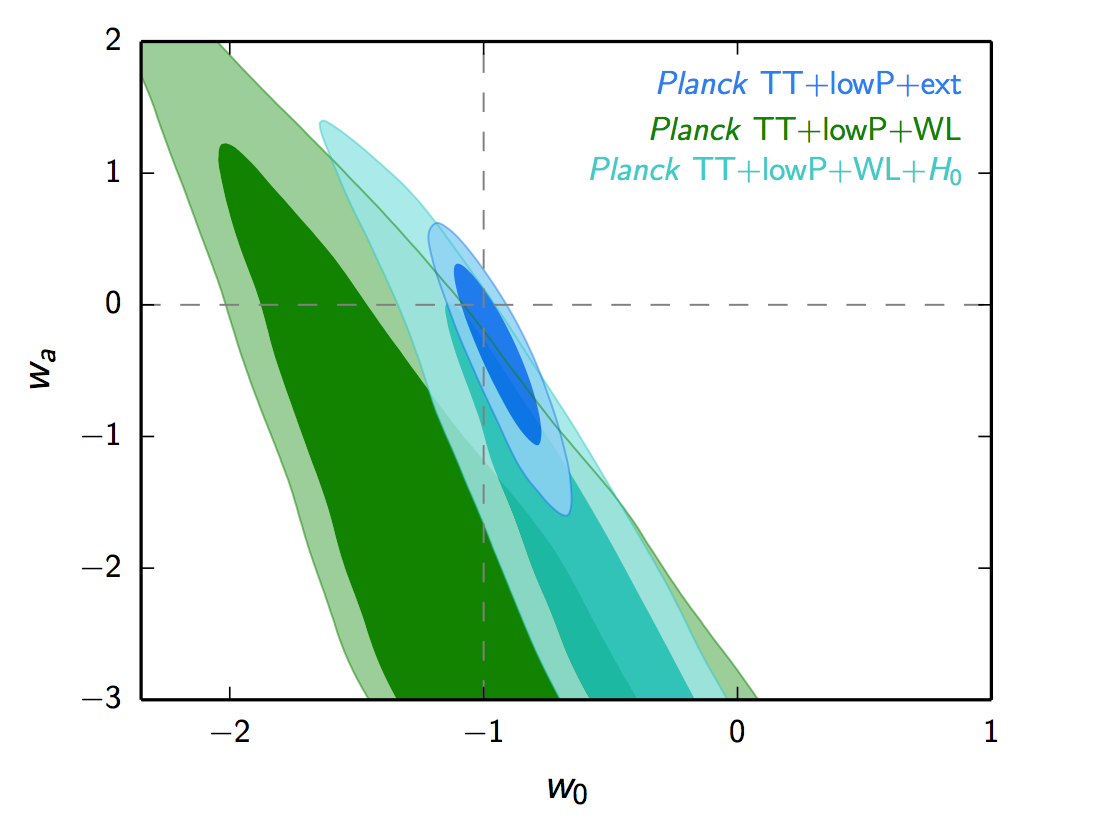} 
\caption{Inferences for the Dark Energy equation of state parameters, $w_0 = w(a_0)$, as compared with its present-day rate of change, $w_a = (\exd w/\exd a)_0$, as given in \cite{Ade:2015xua}. These show broad agreement with a vacuum-energy equation of state, $w = -1$, independent of $a$ (though within relatively large errors). No well-understood examples of stable matter with $w < -1$ are known, but the broader question of whether they might exist remains at present controversial. } \label{Fig:wvswa} 
\end{center}
\end{figure}

The equation of state for the Dark Energy is only weakly constrained, with observations requiring at present both $\rho_{DE 0} \sim 0.7 \, \rho_c > 0$ and $w \lsim -0.7$. The best evidence says $w$ is not changing with time right now, though within large errors. 
The strength of this evidence is shown in Fig.~\ref{Fig:wvswa}, which compares best-fit present-day values for $w$ (called $w_0$ in the figure) and $w_a = \exd w/\exd a$.

If $w$ really is constant it is plausible on theoretical grounds that $w = -1$ and the Dark Energy is simply the Lorentz-invariant vacuum energy density. Although it is not yet known whether the vacuum need be Lorentz invariant to the precision required to draw cosmological conclusions of sufficient accuracy, in the $\Lambda$CDM model it is assumed that the Dark Energy equation of state is $w = -1$.

\subsubsection{Earlier epochs}

Given the present-day cosmic ingredients of the previous section, it is possible to extrapolate their relative abundances into the past in order to estimate what can be said about earlier cosmic environments. This evolution can be complicated when the various components of the cosmic fluid significantly interact with one another (such as for baryons and photons at redshifts larger than about $z \simeq 1100$, as it turns out), but simplifies immensely if the various components of the cosmic fluid do not exchange stress-energy directly with one another. The $\Lambda$CDM model assumes there is no such direct energy exchange between other components and the dark matter and dark energy, and that no exchange exists between the two dark components. 

\begin{figure}[h]
\begin{center}
\includegraphics[width=120mm,height=100mm]{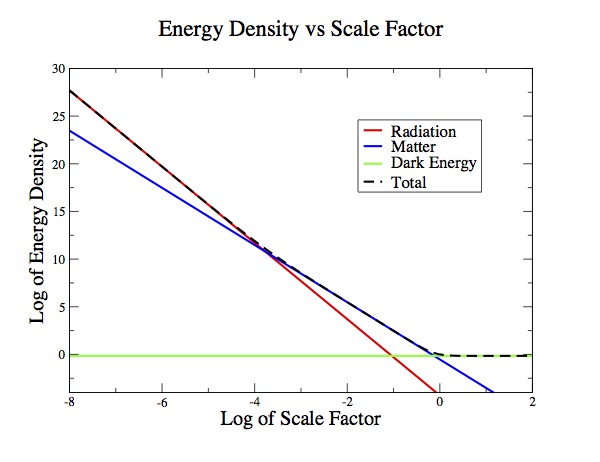} 
\caption{The relative abundance (in the energy density) of radiation, nonrelativistic matter and vacuum energy, vs the size of the universe $a/a_0 = (1+z)^{-1}$. The figure assumes negligible direct energy transfer between these fluids, and shows how this implies each type of fluid dominates during particular epochs. The transition from radiation to matter domination (at redshift $z_{\rm eq} \simeq 3600$) plays an important role in the development of structure in the universe.} \label{FigureEnergy} 
\end{center}
\end{figure}

When the component fluids do not directly exchange stress-energy things simplify because eq.~\pref{EnergyConservationEqn} applies separately to each component individually, dictating the dependence $\rho_i(a)$ and $p_i(a)$ for each of them, as follows:
\begin{itemize}
\item {\bf Radiation:} For photons (and relict neutrinos of sufficiently small mass compared with temperature) we have $w = 1/3$ and so
$\rho(a)/\rho_0 = (a_0/a)^4$;
\item {\bf Non-relativistic Matter:} For both ordinary matter (baryons and electrons) and for the Dark Matter we have $w = 0$ and so $\rho(a)/\rho_0 = (a_0/a)^3$;
\item {\bf Vacuum Energy:} Assuming the Dark Energy has the equation of state $w = -1$ we have $\rho(a) = \rho_0$ for all $a$.
\end{itemize}
This implies the total energy density and pressure have the form
\bea \label{MixedFluidrhoandp}
    \rho(a) &=& \rho_{DE 0} + \rho_{M 0} \left( \frac{a_0}{a}
    \right)^3 + \rho_{R 0} \left( \frac{a_0}{a} \right)^4 \nn\\
    p(a) &=& - \rho_{DE 0} + \frac13 \, \rho_{R 0} \left(
    \frac{a_0}{a} \right)^4 \,,
\eea
showing how the relative contribution of each component within the total cosmic fluid changes as it responds differently to the expansion of the universe (see Fig. \ref{FigureEnergy}).

As the universe is run backwards to smaller sizes it is clear that the Dark Energy becomes less and less important, while relativistic matter becomes more and more important. Although the Dark Energy is now the dominant contribution to $\rho$ and non-relativistic matter is the next most abundant, when extrapolated backwards they switch roles, so $\rho_{M}(a) > \rho_{DE}(a)$, relatively recently, at a redshift
\be
    1+z = \frac{a_0}{a} > \left( \frac{\Omega_{DE 0}}{\Omega_{M 0}}
    \right)^{1/3} \simeq \left( \frac{0.7}{0.3} \right)^{1/3} \simeq 1.3 \,.
\ee
In the absence of Dark Matter the energy density in baryons alone would become larger than the Dark Energy density at a slightly earlier epoch
\be
    1+z > \left( \frac{\Omega_{DE 0}}{\Omega_{B 0}}
    \right)^{1/3} \simeq \left( \frac{0.7}{0.04} \right)^{1/3} \simeq 2.6 \,.
\ee

For times earlier than this the dominant component of the energy density is due to non-relativistic matter, and this remains true back until the epoch when the energy density in radiation became comparable with that in non-relativistic matter. Since $\rho_R \propto a^{-4}$ and $\rho_M \propto a^{-3}$ radiation-matter equality occurs when $z = z_{\rm eq}$ with
\be \label{DMcross}
    1+z_{\rm eq}  = \frac{\Omega_{M 0}}{\Omega_{R 0}}
     \simeq \frac{0.3}{8.4 \times 10^{-5}} \simeq 3600 \,.
\ee
This crossover would have occurred much later in the absence of Dark Matter, since the radiation energy density equals the energy density in baryons when
\be \label{Bcross}
    1+z  = \frac{\Omega_{B 0}}{\Omega_{R 0}}
    \simeq \frac{0.04}{8.4 \times 10^{-5}} \simeq 480 \,.
\ee

Using the dependence of $\rho$ on $a$ in the Friedmann equation then gives $H$ as a function of $a$
\be \label{MixedFluidHvsa}
    H(a) = H_0 \left[ \Omega_{DE 0} + \Omega_{\kappa 0}
    \left( \frac{a_0}{a} \right)^2 + \Omega_{M 0}
    \left( \frac{a_0}{a} \right)^3 + \Omega_{R 0}
    \left( \frac{a_0}{a} \right)^4\right]^{1/2} \,, \ee
where (as before) $\Omega_f = \rho_f/\rho_c$ for $f = \hbox{radiation($R$), matter ($M$) or vacuum ($DE$)}$. The critical density is defined by $\rho_c := 3H^2M_p^2$ and the subscript `0' denotes the present epoch. Finally, eq.~\pref{MixedFluidHvsa} defines the curvature contribution to $H$ as
\be
    \Omega_{\kappa } :=  - \, \frac{\kappa}{(H a)^2} \,.
\ee

\begin{figure}[h]
\begin{center}
\includegraphics[width=80mm,height=60mm]{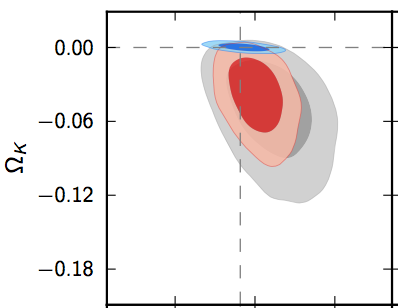} \\
\hspace*{0.8in}\includegraphics[width=60mm,height=15mm]{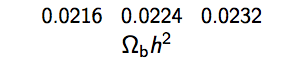} 
\caption{Constraints on the universe's spatial curvature, $\Omega_\kappa$, and its baryon abundance, $\Omega_b$, obtained from CMB observations \cite{Ade:2015xua}. As usual $h = H/(100 $km/sec/Mpc)$\simeq 0.67$. Different coloured ellipses correspond to fits with different combinations of data sets.} \label{FigureOmegaKFit} 
\end{center}
\end{figure}

As mentioned earlier, observations of the CMB constrain the present-day value for $\Omega_{\kappa 0}$, because they tell us about the overall geometry of space through which photons move on their way to us from where they were last scattered by primordial Hydrogen. These observations indicate the universe is close to spatially flat ({\em i.e.}~that $\kappa$ is consistent with $0$). Quantitatively, these CMB observations tell us that $\Omega_{\kappa0}$ is at most of order $10\%$, and so the Friedmann equation implies $\Omega_0 = \Omega_{DE 0} + \Omega_{m 0} + \Omega_{r0} \simeq 1$ (and so $\rho_{DE} + \rho_m + \rho_r = \rho_c$). Joint constraints on $\Omega_\kappa$ and $\Omega_b = \Omega_{B0}$ are shown in Figure \ref{FigureOmegaKFit}.

Because $\kappa/a^2$ falls more slowly with increasing $a$ than does either $\rho_m \propto 1/a^3$ or $\rho_r \propto 1/a^4$, the relatively small size of $\Omega_{\kappa 0}$ implies the $\kappa/a^2$ term contributes negligibly the further one moves to the remote past. This makes it a very good approximation to take $\kappa = 0$ when discussing the very early universe.

In principle \pref{MixedFluidHvsa} can be inserted into the Friedmann equation and integrated to obtain $a(t)$. Although in general this dependence must be obtained numerically, many of its features follow on simple analytic grounds because for most epochs there is only a single component of the cosmic fluid that dominates the total energy density. For instance, for redshifts larger than several thousand $a(t) \propto t^{1/2}$ should be a good approximation, as appropriate for the expansion in a universe filled purely by radiation. 

Once $a_0/a$ falls below $3600$ there should be a brief transition to the time dependence appropriate for a universe dominated by non-relativistic matter and so $a \propto t^{2/3}$. This should apply right up to the very recent past, when $a/a_0$ is around 0.8, after which there is a transition to vacuum-energy domination, during which the universal expansion accelerates to become exponential with $t$. 

In all likelihood we are at present still living in the transition period from matter to vacuum-energy domination. And when $\kappa = 0$ it is also possible to give simple analytic expressions for the time dependence of $a$ in transition regions like this. Neglecting radiation during the matter/dark-energy transition gives a Friedmann equation of the form
\be \label{mccFE}
 \left( \frac{\dot a}{a} \right)^2
 = H^2_{\scriptscriptstyle DE} \left[ 1 + \left( \frac{a_{e}}{a}
 \right)^3 \right] \,,
\ee
where $H_{\scriptscriptstyle DE}^2 = 8 \pi G \rho_{\scriptscriptstyle DE}/3$ is the (constant) Hubble
scale during the pure dark-energy epoch and $a_{e}$ is the value of the scale factor when the energy densities of the matter and dark energy are equal to one another. Integrating this equation (assuming $\dot a > 0$), with the boundary condition that $a = 0$ when $t = 0$ then gives the solution
\be \label{mccsoln}
 a(t) = a_0 \sinh^{2/3}\left( \frac{3 H_{\scriptscriptstyle DE} t}{2} \right) \,,
\ee
where $a_0$ is a constant. Notice that this approaches $a/a_0 \propto \exp(H_{de}t)$ if $H_{\scriptscriptstyle DE} \,t \gg 1$, as appropriate for Dark Energy domination, while for $H_{\scriptscriptstyle  DE} \,t \ll 1$ it instead becomes $a/a_0 \propto t^{2/3}$, as appropriate for a matter-dominated epoch.

\subsubsection{Thermal evolution}

The Hot Big Bang theory of cosmology starts with the idea that the universe was once small and hot enough that it contained just a soup of elementary particles, in order to see if this leads to a later universe that we recognize in cosmological observations. This picture turns out to describe well many of the features we see around us, which are otherwise harder to understand. 

This type of hot fluid cools as the universe expands, leading to several types of characteristic events whose late-time signatures provide evidence for the validity of the Hot Big Bang picture. The first type of characteristic event is the departure from equilibrium that every species of particle always experiences eventually once its particle density becomes too low for particles to find one another frequently enough to maintain equilibrium. 

The second type of characteristic event is the formation of bound states. At finite temperature the net abundance of bound states (like atoms or nuclei, say) is fixed by detailed balance: the competition between reactions (like $e^- p \to H \gamma$) that form the bound states (in this case Hydrogen) and the inverse reactions (like $H \gamma \to e^- p$) that dissociate them. Once the temperature falls below the binding energy of a bound state the typical collision energy falls below the threshold required for dissociation and so the abundance of the bound state grows until the constituents eventually become sufficiently rare that the formation reactions also effectively turn off the production processes. Once this happens the bound-state abundance freezes and for the purposes of later cosmology these bound states can be regarded as being part of the inventory of `elementary' particles during later epochs. 

There is concrete evidence that the formation of bound states took place at least twice in the early universe. The earliest case happened during the epoch of primordial nucleosynthesis, at redshift $z \simeq 10^{10}$, when temperatures were in the MeV regime and protons and neutrons got cooked into light nuclei. The evidence that this occurred comes from the agreement between the primordial abundances of light nuclear isotopes with the results of precision calculations of their formation rates. This agreement is nontrivial because the total formation rate for each nuclear isotope depends on the density of protons and neutrons at the time, and the same value for the baryon density gives successful agreement between theory and observations for ${}^2$H, ${}^3$He, ${}^4$He and ${}^7$Li. The consistency of these calculations both tells us that this picture of their origins is likely right, and the total density of baryons throughout the universe at this time. 

The second important epoch during which bound states formed is the epoch of `recombination', at redshifts around $z \simeq 1100$. At this epoch the temperature of the cosmic fluid is around 1000 K and electrons and nuclei combine to form electrically neutral atoms (like H or He). The evidence that this occurred comes from the existence and properties of the CMB itself. Before neutral atoms formed the charged electrons and protons very efficiently scattered photons, making the universe at that time opaque to light. But this scattering stopped after atoms formed, leaving a transparent universe in which all the photons present in the hot gas remain but no longer scatter very often. Indeed it is this bath of primordial photons, now redshifted down to microwave wavelengths and currently being detected, that we call the CMB.

\begin{figure}[h]
\begin{center}
\includegraphics[width=100mm,height=80mm]{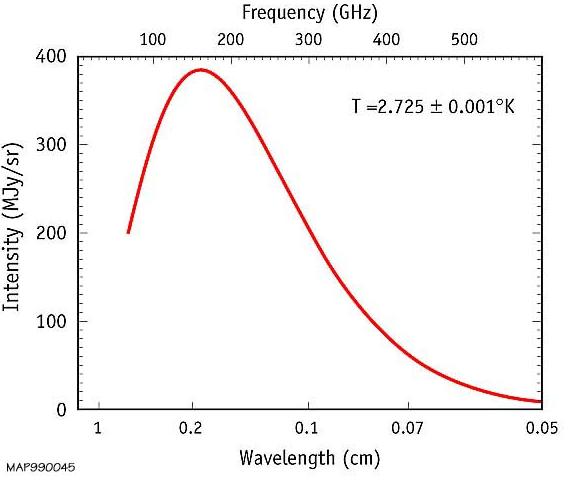} 
\caption{The FIRAS measurement of the thermal distribution of the
  CMB photons. The experimental points lie on the theoretical curve,
  with errors which are smaller than the width of the curve.} \label{FigureFiras} 
\end{center}
\end{figure}

The distribution of these CMB photons has a beautiful thermal form as a function of the present-day photon wavelength, $\lambda_0$, as shown in Fig. \ref{FigureFiras}. The temperature of this distribution is measured as a function of direction in the sky, $T_\gamma(\theta,\phi)$, and it is the angular average of this measured temperature,
\be
    T_{\gamma 0} = \langle T_\gamma \rangle = \frac{1}{4\pi}
    \int T_\gamma(\theta,\phi) \, \sin\theta \; \exd\theta
    \,\exd \phi
    = 2.2725 \; \hbox{K} \,,
\ee
that is used above as the present temperature of the relic photons.

The starting point for making such a thermal description precise is a summary of the various types of particles that are believed to be `elementary' at the temperatures of interest. The highest temperature for which there is direct observational evidence the universe attained in the past is $T \sim 10^{10}$ K, which corresponds to thermal energies of order 1 MeV. The elementary particles which might be expected to be found within a soup having this temperature are the following.

\begin{itemize}
\item {\bf Photons ($\gamma$):} are bosons and have no electric charge or mass, and can be singly emitted and absorbed by any electrically-charged particles.
\item {\bf Electrons and Positrons ($e^\pm$):} are fermions with charge $\pm e$ and masses equal numerically to $m_e = 0.511$ MeV. Because the positron, $e^+$, is the antiparticle for the electron, $e^-$, (and vice versa), these particles can completely annihilate into photons through the reaction
\be
    e^+ + e^- \leftrightarrow 2 \gamma \,,
\ee
and do so once the temperature falls below the electron mass.
\item {\bf Protons ($p$):} are fermions with charge $+e$ and mass $m_p = 938$ MeV. Unlike all of the other particles
described here, the proton and neutron can take part in the strong interactions, for example experiencing reactions like
\be
    p + n \leftrightarrow D + \gamma \,,
\ee
in which a proton and neutron combine to produce a deuterium nucleus. The photon which appears in this expression simply carries off any excess energy which is released by the reaction.
\item {\bf Neutrons ($n$):} are electrically neutral fermions with mass $m_n = 940$ MeV. Like protons, neutrons participate in
the strong interactions. Isolated neutrons are unstable, and left to themselves decay through the weak interactions into a
proton, an electron and an electron-antineutrino:
\be
    n \to p + e^- + \overline{\nu}_e \,.
\ee
\item {\bf Neutrinos and Anti-neutrinos ($\nu_e$, $\overline{\nu}_e$, $\nu_\mu$, $\overline{\nu}_\mu$, $\nu_\tau$, $\overline{\nu}_\tau$):} are fermions that are electrically neutral, and have been found to have nonzero masses whose precise values are not known, but which are known to be smaller than 1 eV.
\item {\bf Gravitons ($G$):} are electrically neutral bosons that mediate the gravitational force in the same way that photons do for the electromagnetic force. Gravitons only interact with other particles with gravitational strength, which is much weaker that the strength of the other interactions. As a result they turn out never to be in thermal equilibrium for any of the temperatures to which we have observational access in cosmology.
\end{itemize}
To these must be added whatever makes up the Dark Matter, provided temperatures and interactions are such that the Dark Matter can be regarded to be in thermal equilibrium.

How would the temperature of a bath of these particles evolve on thermodynamic grounds as the universe expands? The first step asks how the temperature is related to $a$ (and so also $t$), in order to quantify the rate with which a hot bath cools due to the universal expansion.

\medskip\noindent{\bf Relativistic Particles}

\medskip\noindent
The energy density and pressure for a gas of relativistic particles (like photons) when in thermal equilibrium at temperature $T_R$ are given by
\be
    \rho_R = a_B \, T_R^4 \qquad \hbox{and} \qquad
    p_R = \frac13 \, a_B \, T_R^4 \,,
\ee
where $a_B$ is $g/2$ times the Stefan-Boltzmann constant and $g$ counts the number of internal (spin) states of the particles of interest (and so $g=2$ for a gas of photons). 

Combining this with energy conservation, which says $\rho_R \propto (a_0/a)^4$, shows that the product $a T$ is constant, and so
\be \label{RadiationTvsa}
    T_R(a) = T_{R0} \left( \frac{a_0}{a} \right) = T_{R0} (1+z) \,.
\ee
This is equivalent to the statement that the expansion is adiabatic, since the entropy per unit volume of a relativistic gas is $s_R \propto T_R^3$, and so the total entropy in this gas is
\be
    S_R \propto s_R \, a^3 \propto (T_R \, a)^3 =  \hbox{constant} \,.
\ee

Although the relation $T \propto a^{-1}$ is derived above assuming thermal equilibrium, it can continue to hold (for relativistic particles) once the particles become insufficiently dense to scatter frequently enough to maintain equilibrium. This is because the thermal distribution functions for relativistic particles are functions of the ratio of particle energy divided by temperature: $f(\epsilon, T) \propto (e^{\epsilon/T} - 1)^{-1}$. Because relativistic particles have energies $\epsilon(\bfp) = |\bfp|$, where the physical momentum $\bfp$ is related to co-moving momentum by $\bfp = \bfk/a$, their energies redshift $\epsilon \propto a^{-1}$ with the universal expansion. 

This ensures that the distributions remain in the thermal form for all $t$, provided that their temperature is also regarded as falling with $T \propto a^{-1}$, so that $\epsilon/T$ is time-independent. For this reason it makes sense to continue to regard the CMB photon temperature to be falling with $T_R \propto a^{-1}$ even though photons stopped interacting frequently enough to remain in equilibrium once protons and electrons combined into electrically neutral atoms around redshift $z \simeq 1100$. 

\medskip\noindent{\bf Nonrelativistic Particles}

\medskip\noindent
As mentioned earlier, an ideal gas of non-relativistic particles in thermal equilibrium has a pressure and energy density given instead by
\be
    p_M = n \, T_M \qquad \hbox{and} \qquad
    \rho_M = n \, m + \frac{n \, T_M}{\gamma - 1} \,,
\ee
where $n$ is the number density of particles, $m$ is the particle's rest mass and $\gamma = c_p/c_v$ is its ratio of specific heats, with $\gamma = 5/3$ for a gas of monatomic atoms. 

In order to repeat the previous arguments using energy conservation to infer how $T_M$ evolves with $a$ we must first determine what $n$ depends on. If the total number of particles is conserved, so 
\be
    \frac{\exd}{\exd t} \Bigl[ n \, a^3 \Bigr] = 0 \,,
\ee
then consistency of $n \propto a^{-3}$ with energy conservation, eq.~\pref{EnergyConservationEqn},  implies $T_M$ should satisfy
\be
    \frac{\dot{T}_M}{T_M} + 3(\gamma - 1) \, \frac{\dot{a}}{a} = 0
    \,,
\ee
and so
\be
    T_M = T_{M0} \left( \frac{a_0}{a} \right)^{3(\gamma - 1)} =
    T_{M0} (1+z)^{3(\gamma-1)} \,.
\ee
For example, for a monatomic gas with $\gamma = 5/3$ this implies $T_M \propto (1+z)^2\propto a^{-2}$, as also would be expected for an adiabatic expansion given that the entropy density for such a fluid varies with $T_M$ like $s_M \propto (m T_M)^{3/2}$.

When a nonrelativistic species of particle falls out of equilibrium its energy (because it is nonrelativistic) is dominated by its rest-mass: $\epsilon(\bfp) \simeq m$. Because of this $\epsilon$ does not redshift and so the distribution of particles remains frozen at the fixed temperature, $T_f$, where equilibrium first broke down. 

\medskip\noindent{\bf Multi-component fluids}

\medskip\noindent
The previous examples assume negligible energy exchange between these different components, which in particular also precludes them being in thermal equilibrium with one another (allowing their respective temperatures free to evolve independently of one another). But what happens when several components of the fluid {\it are} in thermal equilibrium with one another? This situation actually happens for $z > 1100$ when non-relativistic protons and neutrons (or nuclei) are in equilibrium with relativistic photons, electrons and neutrinos. 

To see how this works, we now repeat the previous arguments for a fluid which consists of both relativistic and non-relativistic components, coexisting in mutual thermal equilibrium at a common temperature, $T$. In this case the energy density and pressure are given by 
\be
    p = n \, T + \frac13 \, a_B \, T^4 \qquad \hbox{and} \qquad
    \rho = n \, m + \frac{n \, T}{\gamma - 1} + a_B \, T^4\,.
\ee
Inserting this into the energy conservation equation, as above, leads to the result
\be \label{MixedFluidTvsa}
    \frac{\dot{T}}{T} + \left[ \frac{ 1 + \mathfrak{s}}{\mathfrak{s} + \frac13
    \, (\gamma -1)^{-1}} \right] \, \frac{\dot{a}}{a} = 0 \,,
\ee
where
\be
    \mathfrak{s} \equiv \frac{4 a_B \, T^3}{3n} = 74.0
    \, \left[ \frac{(T/\hbox{deg})^3}{n/\hbox{cm}^{-3}}
    \right]\,,
\ee
is the relativistic entropy per non-relativistic gas particle. For example, if the relativistic gas consists of photons, then the number of photons per unit volume is $n_\gamma = [30 \, \zeta(3)/\pi^4] a_B T^3 = 3.7 \, a_B T^3$, and so $\mathfrak{s} = 0.37 (n_\gamma/n)$.

Eq.~\pref{MixedFluidTvsa} shows how $T$ varies with $a$, and reduces to the pure radiation result, $T \, a =$ constant, when $\mathfrak{s} \gg 1$ and to the non-relativistic matter result, $T \, a^{3(\gamma - 1)} =$ constant, when $\mathfrak{s} \ll 1$. In general, however, this equation has more complicated solutions because $\mathfrak{s}$ need not be a constant. Given that particle conservation implies $n \propto a^{-3}$, we see that the time-dependence of $\mathfrak{s}$ is given by $\mathfrak{s} \propto (T \, a)^3$.

We are led to the following limiting behaviour. If, initially, $\mathfrak{s} = \mathfrak{s}_0 \gg 1$ then at early times $T \propto a^{-1}$ and so  $\mathfrak{s}$ remains approximately constant (and large). For such a gas the common temperature of the relativistic and non-relativistic fluids continues to fall like $T \propto a^{-1}$. In this case the high-entropy relativistic fluid controls the temperature evolution and drags the non-relativistic temperature along with it. Interestingly, it can do so even if $\rho_M \approx n \, m$ is larger than $\rho_R = a_B \, T^4$, as can easily happen when $m \gg T$. In practice this happens until the two fluid components fall out of equilibrium with one another, after which their two temperatures begin to evolve separately according to the expressions given previously.

On the other hand if $\mathfrak{s} = \mathfrak{s}_0 \ll 1$ initially, then $T \propto a^{-3(\gamma-1)}$ and so $\mathfrak{s} \propto a^{3(4 - 3 \gamma)}$. This falls as $a$ increases provided $\gamma > 4/3$, and grows otherwise. For instance, the particularly interesting case $\gamma = 5/3$ implies $T \propto a^{-2}$ and so $\mathfrak{s} \propto a^{-3}$. We see that if $\gamma > 4/3$, then an initially small $\mathfrak{s}$ gets even smaller still as the universe expands, implying the temperature of both radiation and matter continues to fall like $T \propto a^{-3(\gamma-1)}$. If, however, $1< \gamma < 4/3$, an initially small $\mathfrak{s}$ can grow even as the temperature falls, until the fluid eventually crosses over into the relativistic regime for which $T \propto a^{-1}$ and $\mathfrak{s}$ stops evolving.

\subsection{An early accelerated epoch}

This section now switches from a general description of the $\Lambda$CDM model to a discussion about the peculiar initial conditions on which its success seems to rely. This is followed by a summary of the elements of some simple single-field inflationary models, and why their proposal is motivated as explanations of the initial conditions for the later universe.

\subsubsection{Peculiar initial conditions}

The $\Lambda$CDM model describes well what we see around us, provided that the universe is started off with a very specific set of initial conditions. There are several properties of these initial conditions that seem peculiar, as is now summarized. 

\medskip\noindent{\bf Flatness problem}

\medskip\noindent
The first problem concerns the observed spatial flatness of the present-day universe. As described earlier, observations of the CMB indicate that the quantity $\kappa/a^2$ of the Friedmann equation, eq.~\pref{FriedmannEqn}, is at present consistent with zero. What is odd about this condition is that this curvature term tends to grow in relative importance as the universe expands, so finding it to be small now means that it must have been {\em extremely} small in the remote past.

More quantitatively, it is useful to divide the Friedmann equation by $H^2(t)$ to give
\be
    1 + \frac{\kappa}{(a H)^2} = \frac{8 \pi G \rho}{3 H^2} =:  \Omega(a) \,,
\ee
where (as before) the final equality defines $\Omega(a)$. The problem arises because the product $aH$ decreases with time during both matter and radiation domination. For instance, observations indicate that at present $\Omega = \Omega_0$ is unity to within about 10\%, and since during the matter-dominated era the product $(aH)^2 \propto a^{-1}$ it follows that at the epoch $z_{\rm eq} \simeq 3600$ of radiation-matter equality we must have had
\be
    \Omega(z_{\rm eq}) -1 = \Bigl( \Omega_0 - 1 \Bigr) \left( \frac{a}{a_0} \right)= \frac{ \Omega_0 - 1 }{ 1+z_{\rm
    eq} } \simeq \frac{0.1}{3600} \simeq 2.8 \times 10^{-5} \,.
\ee
So $\Omega-1$ had to be smaller than a few tens of a millionth at the time of radiation-matter equality in order to be of order 10\% now.

And it only gets worse the further back one goes, provided the extrapolation back occurs within a radiation- or matter-dominated era (as seems to be true at least as far back as the epoch of nucleosynthesis). Since during radiation-domination we have $(aH)^2 \propto a^{-2}$ and the redshift of nucleosynthesis is $z_{\rm BBN} \sim 10^{10}$ it follows that at this epoch one must require
\be \label{OmegaBBNEqn}
    \Omega(z_{\rm BBN}) -1 = \Bigl[ \Omega(z_{\rm eq}) - 1 \Bigr]
    \left( \frac{1 + z_{\rm eq}}{1+z_{\rm BBN}} \right)^{2}
    = \frac{0.1}{3600} \left( \frac{3600}{10^{10}} \right)^2
    \approx 3.6 \times 10^{-18} \,,
\ee
requiring $\Omega$ to be unity with an accuracy of roughly a part in $10^{18}$. The discomfort of having the success of a theory hinge so sensitively on the precise value of an initial condition in this way is known as the Big Bang's {\sl Flatness Problem}.

\medskip\noindent{\bf Horizon problem}

\medskip\noindent
Perhaps a more serious question asks why the initial universe can be so very homogeneous. In particular, the temperature fluctuations of the CMB only arise at the level of 1 part in $10^5$, and the question is how this temperature can be so incredibly uniform across the sky. 

Why is this regarded as a problem? It is not uncommon for materials on earth to have a uniform temperature, and this is usually understood as a consequence of thermal equilibrium. An initially inhomogeneous temperature distribution equilibrates by having heat flow between the hot and cold areas, until everything eventually shares a common temperature.

\begin{figure}[h]
\begin{center}
\includegraphics[width=110mm,height=60mm]{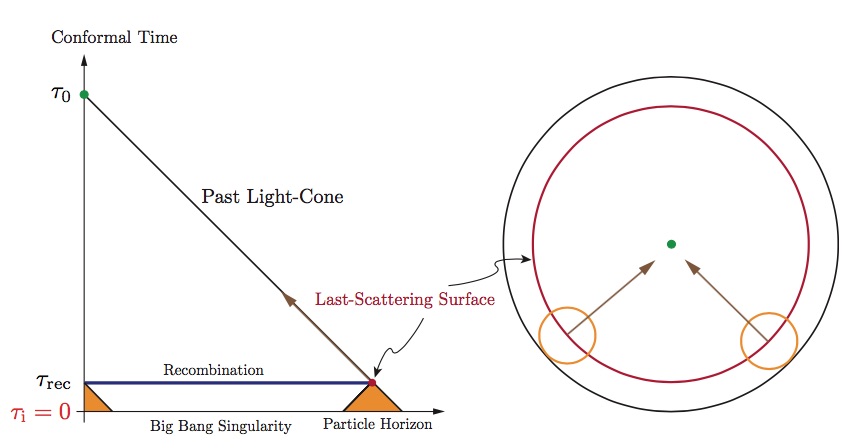} 
\caption{A conformal diagram illustrating how there is inadequate time in a radiation-dominated universe for there to be a causal explanation for the correlation of temperature at different points of the sky in the CMB. (Figure taken from \cite{Baumann}.)} \label{ConformalDiag} 
\end{center}
\end{figure}

The same argument is harder to make in cosmology because in the Hot Big Bang model the universe generically expands so quickly that there has not been enough time for light to travel across the entire sky to bring everyone the news as to what the common temperature is supposed to be. This is easiest to see using conformal coordinates, as in \pref{ConformalCoords}, since in these coordinates it is simple to identify which regions can be connected by light signals. In particular, radially directed light rays travel along lines $\exd \ell = \pm \exd \tau$, which can be drawn as straight lines of slope $\pm 1$ in the $\tau-\ell$ plane, as in Figure \ref{ConformalDiag}. The problem is that $a(\tau)$ reaches zero in a finite conformal time (which we can conventionally choose to happen at $\tau = 0$), since $a(\tau) \propto \tau$ during radiation domination and $a(\tau) \propto \tau^2$ during matter domination. Redshift $z_{\rm rec} \simeq 1100$ (the epoch of recombination, at which the CMB photons last sampled the temperature of the Hydrogen gas with which they interact) is simply too early for different directions in the sky to have been causally connected in the entire history of the universe up to that point. 

To pin this down quantitatively, let us assume that the universe is radiation-dominated for all points earlier than the epoch of radiation-matter equality, $t_{\rm eq}$, so the complete evolution of $a(t)$ until recombination is
\be
  a(t) \simeq \left\{ \begin{array}{l}
  a_{\rm eq} (t/t_{\rm eq})^{1/2} \qquad \hbox{for $0 < t < t_{\rm eq}$}  \\
  a_{\rm eq} (t/t_{\rm eq})^{2/3} \qquad \hbox{for $ t_{\rm eq} < t < t_{\rm rec}$}  \;.
\end{array} \right.
\ee
(The real evolution does not have a discontinuous derivative at $t = t_{\rm eq}$, but this inaccuracy is not important for the argument that follows.) The maximum proper distance, measured at the time of recombination, that a light signal could have travelled by the time of recombination, $t_{\rm rec}$, then is
\bea
    D_{\rm rec} &=& a_{\rm rec} \left[ \int_0^{t_{\rm eq}}  \frac{\exd \hat t}{a(\hat t)} + \int_{t_{\rm eq}}^{t_{\rm rec}}  \frac{\exd \hat t}{a(\hat t)}\right]   = \frac{a_{\rm rec} t_{\rm eq}}{a_{\rm eq}} \left[ 3 \left( \frac{t_{\rm rec}}{t_{\rm eq}} \right)^{1/3} -1 \right]  \nn\\
    &=&  \frac{2}{H_{\rm eq}^+} \left( \frac{a_{\rm rec}}{a_{\rm eq}} \right)^{3/2} \left[ 1 - \frac13 \, \left( \frac{a_{\rm eq}}{a_{\rm rec}} \right)^{1/2} \right] \simeq \frac{1.6}{H_{\rm rec}} \,,
\eea
where $H_{\rm eq}^+ =  2/(3t_{\rm eq})$ denotes the limit of the Hubble scale as $t \to t_{\rm eq}$ on the matter-dominated side. The approximate equality in this expression uses $H \propto a^{-3/2}$ during matter domination as well as using the redshifts $z_{\rm rec} \simeq 1100$ and $z_{\rm eq} \simeq 3600$ (as would be true in the $\Lambda$CDM model) to obtain $a_{\rm eq}/a_{\rm rec} \simeq 1100/3600 \simeq 0.31$.

To evaluate this numerically we use the present-day value for the Hubble constant, $H_0 \simeq 70$ km/sec/Mpc --- or (keeping in mind our units for which $c=1$), $H_0^{-1} \simeq 13$ Gyr $\simeq 4$ Gpc. This then gives $H_{\rm rec}^{-1} \simeq H_0^{-1} (a_{\rm rec}/a_0)^{3/2} \simeq 3 \times 10^{-5} H_0^{-1} \simeq 0.1$ Mpc, if we use $a_0/a_{\rm rec} = 1+ z_{\rm rec} \simeq 1100$, and so $D_{\rm rec} \simeq 0.2$ Mpc.

Now CMB photons arriving to us from the surface of last scattering left this surface at a distance from us that is now of order
\be
    R_0 = a_0 \int_{t_{\rm rec}}^{t_0} \frac{\exd \hat t}{a(\hat t)}
    = 3 t_0 - 3 t_0^{2/3} t_{\rm rec}^{1/3}
    = \frac{2}{H_0} \left[ 1 - \left( \frac{a_{\rm rec}}{a_0}
    \right)^{1/2} \right] \,,
\ee
again using $a \propto t^{2/3}$ and $H \propto a^{-3/2}$, and so $R_0 \simeq 2/H_0 \simeq 8$ Gpc. So the angle subtended by $D_{\rm rec}$ placed at this distance away (in a spatially-flat geometry) is really $\theta \simeq D_{\rm rec}/R_{\rm rec}$ where $R_{\rm rec} = (a_{\rm rec}/a_0) R_0 \simeq 7$ Mpc is its distance {\it at the time of last scattering}, leading to\footnote{This estimate is related to the quantity $\theta_{\scriptscriptstyle MC}$ in the table of Fig.~\ref{Fig:LCDMparams}.} $\theta \simeq 0.2/7 \simeq 1^o$.  Any two directions in the sky separated by more than this angle (about twice the angular size of the Moon, seen from Earth) are so far apart that light had not yet had time to reach one from the other since the universe's beginning. 

How can all the directions we see have known they were all to equilibrate to the same temperature? It is very much as if we were to find a very uniform temperature distribution, {\it immediately} after the explosion of a very powerful bomb.

\medskip\noindent{\bf Defect problem}

\medskip\noindent
Historically, a third problem --- called the `Defect' (or `Monopole') Problem is also used to motivate changing the extrapolation of radiation domination into the remote past. A defect problem arises if the physics of the much higher energy scales relevant to the extrapolation involves the production of topological defects, like domain walls, cosmic strings or magnetic monopoles. Such defects are often found in Grand Unified theories; models proposed to unify the strong and electroweak interactions as energies of order $10^{15}$ GeV.   

These kinds of topological defects can be fatal to the success of late-time cosmology, depending on how many of them survive down to the present epoch. For instance if the defects are monopoles, then they typically are extremely massive and so behave like non-relativistic matter. This can cause problems if they are too abundant because they can preclude the existence of a radiation dominated epoch, because their energy density falls more slowly than does radiation as the universe expands. 

Defects are typically produced with an abundance of one per Hubble volume, $n_d(a_f) \sim H_f^3$, where $H_f = H(a_f)$ is the Hubble scale at their epoch of formation, at which time $a = a_f$. Once produced, their number is conserved, so their density at later times falls like $n_d(a) = H_f^3 (a_f/a)^3$. Consequently, at present the number surviving within a Hubble volume is $n_d(a_0)H_0^{-3} = (H_f \,a_f/H_0 \,a_0)^3$. 

Because the product $aH$ is a falling function of time, the present-day abundance of defects can easily be so numerous that they come to dominate the universe well before the nucleosynthesis epoch.\footnote{Whether they do also depends on their dimension, with magnetic monopoles tending to be more dangerous in this regard than are cosmic strings, say.} This could cause the universe to expand (and so cool) too quickly as nuclei were forming, and so give the wrong abundances of light nuclei. Even if not sufficiently abundant during nucleosynthesis, the energy density in relict defects can be inconsistent with measures of the current energy density.

This is clearly more of a hypothetical problem than are the other two, since whether there is a problem depends on whether the particular theory for the high-energy physics of the very early universe produces these types of defects or not. It can be fairly pressing in Grand Unified models since in these models the production of magnetic monopoles can be fairly generic.

\subsubsection{Acceleration to the rescue}

The key observation when trying to understand the above initial conditions is that they only seem unreasonable because they are based on extrapolating into the past assuming the universe to be radiation (or matter) dominated (as would naturally be true if the $\Lambda$CDM model were the whole story). This section argues that these initial conditions can seem more reasonable if a different type of extrapolation is used; in particular if there were an earlier epoch during which the universal expansion were to accelerate: $\ddot a > 0$ \cite{firstINF, chaotic}.

Why should acceleration help? The key point is that the above initial conditions are a problem because the product $aH$ is a falling function as $a$ increases, for both matter and radiation domination. For instance, for the flatness problem the evolution of the curvature term in the Friedmann equation is $\Omega_\kappa \propto (aH)^{-2}$ and this grows as $a$ grows only because $aH$ decreases with $a$. But if $\ddot a > 0$ then $\dot a = aH$ {\em increases} as $a$ increases, and this can help alleviate the problems. For example, finding $\Omega_\kappa$ to be very small in the recent past would be less disturbing if the more-distant past contained a sufficiently long epoch during which $aH$ grew.

How long is long enough? To pin this down suppose there were an earlier epoch during which the universe were to expand in the same way as during Dark Energy domination, $a(t) \propto e^{Ht}$, for constant $H$. Then $aH = a_0 H \, e^{Ht}$ grows exponentially with time and so even if $Ht$ were of order 100 or less it would be possible to explain why $\Omega_\kappa$ could be as small as $10^{-18}$ or smaller. 

Having $aH$ grow also allows a resolution to the horizon problem. One way to see this is to notice that $a(t) \propto e^{Ht}$ implies $\tau = - H^{-1}e^{-Ht}$ plus a constant (with the sign a consequence of having $\tau$ increase as $t$ does), and so
\be
 a(\tau) = - \, \frac{1}{H(\tau - \tau_0)} \,,
\ee
with $0 < a < \infty$ corresponding to the range $-\infty < \tau < \tau_0$. Exponentially accelerated expansion allows $\tau$ to be extrapolated to arbitrarily negative values, and so allows sufficient time for the two causally disconnected regions of the conformal diagram of Figure \ref{ConformalDiag} to have at one point been in causal contact. 

\begin{figure}[h]
\begin{center}
\includegraphics[width=100mm,height=80mm]{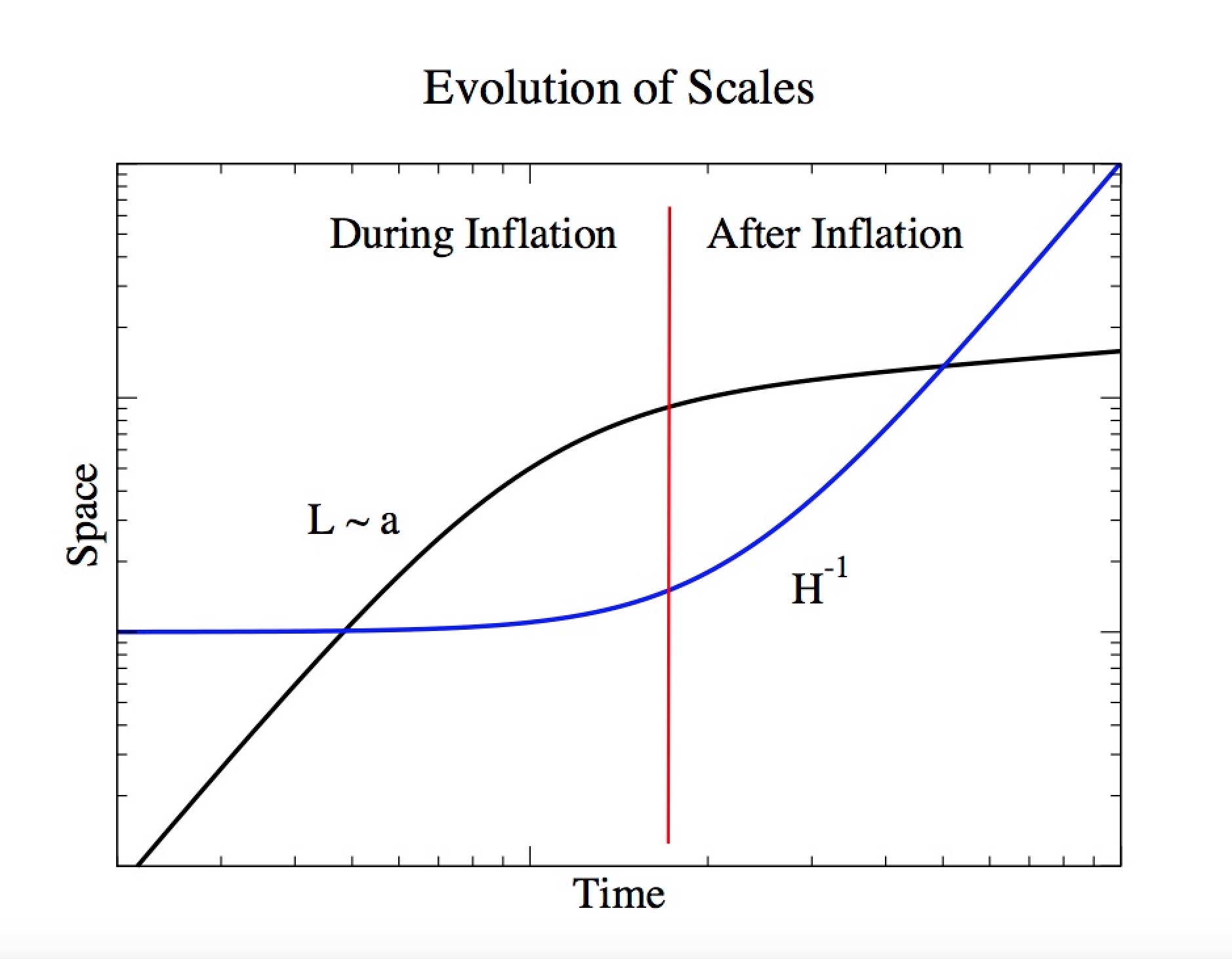} 
\caption{A sketch of the relative growth of physical scales,
$L(t)$, (in black) and the Hubble length, $H^{-1}$, (in blue)
during and after inflation. Horizon exit happens during inflation where the blue and black curves first cross, and this is eventually followed by horizon re-entry where the curves cross again during the later Hot Big Bang era.} \label{FigureScales} 
\end{center}
\end{figure}

Another way to visualize this is to plot physical distance $\lambda (t)\propto a(t)$ and the Hubble radius, $H^{-1}$, against $t$, as in Figure \ref{FigureScales}. Focus first on the right-hand side of this figure, which compares these quantities during radiation or matter domination. During these epochs the Hubble length evolves as $H^{-1} \propto t$ while the scale factor satisfies $a(t) \propto t^p$ with $0<p<1$. Consequently $H^{-1}$ grows more quickly with $t$ than do physical length scales $\lambda(t)$. During radiation or matter domination systems of any given size eventually get caught by the growth of $H^{-1}$ and so `come inside the Hubble scale' as the universe expands. Systems involving larger $\lambda(t)$ do so later than those with smaller $\lambda$. The largest sizes currently visible have only recently crossed inside of the Hubble length, having spent their entire earlier history larger than $H^{-1}$ (assuming always a radiation- or matter-dominated universe).

Having $\lambda > H^{-1}$ matters because physical quantities tend to freeze when their corresponding length scales satisfy $\lambda(t) > H^{-1}$. This then precludes physical processes from acting over these scales to explain things like the uniform temperature of the CMB.  The freezing of super-Hubble scales can be seen, for example, in the evolution of a massless scalar field in an expanding universe, since the field equation $\Box \phi = 0$ becomes in FRW coordinates
\be
  \ddot \phi_k + 3H \, \dot \phi_k + \left( \frac{k}{a} \right)^2 \phi_k = 0 \,,
\ee
where we Fourier expand the field $\phi(x) = \int \exd^3k \; \phi_k \, \exp\left[i \bfk \cdot \bfx \right]$ using co-moving coordinates, $\bfx$. For modes satisfying $2\pi/\lambda = p = k/a \ll H$ the field equation implies $\dot \phi_k \propto a^{-3}$ and so $\phi_k = C_0 + C_1 \int \exd t/a^3$ is the sum of a constant plus a decaying mode. 

Things are very different during exponential expansion, however, as is shown on the left-hand side of Figure \ref{FigureScales}. In this regime $\lambda(t) \propto a(t) \propto e^{Ht}$ grows exponentially with $t$ while $H^{-1}$ remains constant. This means that modes that are initially smaller than the Hubble length get stretched to become larger than the Hubble length, with the transition for a specific mode of length $\lambda(t)$ occurring at the epoch of `Hubble exit', $t= t_{\rm he}$, defined by $2\pi/\lambda (t_{\rm he})= p_{\rm he} = k/a(t_{\rm he}) = H$. In this language it is because the criterion for Hubble exit and entry is $k = aH$ that the growth or shrinkage of $aH$ is relevant to the horizon problem.

How much expansion is required to solve the horizon problem? Choosing a mode $\phi_k$ that is only now crossing the Hubble scale tells us that $k = a_0 H_0$. This same mode would have crossed the horizon during an exponentially expanding epoch when $k = a_{\rm he} H_\ssI$, where $H_\ssI$ is the constant Hubble scale during exponential expansion. So clearly $a_0 H_0 = a_{\rm he} H_\ssI$ where $t_{\rm he}$ is the time of Hubble exit for this particular mode. To determine how much exponential expansion is required solve the following equation for $N_e := \ln(a_{\rm end}/a_{\rm he})$, where $a_{\rm end}$ is the scale factor at the end of the exponentially expanding epoch:
\be
  1 = \frac{a_{\rm he} H_\ssI}{a_0 H_0} = \left( \frac{a_{\rm he} H_\ssI}{a_{\rm end} H_\ssI} \right) \left( \frac{a_{\rm end} H_\ssI}{a_{\rm eq} H_{\rm eq}} \right) \left( \frac{a_{\rm eq} H_{\rm eq}}{a_0 H_0} \right)= e^{-N_e} \left( \frac{a_{\rm eq}}{a_{\rm end}} \right)\left( \frac{a_0}{a_{\rm eq}} \right)^{1/2} \,.
\ee
This assumes (for the purposes of argument) that the universe is radiation dominated right from $t_{\rm end}$ until radiation-matter equality, and uses $aH \propto a^{-1}$ during radiation domination and $aH \propto a^{-1/2}$ during matter domination. $N_e = H_\ssI(t_{\rm end} - t_{\rm he})$ is called the number of $e$-foldings of exponential expansion and is proportional to how long exponential expansion lasts

Using, as above, $(a_{\rm eq} H_{\rm eq})/(a_0 H_0) = (a_0/a_{\rm eq})^{1/2} \simeq 60$, and $(a_{\rm eq} H_{\rm eq})/(a_{\rm end} H_{\rm end}) = a_{\rm end}/a_{\rm eq} = T_{\rm eq}/T_M$ with $T_{\rm eq} \sim 3$ eV, and assuming the energy density of the exponentially expanding phase is transferred perfectly efficiently to produce a photon temperature $T_M$ then leads to the estimate
\be \label{NIFromHorizonEqn}
    N_e \sim \ln \left[(3 \times 10^{23}) \times 60 \right]
    + \ln\left( \frac{T_M}{10^{15} \; \hbox{GeV}} \right)
    \approx 58 + \ln\left( \frac{T_M}{10^{15} \;
    \hbox{GeV}} \right)\,.
\ee

Roughly 60 $e$-foldings of exponential expansion can provide a framework for explaining how causal physics might provide the observed correlations that are observed in the CMB over the largest scales, even if the energy densities involved are as high as $10^{15}$ GeV. We shall see below that life is even better than this, because in addition to providing a {\sl framework} in which a causal understanding of correlations could be solved, inflation itself can provide the {\sl mechanism} for explaining these correlations (given an inflationary scale of the right size).

\subsubsection{Inflation or a bounce?}

An early epoch of near-exponential accelerated expansion has come to be known as an `inflationary' early universe. Acceleration within this framework speeds up an initially expanding universe to a higher expansion rate.  However, an attentive reader may notice that although acceleration is key to helping with $\Lambda$CDM's initial condition issues, there is no {\em a priori} reason why the acceleration must occur in an initially expanding universe, as opposed (say) to one that is initially contracting. Models in which one tries to solve the problems of $\Lambda$CDM by having an initially contracting universe accelerate to become an expanding one are called `bouncing' cosmologies. 

Since it is really the acceleration that is important, bouncing models should in principle be on a similar footing to inflationary ones. In what follows only inflationary models are considered, for the following reasons:

\medskip\noindent{\bf Validity of the semiclassical methods}

\medskip\noindent
Predictions in essentially all cosmological models are extracted using semiclassical methods: one typically writes down the action for some system and then explores its consequences by solving its classical equations of motion. So a key question for all such models is the identification of the small parameter (or parameters) that suppresses quantum effects and so controls the underlying semiclassical approximation. In the absence of such a control parameter classical predictions need not capture what the system really does. Such a breakdown of the semiclassical approximation really means that the `theory error' in the model's predictions could be arbitrarily large, making comparisons to observations essentially meaningless.

A reason sometimes given for not pinning down the size of quantum corrections when doing cosmology is that gravity plays a central role, and we do not yet know the ultimate theory of quantum gravity. Implicit in this argument is the belief that the size of quantum corrections is incalculable without such an ultimate theory, perhaps because of the well-known divergences in quantum predictions due to the non-renormalizability of General Relativity \cite{NRGR}. But experience with non-renormalizable interactions elsewhere in physics tells us that quantum predictions can sometimes be made, provided one recognizes they involve an implicit low-energy/long-distance expansion relative to the underlying physical scale set by the dimensionful non-renormalizable couplings. Because of this the semiclassical expansion parameter in such theories is usually the ratio between this underlying short-distance scale and the distances of interest in cosmology (which, happily enough, aims at understanding the largest distances on offer). Effective field theories provide the general tools for quantifying these low-energy expansions, and this is why EFT methods are so important for any cosmological studies.

As is argued in more detail in \S\ref{Sec:EFT}, the semiclassical expansion in cosmology is controlled by small quantities like $(\lambda M_p)^{-2}$ where $\lambda$ is the smallest length scale associated with the geometry of interest. In practice it is often $\lambda \sim H^{-1}$ that provides the relevant scale in cosmology, particularly when all geometrical dimensions are similar in size. So a rule of thumb generically asks the ratio $H^2/M_p^2$ to be chosen to be small:
\be \label{HllMpsq}
  \frac{H^2}{M_p^2} \propto \frac{\rho}{M_p^4} \ll 1 \,,
\ee
as a necessary condition\footnote{The semiclassical criterion can be stronger than this, though this can often only be quantified within the context of a specific proposal for what quantum gravity is at the shortest scales. For instance, if it is string theory that takes over at the shortest scales then treatment of cosmology using a field theory -- rather than fully within string theory -- requires \pref{HllMpsq} be replaced by the stronger condition $H^2/M_s^2 \ll 1$, where $M_s \ll M_p$ is the string scale, set for example by the masses of the lightest string excited states.} for quantum cosmological effects to be suppressed.

For inflationary models $H$ is usually at its largest during the inflationary epoch, with geometrical length scales only increasing thereafter, putting one deeper and deeper into the semiclassical domain. It is a big plus for these models that they can account for observations while wholly remaining within the regime set by \pref{HllMpsq}, and this is one of the main reasons why they receive so much attention. 

For bouncing cosmologies the situation can be more complicated. The smallest geometrical scale $\lambda$ usually occurs during the epoch near the bounce, even though $H^{-1}$ itself usually tends to infinity there. In models where $\lambda$ becomes comparable to $M_p^{-1}$ (or whatever other scale -- such as the string length scale, $M_s^{-1} \gg M_p^{-1}$ -- that governs short-distance gravity), quantum effects during the bounce need not be negligible and the burden on proponents is to justify why semiclassical predictions actually capture what happens during the bounce. 

\medskip\noindent{\bf Difficulty of achieving a semiclassically large bounce}

\medskip\noindent
Another issue arises even if the scale $\lambda$ during a bounce does remain much larger than the more microscopic scales of gravity. In this regime the bounce can be understood purely within the low-energy effective theory describing the cosmology, for which General Relativity should be the leading approximation. But (when $\kappa = 0$) the Friedmann equation for FRW geometries in General Relativity states that $H^2 = {\rho}/{3M_p^2}$, and so $\rho$ must pass through zero at the instant where the contracting geometry transitions to expansion (since $H = \dot a/a$ vanishes at this point). Furthermore, using \pref{FriedmannEqn} and \pref{AccEq}, it must also be true that 
\be
   \dot H = \frac{\ddot a}{a} - H^2 = - \,\frac{1}{2M_p^2} (\rho + p) > 0 \,,
\ee
at this point in order for $H$ to change sign there, which means the dominant contributions to the cosmic fluid must satisfy $\rho + p < 0$ during the bounce.\footnote{This is usually phrased as a violation of the `null-energy' condition, which states that $T_{\mu\nu} n^\mu n^\nu \ge 0$ for all null vectors $n^\mu$.}

Although there are no definitive no-go theorems, it has proven remarkably difficult to find a convincing physical system that both satisfies the condition $\rho + p < 0$ and does not also have other pathologies, such as uncontrolled runaway instabilities. For instance within the class of multiple scalar field models for which the lagrangian density is ${\cal L} = \sqrt{-g} \Bigl[  \frac12 \, G_{ij}(\phi) \, \partial_\mu \phi^i \, \partial^\mu \phi^j + V(\phi) \Bigr]$ we have $\rho + p = G_{ij}(\phi) \, \dot \phi^i \, \dot \phi^j$ and so $\rho + p < 0$ requires the matrix of functions $G_{ij}(\phi)$ to have a negative eigenvalue. But if this is true then there is always a combination of fields for which the kinetic energy is negative (what is called a `ghost'), and so is unstable towards the development of arbitrarily rapid motion. Such a negative eigenvalue also implies the gradient energy $\frac12\, G_{ij} \nabla \phi^i \cdot \nabla \phi^j$ is also unbounded from below, indicating instability towards the development of arbitrarily short-wavelength spatial variations.

\pagebreak

\medskip\noindent{\bf Phenomenological issues}

\medskip\noindent
In addition to the above conceptual issues involving the control of predictions, there are also potential phenomenological issues that bouncing cosmologies must face. Whereas expanding geometries tend to damp out spatially varying fluctuations -- such as when gradient energies involve factors like $(k/a)^2$ that tend to zero as $a(t)$ grows -- the opposite typically occurs during a contracting epoch for which $a(t)$ shrinks. This implies that inhomogeneities tend to grow during the pre-bounce contraction --- even when the gradient energies are bounded from below --- and so a mechanism must be provided for why we emerge into the homogeneous and isotropic later universe seen around us in observational cosmology. 

It is of course important that bouncing cosmologies be investigated, not least in order to see most fully what might be required to understand the flatness and horizon problems. Furthermore it is essential to know whether there are alternative observational implications to those of inflation that might be used to marshal evidence about what actually occurred in the very early universe. But within the present state of the art inflationary models have one crucial advantage over bouncing cosmologies: they provide concrete semiclassical control over the key epoch of acceleration on which the success of the model ultimately relies. Because of this inflationary models are likely to remain the main paradigm for studying pre-$\Lambda$CDM extrapolations, at least until bouncing cosmologies are developed to allow similar control over how primordial conditions get propagated to the later universe through the bounce.  

\subsubsection{Simple inflationary models}

So far so good, but what kind of physics can provide both an early period of accelerated expansion and a mechanism for ending this expansion to allow for the later emergence of the successful Hot Big Bang cosmology? 

Obtaining the benefits of an accelerated expansion requires two things: $(i)$ some sort of physics that hangs the universe up for a relatively long period with an accelerating equation of state, $p < - \frac13 \, \rho < 0$; {\sl and} $(ii$) some mechanism for ending this
epoch to allow the later appearance of the radiation-dominated epoch within which the usual Big Bang cosmology starts. Although a number of models exist that can do this, none yet seems completely compelling. This section describes some of the very simplest such models.

The central requirement is to have some field temporarily dominate the universe with potential energy, and for the vast majority of models this new physics comes from the dynamics of a scalar field, $\varphi(x)$, called the `inflaton'. This field can be thought of as an order parameter characterizing the dynamics of the vacuum at the very high energies likely to be relevant to inflationary cosmology. Although the field $\varphi$ can in principle depend on both position and time, once inflation gets going it rapidly smooths out spatial variations, suggesting the study of homogeneous configurations: $\varphi = \varphi(t)$.

\subsubsection*{Higgs field as inflaton}

No way is known to obtain a viable inflationary model simply using the known particles and interactions, but a minimal model \cite{HiggsInf} does use the usual scalar Higgs field already present in the Standard Model as the inflaton, provided it is assumed to have a nonminimal coupling to gravity of the form $\delta {\cal L} = - \xi \, \sqrt{-g} \; ({\cal H}^\dagger {\cal H}) \, R$, where ${\cal H}$ is the usual Higgs doublet and $R$ is the Ricci scalar. Here $\xi$ is a new dimensionless coupling, whose value turns out must be of order $10^4$ in order to provide a good description of cosmological observations. Inflation in this case turns out to occur when the Higgs field takes trans-Planckian values, ${\cal H}^\dagger {\cal H} > M_p^2$, assuming $V$ remains proportional to $({\cal H}^\dagger {\cal H})^2$ at such large values. 

As argued in \cite{HIPCCrit, HIUnitarity}, although the large values required for both $\xi$ and ${\cal H}^\dagger {\cal H}$ needn't invalidate the validity of the EFT description, they do push the envelope for the boundaries of its domain of validity. In particular, semiclassical expansion during inflation turns out to require the neglect of powers of $\sqrt\xi \,H/M_p$, which during inflation is to be evaluated with $H \sim M_p/\xi$. This means both that the semiclassical expansion is in powers of $1/\sqrt\xi$, and that some sort of new physics (or `UV completion') must intervene at scales $M_p/\sqrt\xi \sim \sqrt\xi\, H$, not very far above inflationary energies. Furthermore, it must do so in a way that also explains why the lagrangian should have the very particular large-field form that is required for inflation. In particular, $V$ must be precisely proportional to the square, $f^2$, of the coefficient of the nonminimal Ricci coupling, $f({\cal H}^\dagger {\cal H}) R$, at trans-Planckian field values, since this is ultimately what ensures the potential is flat when expressed in terms of canonically normalized variables in this regime. There are no known proposals for UV completions that satisfy all of the requirements, although conformal or scale invariance seems likely to be relevant \cite{NonminScale}.

This example raises a more general point that is worth noting in passing: having trans-Planckian fields during inflation need not in itself threaten the existence of a controlled low-energy EFT description. The reason for this --- as is elaborated in more detail in \S\ref{Sec:EFT} below --- is that the EFT formulation is ultimately a low-energy expansion and so large fields are only dangerous if they also imply large energy densities. Using an EFT to describe trans-Planckian field evolution need not be a problem so long as the evolution satisfies $H \ll M$ at the field values of interest, where $M \lsim M_p$ is the scale of the physics integrated out to obtain the EFT in question. The condition $H \ll M$ becomes $V \ll M_p^4$ if it happens that $M \sim M_p$. (In any explicit example the precise conditions for validity of EFT methods are obtained using power-counting arguments along the lines of those given in \S\ref{Sec:EFT} below.)

\subsubsection*{New field as inflaton}

The simplest models instead propose a single new relativistic scalar field, $\varphi$, and design its dynamics through choices made for its potential energy, $V(\varphi)$. Taking 
\be
  {\cal L} = \sqrt{-g} \left[ \frac12 \, \partial_\mu \varphi \, \partial^\mu \varphi + V(\varphi) \right] \,,
\ee
the inflaton field equation becomes $\Box \varphi = V'(\varphi)$, which for homogeneous configurations $\varphi(t)$ reduces in an FRW geometry to
\be \label{ScalarFieldEqn}
    \ddot{\varphi} + 3 H \dot{\varphi} + V' = 0 \,,
\ee
where $V' = \exd V/\exd\varphi$. 

The Einstein field equations are as before, but with new $\varphi$-dependent contributions to the energy density and pressure: $\rho = \rho_{\rm rad} + \rho_{\rm m} + \rho_\varphi$ and $p = \frac13 \, \rho_{\rm rad} + p_\varphi$, where 
\be \label{Rhoandpvarphi}
    \rho_\varphi = \frac12 \, \dot{\varphi}^2 + V(\varphi) \qquad
    \hbox{and} \qquad
    p_\varphi = \frac12 \, \dot{\varphi}^2 - V(\varphi) \,.
\ee
The Dark Energy of the present-day epoch is imagined to arise by choosing $V$ so that its minimum satisfies $\rho_{DE} = V(\varphi_{\rm min})$. Inflation is imagined to occur when $\varphi$ evolves slowly through a region where $V(\varphi) \gg V(\varphi_{\rm min})$ is very large, and ends once $\varphi$ rolls down towards its minimum.

With these choices energy conservation for the $\varphi$ field --- $\dot\rho_\varphi + 3 (\dot{a}/a) (\rho_\varphi + p_\varphi) = 0$ follows from the field equation, eq.~\pref{ScalarFieldEqn}. Some couplings must also exist between the $\varphi$ field and ordinary Standard Model particles in order to provide a channel to transfer energy from the inflaton to ordinary particles, and so reheat the universe as required for the later Hot Big Bang cosmology. But $\varphi$ is not imagined to be in thermal equilibrium with itself or with the other kinds of matter during inflation or at very late times, and this can be self-consistent if the coupling to other matter is sufficiently weak and if the $\varphi$ particles are too heavy to be present once the cosmic fluid cools to the MeV energies and below (for which we have direct observations). 

\subsubsection*{Slow-Roll Inflation}

To achieve an epoch of near-exponential expansion, we seek a solution to the above classical field equations for $\varphi(t)$ in which the Hubble parameter, $H$, is approximately constant. This is ensured if the total energy density is dominated by $\rho_\varphi$, with $\rho_\varphi$ also approximately constant. As we have seen, energy conservation implies the pressure must then satisfy $p_\varphi \approx - \rho_\varphi$. Inspection of eqs.~\pref{Rhoandpvarphi} shows that both of these conditions are satisfied if the $\varphi$ kinetic energy is negligible compared with its potential energy: 
\be \label{SmallKE}
   \frac12 \dot{\varphi}^2 \ll V(\varphi) \,,
\ee
since then $p_\varphi \simeq - V(\varphi) \simeq - \rho_\varphi$. So long as $V(\varphi)$ is also much larger than any other energy densities, it would dominate the Friedmann equation and $H^2 \simeq V/(3 M_p^2)$ would then be approximately constant.

What properties must $V(\varphi)$ satisfy in order to allow \pref{SmallKE} to hold for a sufficiently long time? This requires a long period of time where $\varphi$ moves slowly enough to allow {\sl both} the neglect of $\frac12 \, \dot\varphi^2$ relative to $V(\varphi)$ in the Friedmann equation, \pref{FriedmannEqn}, {\sl and} the neglect of $\ddot\varphi$ in the scalar field equation, \pref{ScalarFieldEqn}.

The second of these conditions allows eq.~\pref{ScalarFieldEqn} to be written in the approximate {\sl slow-roll} form,
\be \label{SRScalarFieldEqn}
    \dot{\varphi} \approx - \left( \frac{V'}{3H} \right) \,.
\ee
Using this in \pref{SmallKE} then shows $V$ must satisfy $(V')^2/(9 H^2 V) \ll 1$, leading to the condition that slow-roll inflation requires $\varphi$ must lie in a region for which
\be \label{epsilonParameterDef}
    \epsilon := \frac12 \left( \frac{M_p V'}{V} \right)^2
    \ll 1 \,.
\ee
Physically, this condition requires $H$ to be approximately constant over any given Hubble time, inasmuch as $3 M_p^2 H^2 \simeq V$ implies $6M_p^2 H\dot H \simeq V' \dot \varphi \simeq -(V')^2/3H$ and so 
\be
 -  \frac{\dot H}{H^2} \simeq \frac{(V')^2}{18H^4M_p^2} \simeq \frac{M_p^2(V')^2}{2V^2} = \epsilon\ll 1 \,.
\ee

Self-consistency also demands that if eq.~\pref{SRScalarFieldEqn} is differentiated to compute $\ddot \varphi$ it should be much smaller than $3H\dot \varphi$. Performing this differentiation and demanding that $\ddot \varphi$ remain small (in absolute value) compared with $3 H \dot\varphi$, then implies $|\eta| \ll 1$ where
\be \label{etaParameterDef}
    \eta := \frac{M_p^2 \, V''}{V} \,,
\ee
defines the second slow-roll parameter.  The slow-roll parameters $\epsilon$ and $\eta$ are important \cite{SlowRollParams} because (as shown below) the key predictions of single-field slow-roll inflation for density fluctuations can be expressed in terms of the three parameters $\epsilon$, $\eta$ and the value, $H_\ssI$, of the Hubble parameter during inflation.

Given an explicit shape for $V(\varphi)$ one can directly predict the amount of inflation that occurs after the epoch of Hubble exit (where currently observable scales become larger than the Hubble length). This is done by relating the amount of expansion directly to the distance $\varphi$ traverses in field space during this time. To this end, rewriting eq.~\pref{SRScalarFieldEqn} in terms of $\varphi' \equiv \exd \varphi/\exd a$, leads to
\be
    \frac{\exd\varphi}{\exd a} = \frac{\dot{\varphi}}{\dot{a}} =
    -\, \frac{V'}{3a H^2} = - \, \frac{M_p^2 \, V'}{aV} \,,
\ee
which when integrated between horizon exit, $\varphi_{\rm he}$, and final value, $\varphi_{\rm end}$, gives the amount of expansion during inflation as $a_{\rm end}/a_{\rm he} = e^{N_e}$, with
\be \label{NIvsPhiEqn}
    N_e = \int_{a_{\rm he}}^{a_{\rm end}} \frac{\exd a}{a}
    = \int_{\varphi_{\rm end}}^{\varphi_{\rm he}} \exd
    \varphi \left( \frac{V}{M_p^2 \, V'} \right)
    = \frac{1}{M_p} \int_{\varphi_{\rm end}}^{\varphi_{\rm he}}
    \frac{\exd \varphi}{\sqrt{2 \epsilon}} \,.
\ee
In these expressions $\varphi_{\rm end}$ can be defined by the point where the slow-roll parameters are no longer small, such as where $\epsilon \simeq \frac12$. Then this last equation can be read as defining $\varphi_{\rm end}(N_e)$, as a function of the desired
number of $e$-foldings between the the epoch of horizon exit and the end of inflation, since this is this quantity constrained to be large by the horizon and flatness problems. 

Notice also that if $\epsilon$ were approximately constant during inflation, then eq.~\pref{NIvsPhiEqn} implies that $N_e \approx (\varphi_{\rm he} - \varphi_{\rm end})/(\sqrt{2\epsilon} \, M_p)$. In such a case $\varphi$ must traverse a range of order $N_e M_p \sqrt{2\epsilon}$ between $\varphi_{\rm he}$ and $\varphi_{\rm end}$. This is larger than order $M_p$ provided only that $1 \gg \epsilon \gsim 1/N_e^2$, showing why Planckian fields are often of interest for inflation \cite{Lyth}.

It is worth working through what these formulae mean in a few concrete choices for the shape of the scalar potential. 

\subsubsection*{Example I:  Quadratic model}

The simplest example of an inflating potential \cite{chaotic, msqphisq} chooses $\varphi$ to be a free massive field, for which
\be \label{PotentialDef}
    V = \frac12 \, m^2 \, \varphi^2 \,,
\ee
and so $V' = m^2 \, \varphi$ and $V'' = m^2$, leading to slow-roll parameters of the form
\be
    \epsilon = \frac12 \, \left(
    \frac{2 M_p}{\varphi} \right)^2
    \qquad \hbox{and} \qquad
    \eta = \frac{2 M_p^2}{\varphi^2}  \,,
\ee
and so $\epsilon = \eta$ in this particular case, and slow roll requires $\varphi \gg M_p$. The scale for inflation in this field range is $V = \frac12 \, m^2 \, \varphi^2$ and so $H_I^2 \simeq m^2\,\varphi^2/(6 \, M_p^2)$. We can ensure $H_I^2/M_p^2 \ll 1$ even if $\varphi \gg M_p$ by choosing $m/M_p$ sufficiently small. Observations will turn out to require $\epsilon \sim \eta \sim 0.01$ and so the regime of interest is $\varphi_{\rm he} \sim 10 M_p$, and so $H_I/M_p \ll 1$ requires $m/M_p \ll 0.1$. 

In this large-field regime $\varphi$ (and so also $V$ and $H$) evolves only very slowly despite there being no nearby stationary point for $V$ because Hubble friction slows $\varphi$'s slide down the potential. Since $\varphi$ evolves towards smaller values, eventually slow roll ends once $\eta$ and $\epsilon$ become $O(1)$. Choosing $\varphi_{\rm end}$ by the condition $\epsilon(\varphi_{\rm end}) = \eta(\varphi_{\rm end}) = \frac12$ implies $\varphi_{\rm end} = 2 M_p$. The number of $e$-foldings between horizon exit and $\varphi_{\rm end} = 2 M_p$ is then given by eq.~\pref{NIvsPhiEqn}, which in this instance becomes
\be \label{NIvsPhiEqnLF}
    N_e    = \int_{2M_p}^{\varphi_{\rm he}} \exd
    \varphi \left( \frac{\varphi}{2M_p^2} \right)
    = \left(\frac{\varphi_{\rm he}}{2 M_p} \right)^2 - 1 \,,
\ee
and so obtaining $N_e \sim 63$ $e$-foldings (say) requires choosing $\varphi_{\rm he} \sim 16 \, M_p$. In particular $\epsilon_{\rm he} := \epsilon(\varphi_{\rm he})$ and $\eta_{\rm he} := \eta(\varphi_{\rm he})$ can be expressed directly in terms of $N_e$, leading to
\be 
  \epsilon_{\rm he} = \eta_{\rm he} = \frac{1}{2(N_e+1)} \,,
\ee
which are both of order $10^{-2}$ for $N_e \simeq 60$. As seen below, the prediction $\epsilon = \eta$ is beginning to be disfavoured by cosmological observations.

\subsubsection*{Example II:  pseudo-Goldstone axion}

The previous example shows how controlled inflation requires the inflaton mass to be small compared with the scales probed by $\varphi$. Small masses arise because the condition $|\eta| \ll 1$ implies the inflaton mass satisfies $m^2 \sim |V'' | \sim |\eta \,V/M_p^2| \ll V/M_p^2 \simeq 3H^2$. Consequently $m$ must be very small compared with $H$, which itself must be Planck suppressed compared with other scales (such as $v \sim V^{1/4}$) during inflation. From the point of view of particle physics such small masses pose a puzzle because it is fairly uncommon to find interacting systems with very light spinless particles in their low-energy spectrum.\footnote{From an EFT perspective having a light scalar requires the coefficients of low-dimension effective interactions like $\phi^2$ to have unusually small coefficients like $m^2$ rather than being as large as the (much larger) UV scales $M^2$.} 

The main exceptions to this statement are Goldstone bosons for the spontaneous breaking of continuous global symmetries since these are guaranteed to be massless by Goldstone's theorem. This makes it natural to suppose the inflaton to be a pseudo-Goldstone boson ({\em i.e.}~a would-be Goldstone boson for an approximate symmetry, much like the pions of chiral perturbation theory). In this case Goldstone's theorem ensures the scalar's mass (and other couplings in the scalar potential) must vanish in the limit the symmetry becomes exact, and this `protects' it from receiving generic UV-scale contributions.  For abelian broken symmetries this shows up in the low-energy EFT as an approximate shift symmetry under which the scalar transforms inhomogeneously: $\varphi \to \varphi + $constant. 

If the approximate symmetry arises as a $U(1)$ phase rotation for some microscopic field, and if this symmetry is broken down to discrete rotations, $Z_\ssN \subset U(1)$, then the inflaton potential is usually trigonometric \cite{NatInf}:
\be \label{PotentialDef}
    V = V_0 + \Lambda^4 \left[ 1 - \cos \left( \frac{\varphi}{f} \right) \right] = V_0 + 2\Lambda^4 \sin^2 \left( \frac{\varphi}{2f} \right)  \,,
\ee
for some scales $V_0$, $\Lambda$ and $f$. Here $V_0$ is chosen to agree with $\rho_{DE}$ and because $\rho_{DE}$ is so small the parameter $V_0$ is dropped in what follows. The parameter $\Lambda$ represents the scale associated with the explicit breaking of the underlying $U(1)$ symmetry while $f$ is related to the size of its spontaneous breaking. The statement that the action is approximately invariant under the symmetry is the statement that $\Lambda$ is small compared with UV scales like $f$. Expanding about the minimum at $\varphi= 0$ reveals a mass of size $m = \Lambda^2/f \ll \Lambda \ll f$, showing the desired suppression of the scalar mass.

With this choice $V' = (\Lambda^4/f) \sin(\varphi/f)$ and $V'' = (\Lambda^4/f^2) \cos(\varphi/f)$, leading to slow-roll parameters of the form
\be \label{SlowRollNatInf}
    \epsilon = \frac{M_p^2}{2f^2} \, \cot^2 \left(
    \frac{\varphi}{2f} \right)
    \qquad \hbox{and} \qquad
    \eta = \frac{M_p^2}{2f^2} \left[ \cot^2 \left( \frac{\varphi}{2f} \right) - 1 \right] \,,
\ee
and so $\eta = \epsilon - (M_p^2/2f^2)$. Notice that in the limit $M_p \lsim \varphi \ll f$ these go over to the $m^2 \varphi^2$ case examined above, with $m = \Lambda^2/f$. 

Slow roll in this model typically requires $f \gg M_p$. This can be seen directly from \pref{SlowRollNatInf} for generic $\varphi \simeq f$, but also follows when $\varphi \ll f$ because in this case the potential is close to quadratic and slow roll requires $M_p \ll \varphi \ll f$. The scale for inflation is $V \simeq \Lambda^4$ and so $H_I \sim \Lambda^2/ M_p$. This ensures $H_I^2/M_p^2 \ll 1$ follows from the approximate-symmetry limit which requires $\Lambda \ll M_p$. The condition $\epsilon \sim 0.01$ is arranged by choosing $f \sim 10 M_p$, but once this is done the prediction $\epsilon \simeq \eta$ is in tension with recent observations.

The number of $e$-foldings between horizon exit and $\varphi_{\rm end}$ is again given by eq.~\pref{NIvsPhiEqn}, so
\be \label{NIvsPhiEqnLF}
    N_e    =  \frac{2f}{M_p^2}  \int_{\varphi_{\rm end}}^{\varphi_{\rm he}} \exd
    \varphi \; \tan \left( \frac{\varphi}{2f} \right) 
    = \left(\frac{2f}{M_p} \right)^2  \ln \left| \frac{ \sin(\varphi_{\rm he}/2f) }{\sin(\varphi_{\rm end}/2f) }\right| \,,
\ee
which is only logarithmically sensitive to $\varphi_{\rm he}$, but which can easily be large due to the condition $f \gg M_p$. 

While models such as this do arise generically from UV completions like string theory \cite{UVaxion}, axions in string theory typically arise with $f \ll M_p$ \cite{StringAxionsHard}, making the condition $f \gg M_p$ tricky to arrange \cite{Nflation}.

\subsubsection*{Example III:  pseudo-Goldstone dilaton}

Another case where the inflaton mass is protected by an approximate shift symmetry arises when it is a pseudo-Goldstone boson for a scaling symmetry of the underlying UV theory. Such `accidental' scale symmetries turn out to be fairly common in explicit examples of UV completions because scale invariances are automatic consequences of higher-dimensional supergravities \cite{SugraScaling}. Because it is a scaling symmetry the same arguments leading to trigonometric potentials for the compact $U(1)$ rotations instead in this case generically lead to exponential potentials \cite{LargeField}. 

In this case the form expected for the scalar potential during the inflationary regime would be
\be \label{PotentialDef}
    V = V_0 - V_1 e^{- {\varphi}/{f} } + \cdots    \,,
\ee
for some scales $V_0$, $V_1$ and $f$. Our interest is in the regime $\varphi \gg f$ and in this regime $V_0$ dominates, and so is chosen as needed for inflationary cosmology, with $H_I^2 \simeq V_0/(3M_p^2)$. Control over the semiclassical limit requires $V_0 \ll M_p^4$.

With this choice the relevant potential derivatives are $V' \simeq (V_1/f) \, e^{-\varphi/f}$ and $V'' \simeq -(V_1/f^2 ) \, e^{-\varphi/f}$ leading to slow-roll parameters of the form
\be \label{SlowRollNatInf2}
    \epsilon \simeq \frac{1}{2}  \left( \frac{M_p V_1}{f V_0} \right)^2 \, e^{-2\varphi/f}    \qquad \hbox{and} \qquad
    \eta \simeq -\left( \frac{M_p^2 V_1}{f^2 V_0} \right) e^{-\varphi/f} \,,
\ee
and so 
\be
 \epsilon =  \frac12 \left( \frac{f}{M_p} \right)^2 \eta^2 \,. 
\ee
The number of $e$-foldings between horizon exit and $\varphi_{\rm end}$ is again given by eq.~\pref{NIvsPhiEqn}, so
\be \label{NIvsPhiEqnLF}
    N_e    =  \left( \frac{fV_0}{M_p^2 V_1} \right) \int_{\varphi_{\rm end}}^{\varphi_{\rm he}} \exd
    \varphi \; e^{\varphi/f}  
    = \left(\frac{f^2 V_0}{M_p^2 V_1} \right)  \left[ e^{\varphi_{\rm he}/f} - e^{\varphi_{\rm end}/f} \right] \,,
\ee
which can easily be large so long as $\varphi_{\rm he} \gg f$ and $\varphi_{\rm end}/f$ is order unity. 

Notice that $\epsilon$ and $\eta$ are generically small whenever $\varphi \gg f$, even if $V_1 \sim V_0$, so there is no need to require $f$ be larger than $M_p$ to ensure a slow roll. Typical examples of underlying UV theories (see below) give $f \sim M_p$, in which case $\epsilon \simeq \eta^2$. It turns out that this prediction provides better agreement with experiment than $\epsilon \simeq \eta$ does, and (as seen below) the generic expectation that $\epsilon \sim \eta^2$ has  potentially interesting observational consequences for measurements of primordial gravitational waves because it relates the as-yet-unmeasured tensor-to-scalar ratio, $r \lsim 0.07$, to the observed spectral tilt, $n_s \simeq 0.96$, giving the prediction $r_{\rm th} \simeq (n_s - 1)^2 \simeq 0.002$.

Interestingly, many successful inflationary models can be recast into this exponential form, usually with specific values predicted for $f$. The earliest instance using an exponential potential \cite{Goncharov} came from a supergravity example with $f = \sqrt{\frac16} \,M_p$, with a nonlinearly realized $SU(1,1)$ symmetry. Such symmetries are now known to arise fairly commonly when dimensionally reducing higher-dimensional supersymmetric models \cite{SugraScaling, LargeField}. This early supergravity example foreshadows the results from a class of explicit higher-dimensional UV completions within string theory \cite{FibreInf}, which reduce to the above with $f = \sqrt3\, M_p$, while the first extra-dimensional examples of this type \cite{ExpRadius} gave $f = \sqrt2 \; M_p$.  

In fact, the Higgs-inflation model described earlier can also be recast to look like a scalar field with an exponential potential of the form considered here, once it is written with canonically normalized fields. The prediction in this case is $f = \sqrt{\frac32}\, M_p$. The same is true for another popular model that obtains inflation using curvature-squared interactions \cite{Starobinsky}, for which again $f = \sqrt{\frac32}\, M_p$. Although both of these models are hard to obtain in a controlled way from UV completions directly, their formulation in terms of exponential potentials may provide a way to do so through the back door.

\section{Cosmology: Fluctuations}

This section repeats the previous discussion of $\Lambda$CDM cosmology and its peculiar initial conditions, but extends it to the properties of fluctuations about the background cosmology.

\subsection{Structure formation in $\Lambda$CDM}

Previous sections show that the universe was very homogeneous at
the time of photon last scattering, since the temperature
fluctuations observed in the distribution of CMB photons have an
amplitude $\delta T/T \sim 10^{-5}$. On the other hand the
universe around us is full of stars and galaxies and so is far
from homogeneous. How did the one arise from the other?

The basic mechanism for this in the $\Lambda$CDM model is based on gravitational instability: the gravitational force towards an initially
over-dense region acts to attract even more material towards this
region, thereby making it even more dense. This process can feed
back on itself until an initially small density perturbation
becomes dramatically amplified, such as into a star. This section
describes the physics of this instability, in the very early
universe when the density contrasts are small enough to be
analyzed perturbatively in the fluctuation amplitude. The
discussion follows that of ref.~\cite{Mukhanov, RobertMexico}.

\subsubsection{Nonrelativistic Density Perturbations}
\label{Sec:growthRadDom}

We start with the discussion of gravitational instability for non-relativistic fluids, both for simplicity and since this is the sector that actually displays the instability in practice. The equations found here provide a self-consistent description of how linearized density fluctuations for non-relativistic matter evolve in a matter-dominated universe (which is the main one relevant for structure growth), and also turn out to capture how non-relativistic density fluctuations grow when the total energy density is dominated by radiation or Dark Energy.

The following equations of motion describe the evolution of a simple non-relativistic fluid with energy density, $\rho$, pressure, $p$, entropy density, $s$, and local fluid velocity $\bfv$. Each equation expresses a local conservation law, 
\bea
    \frac{\partial \rho}{\partial t} + \nabla \cdot (\rho \bfv)
    &=& 0  \qquad \hbox{(energy conservation)} \nn\\
    \rho \left[ \frac{\partial \bfv}{\partial t} + (\bfv \cdot
    \nabla) \bfv \right] + \nabla p + \rho \nabla \phi &=& 0
    \qquad \hbox{(momentum conservation)}  \\
    \frac{\partial s}{\partial t} +  \nabla \cdot (s \bfv) &=& 0
    \qquad \hbox{(entropy conservation)} \nn \\
    \nabla^2 \phi - 4 \pi G \rho &=& 0 \qquad \hbox{(universal
    gravitation)} \,, \nn
\eea
and they are imagined supplemented by an equation of state, $p = p(\rho,s)$. Here $\phi$ denotes the local Newtonian gravitational potential. Because they are nonrelativistic these equations are expected to break down for super-Hubble modes, for which $k/a \lsim H$ and the proscription against motion faster than light plays an important role. 

For cosmological applications expand about a homogeneously
and radially expanding background fluid configuration. For these
purposes consider a fluid background for which $\bfv_0 =
H(t) \, \bfr$, where $H(t)$ is assumed a given function of $t$. In
this case $\nabla \cdot \bfv_0 = 3 H(t)$. This flow is motivated
by the observation that it corresponds to the proper velocity if
particles within the fluid were moving apart from one another
according to the law $\bfr(t) = a(t) \, \bfy$, with $\bfy$ being a
time-independent co-moving coordinate. In this case $\bfv_0 :=
\exd \bfr/\exd t = \dot{a} \, \bfy = H(t) \, \bfr(t)$ where $H =
\dot{a}/a$. In this sense $H(t)$ describes the Hubble parameter for the background fluid's expansion.

\medskip\noindent{\bf Background Quantities}

\smallskip\noindent
We now ask what the rest of the background quantities,
$\rho_0(t)$, $p_0(t)$ and $\phi_0(t)$ must satisfy in order to be
consistent with this flow. The equation of energy conservation
implies $\rho_0$ must satisfy
\be
    0 = \dot{\rho}_0 + \nabla \cdot
    (\rho_0 \bfv_0) = \dot{\rho}_0 +
    3 H \, \rho_0 \,,
\ee
and so, given $H = \dot{a}/a$, it follows that $\rho_0 \propto a^{-3}$. Not surprisingly, the density of a non-relativistic expanding fluid necessarily falls with universal expansion in the same way required by the full relativistic treatment.

Using this density in the law for universal gravitation requires
the gravitational potential, $\phi_0$, take the form
\be
    \phi_0 = \frac{2 \pi G \rho_0}{3} \, \bfr^2 \,,
\ee
and so $\nabla \phi_0 = \frac43 \, \pi G \rho_0 \, \bfr$. This
describes the radially-directed gravitational potential that acts
to decelerate the overall universal expansion.

Given this gravitational force, the momentum conservation
equation, using $\dot{\bfv}_0 + (\bfv_0 \cdot \nabla) \bfv_0 = [H
+ \dot{H}/H] \, \bfv_0$ and $\bfv_0 = H \, \bfr$, becomes
\be \label{NRFriedmannEqn}
    \left[ \dot{H} + H^2 + \frac{4 \pi G \rho_0}{3} \right] \bfr =
    0 \,.
\ee
This is equivalent to the Friedmann equation, as is now shown.
Notice that if we take $a \propto t^\alpha$ then $H = \alpha/t$
and $\dot{H} = -\alpha/t^2 = -H^2/\alpha$. This, together with
$\rho_0 \propto a^{-3} \propto t^{-3\alpha}$, is consistent with
eq.~\pref{NRFriedmannEqn} only if $\alpha = 2/3$, as expected for
a matter-dominated universe. Furthermore, with this choice for
$\alpha$ we also have $\dot{H} + H^2 = - \frac12 \,H^2$, and so
eq.~\pref{NRFriedmannEqn} is equivalent to
\be
    H^2 = \frac{8 \pi G}{3} \, \rho_0 \,,
\ee
which is the Friedmann equation, as claimed.  

When studying perturbations we solve the entropy equation by taking $s_0 = 0$. This is done mostly for simplicity, but it is also true that for many situations the thermal effects described by $s_0$ play a negligible role.

\medskip\noindent{\bf Perturbations during matter domination}

\smallskip\noindent
To study perturbations about this background take $\bfv = \bfv_0 + \dbfv$, $\rho = \rho_0 + \drho$, $p = p_0 + \dpr$, $s = \dss$ and $\phi = \phi_0 + \dphi$, and expand the equations of motion to first order in the perturbations. Defining $D_t = \partial/\partial t + \bfv_0 \cdot \nabla$, the linearized equations in this case become
\bea \label{LinearPerts}
    D_t \, \drho + 3 H \, \drho + \rho_0 \nabla \cdot \dbfv
    &=& 0 \nn \\
    \rho_0  (D_t \, \dbfv + H \, \dbfv) + \nabla \dpr +
    \rho_0 \nabla \dphi &=& 0 \\
    D_t \, \dss  &=& 0 \nn \\
    \nabla^2 \dphi - 4 \pi G \drho &=& 0 \,. \nn
\eea
To obtain this form for the momentum conservation equation requires using the equations of motion for the background quantities.

Our interest is in the evolution of $\delta \rho$, and this can be isolated by taking $D_t$ of the first of eqs.~\pref{LinearPerts} and the divergence of the second if these equations, and using the results to eliminate $\delta \bfv$. The remaining equations involve the two basic fluid perturbations, $\drho$ and $\dss$, and imply both $D_t \, \dss = 0$ and
\be
    D_t^2 \, \left( \frac{\drho}{\rho_0} \right)
    + 2 H \, D_t \, \left( \frac{\drho}{\rho_0} \right)
    - c_s^2 \, \nabla^2 \left( \frac{\drho}{\rho_0} \right)
    - 4 \pi G \rho_0 \, \left( \frac{\drho}{\rho_0} \right)
    = \frac{\xi}{\rho_0} \, \delta s \,,
\ee
where
\be
  c_s^2 := \left( \frac{\partial p}{\partial \rho} \right)_{s0} \quad \hbox{and} \quad
  \xi := \left( \frac{\partial p}{\partial s} \right)_{\rho 0} \,.
\ee

In order to analyze the solutions to this equation, it is convenient to change variables to a co-moving coordinate, $\bfy$, defined by $\bfr = a(t) \, \bfy$. In this case, for any function $f = f(\bfr,t)$ we have $(\partial f/\partial t)_\bfy = (\partial f/\partial t)_\bfr + H \bfr \cdot \nabla f = D_t \, f$, and $\nabla f = (1/a) \nabla_y f$. Fourier transforming the perturbations in co-moving coordinates, $\drho/\rho_0 = \delta_k(t) \, \exp[i \bfk \cdot \bfy]$, leads to the following master equation governing density perturbations
\be
    \ddot{\delta}_k + 2H \, \dot{\delta}_k + \left( \frac{c_s^2 \,
    k^2}{a^2} - 4 \pi G \rho_0 \right)  \, \delta_k
    = \left( \frac{\xi}{\rho_0} \right) \,
    \dss \,,
\ee
where the over-dot denotes $\exd/\exd t$.

These equations have solutions whose character depends on the relative size of $k/a$ and the Jeans wave-number,
\be
    k_J^2(t) = \frac{4 \pi G \rho_0(t)}{c_s^2(t)}
    = \frac{3 H^2(t)}{2 \, c_s^2(t)}
    \,,
\ee
with instability occurring once $k/a \ll k_J$. Notice that so long
as $c_s \sim O(1)$ the Jeans length is comparable in size to the
Hubble length, $\ell_J \sim H^{-1}$. For adiabatic fluctuations
($\dss_k = 0$) the above equation implies that the
short-wavelength fluctuations ($k/a \gg k_J$) undergo damped
oscillations of the form
\be \label{DampedOscillations}
    \delta_k(t) \propto a^{-1/2} \, \exp \left[ \pm i k c_s \int^t
    \frac{\exd t'}{a(t')} \right] \,.
\ee
The overall prefactor of $a^{-1/2}$ shows how these oscillations are damped due to the universal expansion, or Hubble friction.

Long-wavelength adiabatic oscillations ($k/a \ll k_J$) exhibit an instability, with the unstable mode growing like a power law of $t$. The approximate solutions are
\be
    \delta_k(t) \propto t^{2/3} \propto a(t) \qquad
    \hbox{and} \qquad \delta_k(t) \propto t^{-1} \propto a^{-3/2}(t) \,,
\ee
with the $\delta_k(t) \propto a \propto t^{2/3}$ solution describing the unstable mode. The instability has power-law rather than exponential growth because the expansion of space acts to reduce the density, and this effect fights the density increase due to gravitational collapse.

Because both the red-shifted wave-number, $k/a$, and the Jeans wave-number, $k_J$, depend on time, the overall expansion of the background can convert modes from stable to unstable (or vice
versa). Whether this conversion is towards stability or instability depends on the the time dependence of $a k_J$, which is governed by the time-dependence of the combination $a H/c_s$. If $a \propto t^\alpha$ then $a H \propto t^{\alpha -1} \propto a^{1-1/\alpha}$, and so $aH$ increases with $t$ if $\alpha > 1$ and decreases with $t$ if $\alpha < 1$. Since $\alpha = 2/3$ for the matter-dominated universe of interest here, it follows that $aH \propto t^{-1/3} \propto a^{-1/2}$, and so {\sl decreases} with $t$. Provided that $c_s$ does not change much, this ensures that in the absence of other influences modes having fixed $k$ pass from being unstable to stable as $a$ increases due to the overall expansion.

\medskip\noindent{\bf Perturbations during radiation and vacuum domination}

\smallskip\noindent
A completely relativistic treatment of density perturbations requires following fluctuations in the matter stress energy as well as in the metric itself (since these are related by Einstein's equations relating geometry and stress-energy). The details of such calculations go beyond the scope of these notes, although some of the main features are described below. But the above considerations suffice to address a result that is an important part of the structure-formation story: the stalling of perturbation growth for nonrelativistic matter during radiation- or vacuum-dominated epochs.

To contrast how fluctuations grow during radiation and matter domination it is instructive to examine the transition from radiation to matter domination. To this end we again track the growth of density fluctuations for non-relativistic matter, $\delta \rho_{m0}/\rho_{m0}$, and do so using the same Fourier-transformed equation as before,
\be \label{deltaeq}
  \ddot \delta_\bfk + 2 H \, \dot \delta_\bfk + \left( \frac{c_s^2 \bfk^2}{a^2}  - 4\pi G \rho_{m0} \right) \delta_\bfk= 0 \,,
\ee
but with $H^2 = 8 \pi G \rho_0/3$ where $\rho_0 = \rho_{m0} + \rho_{r0}$ includes both radiation and matter. In particular, during the transition between radiation and matter domination the Hubble scale satisfies
\be \label{RadDomBackgndH}
 H^2(a) = \frac{8 \pi G \rho_{0}}{3} = \frac{H_{\rm eq}^2}{2} \left[ \left( \frac{a_{\rm eq}}{a} \right)^3 + \left( \frac{a_{\rm eq}}{a} \right)^4 \right] \,,
\ee
where radiation-matter equality occurs when $a = a_{\rm eq}$, at which point $H(a=a_{\rm eq}) = H_{\rm eq}$. 

As described above, any departure from the choice $a(t) \propto t^{2/3}$ --- such as occurs when radiation is non-negligible in $\rho(a)$ --- means that the background momentum-conservation equation, eq.~\pref{NRFriedmannEqn}, is no longer satisfied. Instead the expression for $H$ comes from solving the fully relativistic radiation-dominated Friedmann equation, eq.~\pref{RadDomBackgndH}. But this does not mean that the nonrelativistic treatment of the fluctuations, eqs.~\pref{LinearPerts}, must fail, since the important kinematics and gravitational interactions amongst these perturbations remain the Newtonian ones. To first approximation the leading effect of the radiation domination for these fluctuations is simply to change the expansion rate, as parameterized by $H(a)$ in \pref{RadDomBackgndH}.

For all modes for which the pressure term, $c_s^2\, \bfk^2/a^2$, is negligible, \pref{deltaeq} implies $\delta(x)$ satisfies
\be
  2x(1+x) \, \delta'' + (3x+2) \, \delta' - 3 \, \delta = 0 \,,
\ee
where the rescaled scale factor, $x = a/a_{\rm eq}$, is used as a proxy for time and primes denote differentiation with respect to $x$. As is easily checked, this is solved by $\delta^{(1)} \propto \left( x + \frac23 \right)$, and so the growing mode during matter domination ({\em i.e.} $x \gg 1$) does not also grow during radiation domination ($x \ll 1$).Furthermore, the solution linearly independent to this one can be found using the Frobenius method, and this behaves for $x \ll 1$ ({\em i.e.} deep in the radiation-dominated regime) as $\delta^{(2)} \propto \delta^{(1)}  \ln x + (\hbox{analytic})$, where `analytic' denotes a simple power series proportional to $1 + c_1 x + \cdots$. These solutions show how density perturbations for non-relativistic matter grow at most logarithmically during the radiation-dominated epoch.

A similar analysis covers the case where Dark Energy (modelled as a cosmological constant) dominates in an $\Omega = 1$ universe. In this case $4\pi G \rho_{m0} \sim \Omega_m H^2 \ll H^2$ and so the instability term becomes negligible relative to the first two terms of \pref{deltaeq}. This leads to
\be
  \ddot \delta + 2H \, \dot \delta \simeq 0 \,,
\ee
which has as solution $\dot \delta \propto a^{-2}$. Integrating again gives a frozen mode, $\delta \propto a^0$, and a damped mode that falls as $\delta \propto a^{-2}$ when $H$ is constant (as it is when Dark Energy dominates and $a \propto e^{Ht}$). This shows that non-relativistic density perturbations also stop growing once matter domination ends. 

We are now in a position to summarize how inhomogeneities grow in the late universe, assuming the presence of an initial spectrum of very small primordial density fluctuations. The key observation is that several conditions all have to hold in order for there to be appreciable growth of density inhomogeneities. These conditions are:
\begin{enumerate}
\item Fluctuations of any type do not grow for super-Hubble modes, for which $k/a \ll H$, regardless of what type of matter dominates the background evolution.
\item Fluctuations in nonrelativistic matter can be unstable, growing as $\delta_k \propto a(t)$, but only in a matter-dominated universe ahd for those modes in the momentum window $H \ll (k/a) \ll H/c_s$. 
\item No fluctuations in relativistic matter ever grow appreciably, either inside or outside the Hubble scale. (Although this is not shown explicitly above for relativistic matter, and requires the fully relativistic treatment, the instability window $H \ll k/a \ll H/c_s$ for nonrelativistic fluctuations is seen to close as they become relativistic --- {\em i.e.} as $c_s \to 1$.)
\end{enumerate}

Before pursuing the implications of these conditions for instability, a pause is in order to describe what properties of fluctuations are actually measured.

\subsubsection{The Power Spectrum}

The presence of unstable density fluctuations implies the universe does not remain precisely homogeneous and isotropic once matter domination begins, and so the view seen by observers like us depends on their locations in the universe relative to the fluctuations. For this reason, when comparing with observations it is less useful to try to track the detailed form of a specific fluctuation and instead better to characterize fluctuations by their statistical properties, since these can be more directly applied to observers without knowing their specific place in the universe. In particular we imagine there being an ensemble of density fluctuations, whose phases we assume to be
uncorrelated and whose amplitudes are taken to be random variables.

On the observation side statistical inferences can be made about the probability distribution governing the distribution of fluctuation amplitudes by measuring statistical properties of the matter distribution observed around us. For instance, a useful statistic measures the mass-mass auto-correlation function
\be \label{CorrDef}
    \xi(\bfr - \bfr') \equiv \frac{ \langle \drho(\bfr) \,
    \drho(\bfr') \rangle}{\langle{\rho} \rangle^2} \,,
\ee
which might be measured by performing surveys of the positions of large samples of galaxies.\footnote{A practical complication arises because although galaxies are relatively easy to count, most of the mass density is actually Dark Matter. Consequently assumptions are required to relate these to one another; the usual choice being that the galaxy and mass density functions are related to one another through a phenomenologically defined `bias' factor.} When using \pref{CorrDef} with observations the average $\langle \cdots \rangle$ is interpreted as integration of one of the positions (say, $\bfr'$) over all directions in the sky.\footnote{The density correlation function can also be measured using the temperature fluctuations of the CMB, because these fluctuations can be interpreted as redshifts acquired by CMB photons as they climb out of the gravitational potential wells formed by density fluctuations in nonrelativistic matter.} 

When making predictions $\langle \cdots \rangle$ instead is regarded as an average over whatever ensemble is thought to govern the statistics of the fluctuations $\delta_k$. Fourier transforming $\drho(\bfr) / \langle \rho \rangle = \int \exd^3 k \, \delta_k \, \exp[i \bfk\cdot \bfr]$ in comoving coordinates, as before, allows $\xi(\bfr)$ to be related to the following ensemble average over the Fourier mode amplitudes, $\delta_k$.
\be
    \xi(r) = \int \frac{\exd^3k}{(2\pi)^3} \; \langle |\delta_k|^2 \rangle \,
    \exp[i \bfk \cdot \bfr] = \frac{1}{2 \pi^2} \int_0^\infty \frac{\exd k}{k} \; k^3 \,
    P_{\rho\,}(k) \left( \frac{\sin kr}{kr} \right) \,,
\ee
which defines the density {\sl power spectrum}: $P_{\rho\,}(k) := \langle |\delta_k |^2 \rangle$. 

\begin{figure}[h]
\begin{center}
\includegraphics[width=100mm,height=80mm]{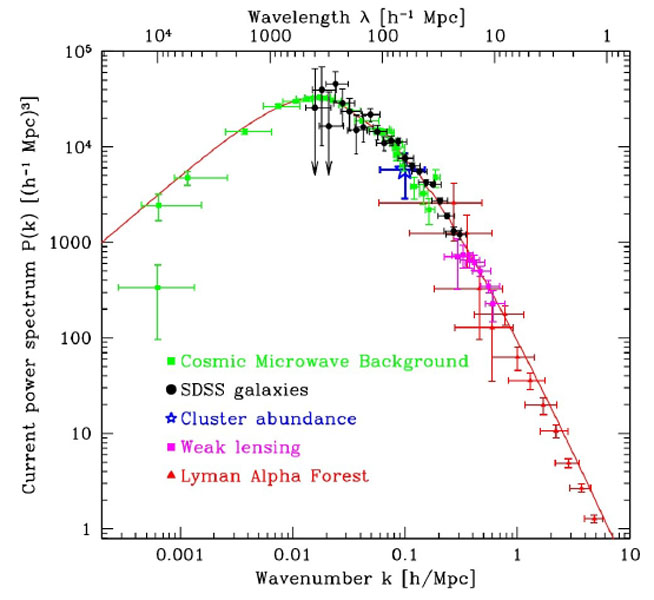} 
\caption{The power spectrum as obtained from measurements of the
  CMB spectrum, together with the SDSS Galaxy Survey, observations of abundances of galaxy clusteres and Lyman-$\alpha$ measurements (taken from \cite{SDSSparams}).} \label{Fig:PowerSpectrum} 
\end{center}
\end{figure}

For homogeneous and isotropic backgrounds $P_{\rho\,}(k)$ depends only on the magnitude $k = |\bfk|$ and not on direction, and this is used above to perform the angular integrations. The average in these expressions is over the ensemble, and it is this average which collapses the right-hand side down to a single Fourier integral. The last equality motivates the definition
\be
 \Delta^2_\rho(k) := \frac{k^3}{2\pi^2} \; P_{\rho\,}(k) \,.
\ee

\begin{figure}[h]
\begin{center}
\includegraphics[width=100mm,height=80mm]{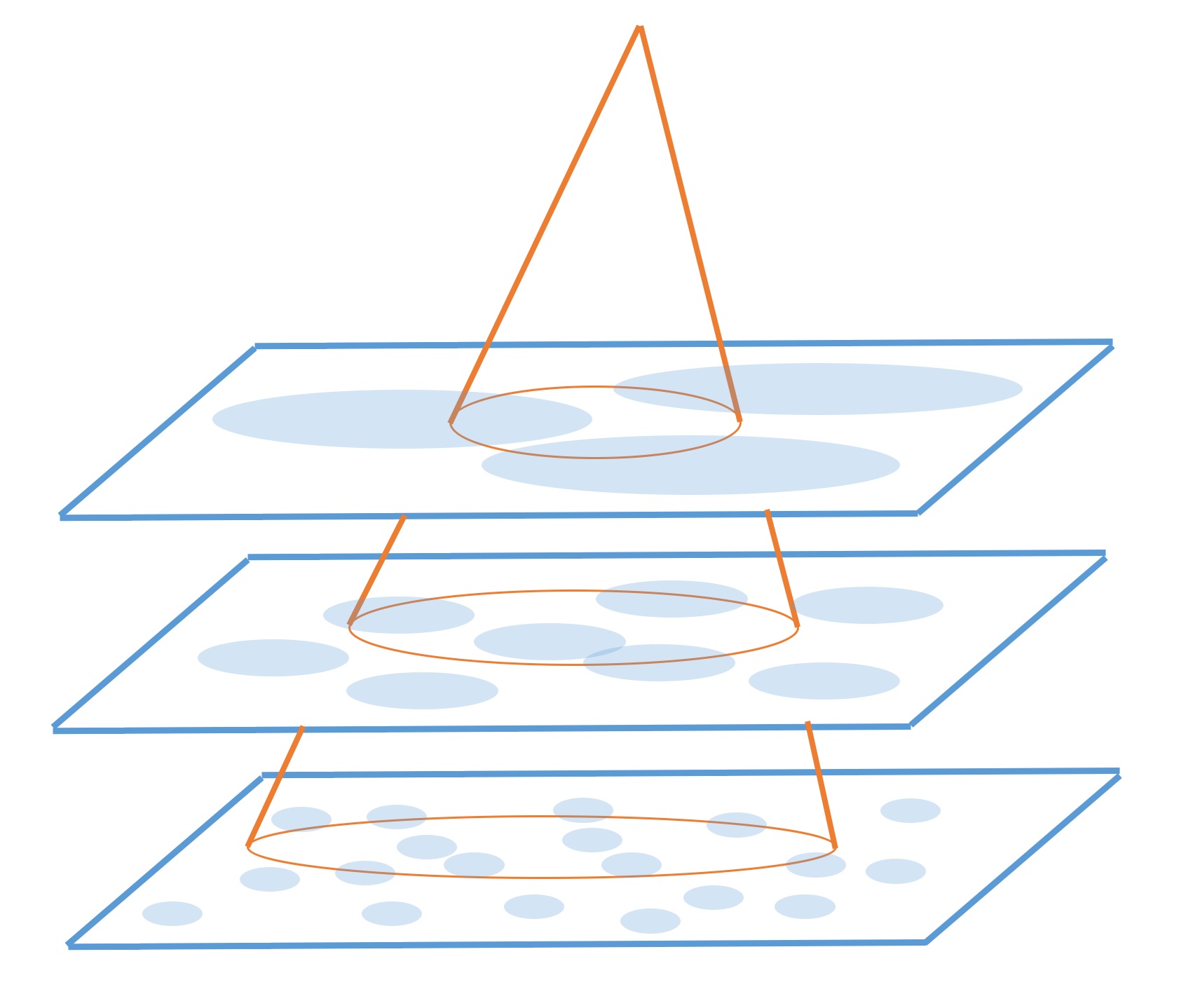} 
\caption{A sketch of several spatial slices intersecting the past light cone of an astronomer on Earth. The orange ovals indicate how the light cone has larger intersections with the spatial slices the further back one looks. The pale blue ovals indicate regions the size of the Hubble distance on each spatial slice. Correlations outside of these ovals (such as the uniformity of the CMB temperature) represent a puzzle for $\Lambda$CDM cosmology. The figure shows how later times (higher slices) have larger Hubble distances, as well as how observations only sample the largest distance scales on the most remote spatial slices. This illustrates why CMB measurements tend to constrain the power spectrum for small $k$ while observations of more nearby objects (like galaxy distributions or the distribution of foreground Lyman-$\alpha$ Hydrogen gas clouds) constrain larger $k$. } \label{LookBack} 
\end{center}
\end{figure}

A variety of observations over the years give the form of $P_{\rho\,}(k)$ as inferred from the distribution of structure around us, with results summarized in Figure \ref{Figure1}. As this figure indicates, inferences about the shape of $P_\rho(k)$ for small $k$ come from measurements of the temperature fluctuations in the CMB; those at intermediate $k$ come from galaxy distributions as obtained through galaxy surveys and those at the largest $k$ come from measurements of the how quasar light is absorbed by intervening Hydrogen gas clouds, the so-called Lyman-$\alpha$ `forest'.  The reasons why different kinds of measurements control different ranges of $k$ are illustrated in Figure \ref{LookBack}, which shows how the distance accessible to observations is correlated with how far back one looks into the universe: measurements of distant objects in the remote past ({\em e.g.}~the CMB) determine the shape of $P_\rho(k)$ for small $k$ while measurements of more nearby objects in the more recent past ({\em e.g.}~the Lyman-$\alpha$ forest) constrain $P_\rho(k)$ for larger $k$. 

The observations summarized in Figure \ref{Fig:PowerSpectrum} are well approximated by the phenomenological formula,
\be
    P(k) = \frac{A \, k^{n_s}}{(1 + \alpha \, k + \beta \, k^2)^2} \,,
\ee
where
\be
    \alpha = 16 \left( \frac{0.5}{\Omega \, h^2} \right) \; \hbox{Mpc}
    \quad \hbox{and} \quad
    \beta = 19 \left( \frac{0.5}{\Omega \, h^2} \right)^2 \; \hbox{Mpc}^2 
    \quad \hbox{and} \quad n_s = 0.97 \,.
\ee
Here $h = H_0/(100 \; \hbox{km/sec/Mpc}) \approx 0.7$, and $\Omega \approx 1$ denotes the present value of $\rho/\rho_c$. Given that $n_s \approx 1$ the observations suggest the power spectrum is close to linear, $P(k) \propto k$ for $k \ll k_{\star} \sim 0.07$ Mpc${}^{-1}$, and $P(k) \propto k^{-3}$ for $k \gg k_{\star}$. The value $k_\star$ here is simply defined to be the place where $P_\rho(k)$ turns over and makes the transition from $P_\rho \propto k$ to $P_\rho \propto k^{-3}$. 

As described below, there are good reasons to believe that the shape of $P_{\rho\,}(k)$ for $k \ll k_\star$ represents the pattern of primordial fluctuations inherited from the very early universe, while the shape for $k > k_\star$ reflects how fluctuations evolve in the later universe. Consequently observations are consistent with primordial fluctuations being close to\footnote{Close to but not equal to. Fits to $\Lambda$CDM cosmology establish $n_s$ is significantly different from 1. } a {\sl Zel'dovich} spectrum, $P_\rho(k) = A \, k$, corresponding to $n_s = 1$. As is seen below, the parameter $n_s$ is predicted to be close to, but not equal to, unity by inflationary models.

For later purposes it proves more convenient to work with the power spectrum for the Newtonian gravitational potential, $\delta\phi$, that is related to $\delta \rho$ by Poisson's equation --- the last of eqs.~\pref{LinearPerts} --- and so $\delta \phi_k \propto \delta_k /k^2$. Because of this relation their power spectra are related by $P_\phi(k) = P_\rho(k) / k^4$ as well as 
\be
 \Delta^2_\phi (k) := \frac{k^3}{2\pi^2} \, P_\phi(k) = \frac{P_\rho(k)}{2\pi^2 k} \propto \left\{ \begin{matrix} k^{n_s-1} & \hbox{if}
    \quad k \ll k_{\star} \\
    k^{n_s-5} & \hbox{if} \quad k \gg k_{\star} \end{matrix}
    \right. \,.
\ee
This last expression also clarifies why the choice $n_s = 1$ is called scale invariant. When $n_s = 1$ the primordial ($k \ll k_\star$)  spectrum for $\Delta^2_\phi(k)$ becomes $k$-independent, as would be expected for a scale-invariant process.

\subsubsection{Late-time structure growth}

Before trying to explain the properties of the primordial part of the power spectrum --- $\Delta^2_\phi(k) \propto A k^{n_s - 1}$ --- a further digression is in order to explain the explanation for why the measured distribution has the peculiar hump-shaped form, bending at $k \simeq k_\star$.  This shape arises due to the processing of density fluctuations by their evolution in the subsequent universe, as is now described.

The key observations go back to the three criteria, given at the end of \S\ref{Sec:growthRadDom}, for when fluctuating modes can grow. These state that the fluctuations that are most important are those involving nonrelativistic matter, although these remain frozen unless the universe is matter dominated and the mode number lies within the interval $H \ll k/a \ll H/c_s$.  These conditions for growth  
superimpose a $k$-dependence on $P_\rho(k)$, for the following reasons. 

The important wave-number $k_\star$ corresponds to the wave-number, $k_{\rm eq}$, for which modes satisfy $k/a \sim H$ at the epoch of radiation-matter equality (which occurs at $z_{\rm eq} = 3600$). Numerically, $k_{\rm eq}$ corresponds to a co-moving wave-number of order $k_{\rm eq} \sim 0.07$ Mpc${}^{-1}$. What is important about this scale is that it divides modes (with $k > k_{\rm eq}$) that re-enter the Hubble scale during radiation domination and those (with $k < k_{\rm eq}$) that re-enter during matter domination. 

Because they re-enter during matter domination, all Dark Matter fluctuation modes with $k < k_{\rm eq}$ are free to begin growing immediately on re-entry and have done so ever since, at least until they either become nonlinear --- when $\delta_k \sim \cO(1)$ --- or the universe reaches the very recent advent of Dark Energy domination. So the present-day power spectrum for these modes reflects the primordial one which was frozen into these modes long ago when they left the Hubble scale in the pre-$\Lambda$CDM era. It is these modes that reveal the primordial distribution
\be
    P(k) \propto k^{n_s} \qquad \hbox{(for $k \ll k_{\rm eq}$)} \,.
\ee

By contrast, those modes with $k \gg k_{\rm eq}$ re-enter the Hubble scale during the radiation-dominated epoch that precedes matter-radiation equality. The amplitude of these modes therefore remain frozen at their values at the time of re-entry, because they are unable to grow while the universe is radiation dominated. Consequently they remain stunted in amplitude relative to their longer-wavelength counterparts while waiting for the universe to become matter-dominated, leading to a suppression of $P_{\rho\,}(k)$ for $k \gg k_{\rm eq}$. 

The {\sl relative} stunting of large-$k$ modes compared to small-$k$ modes can be computed from the information that the unstable modes grow with amplitude $\delta_k(a) \propto a$ during matter-domination. For $k < k_{\rm eq}$ this growth applies as soon as they cross the Hubble scale, while for $k > k_{\rm eq}$ the modes cannot grow in this way until the transition from radiation to matter domination. 
As a result the relative size of two modes, one with $k_0 \ll k_{\rm eq}$ and one with $k \gg k_{\rm eq}$, is
\be \label{delta_kEqn2}
    \frac{\delta_k(a)}{\delta_{k_0}(a)}
    \propto \frac{\delta_k(a_k) (a/a_{\rm eq})}{\delta_{k_0}(a_{k_0}) (a/a_{k_0})} 
    \propto   \frac{\delta_k(a_k) (a/a_{k})}{\delta_{k_0}(a_{k_0}) (a/a_{k_0})}  \,
    \left( \frac{k_{\rm eq}}{k} \right)^2
    \,,
\ee
where $a_k$ denotes the scale factor at the ($k$-dependent) epoch of re-entry, defined by $k = a_k H_k$. The first relation in \pref{delta_kEqn2} uses that modes in the numerator all start growing at the same time (radiation-matter equality), while those in the denominator grow for a $k_0$-dependent amount $a/a_{k_0}$. The second relation then makes the $k$-dependence of the suppression $a_k/a_{\rm eq}$ in the numerator explicit, using the matter-domination evolution $aH \propto a^{-1/2}$ in the re-entry condition to conclude $k = a_k H_k \propto a_k^{-1/2}$ and so $a_k \propto k^{-2}$. 

This leads to the expectation that the power spectrum has the form $P(k) = P_{\rm prim}(k) \, \cT(k)$, where $P_{\rm prim}(k) = \langle | \delta_k(a) |^2 \rangle = \langle | \delta_k(a_k) |^2 \rangle (a/a_k)^2$ is the primordial power spectrum and $\cT(k)$ is the transfer function that expresses the relative stunting of modes for $k \gg k_{\rm eq}$. Keeping in mind that $P(k) \propto |\delta_k|^2$ the above discussion shows we expect $\cT(k) \simeq 1$ for $k \ll k_{\rm eq}$ and $\cT(k) \simeq (k_{\rm eq}/k)^4$ for $k \gg k_{\rm eq}$. Given a primordial distribution $P_{\rm prim}(k) \simeq A k^{n_s}$ this leads to
\be
    P_{\rho\,}(k) \propto \left\{ \begin{matrix} k^{n_s} & \hbox{if}
    \quad k \ll k_{\star} \\
    k^{n_s-4} & \hbox{if} \quad k \gg k_{\star} \end{matrix}
    \right. \,,
\ee
much as is observed.

It is noteworthy that the success of the above argument contains more evidence for the existence of Dark Matter. For many modes $\delta_k \simeq \cO(1)$ occurs before the present epoch, at which point nonlinear gravitational physics is expected to produce the large-scale structure actually seen in galaxy surveys. But the observed isotropy of the CMB implies the amplitude of $\delta_k(a_{\rm rec})$ must have started off very small at the time photons last scattered from the Hydrogen gas at redshift $z_{\rm rec}  \sim 1100$. Given this small start; given the fact that modes cannot grow before matter-radiation equality; and given that instability growth is proportional to $a$, a minimum amount of time is required for fluctuations to become nonlinear early enough to account for the observed distribution of nonlinear structure (like galaxies). Crucially, if Dark Matter did not exist then growth could not  start until redshift $z_{\rm eq}(\hbox{baryons only}) \simeq 480$ --- {\em c.f.}~\pref{Bcross} --- which does not leave enough time. But the presence of Dark Matter moves back the epoch of radiation-matter equality to $z_{\rm eq} \simeq 3600$ --- {\em c.f.}~\pref{DMcross} --- giving sufficient time for nonlinear structure to form at the required scales.

The story of late-time fluctuations is even much richer than the above would lead one to believe, with detailed comparisons between observations and theory. A spectacular example of this is provided by the observation of `baryon acoustic oscillations' (BAO), which are observed correlations between the distribution of galaxies and the distribution of CMB temperature fluctuations. The physical origin of these correlations lies in the coupled late-time evolution of perturbations in the Dark Matter and baryon-radiation fluid. Once fluctuations in the baryon-photon fluid begin to be free to oscillate the local dark Dark Matter evolution acts as a forcing term. This sends out a sound wave in the density of the baryon-photon fluid that initially propagates at a significant fraction of the speed of light, due to the dominance of the photon entropy in this fluid. But the speed of sound for the baryons drops like a rock once the baryons and photons decouple from one another at recombination, causing the sound wave to stall. The resulting correlation has been observed, and its properties again confirm the $\Lambda$CDM model with values for the model parameters consistent with other determinations.

\subsection{Primordial fluctuations from inflation}

The previous discussion shows that fluctuations in the $\Lambda$CDM model also provide a successful description of structure in the universe, but only given the initial condition of a primordial spectrum of fluctuations having a specific power-law form: $P_{\rho\,}(k) \simeq A_s k^{n_s}$ (or $\Delta^2_\phi(k) \simeq A_s k^{n_s-1}$). It again falls to the earlier universe to explain why primordial fluctuations should have this specific form, and why it should be robust against the many poorly understood details governing the physics of this earlier epoch. 

It is remarkable that there is evidence that an earlier period of inflationary expansion can also explain this initial distribution of fluctuations \cite{Fluctuations}. This section provides a sketch of this evidence. Since the modes of interest start off during $\Lambda$CDM outside the Hubble length, $k \ll aH$, and are known to be small, their evolution can be tracked into earlier epochs using linear perturbation theory. Because the modes are super-Hubble in size the treatment must be relativistic, and so involves linearizing the coupled Einstein-matter field equations. The first part of this section sketches how this super-Hubble evolution works, and shows how to relate the primordial fluctuations that re-enter the Hubble scale to those that exit the Hubble scale during the inflationary epoch (see Figure \ref{FigureScales}). 

At first sight this just pushes to problem back to an earlier time, requiring an explanation why a particular pattern of fluctuations should exist during inflation. Even worse, within the classical approximation there is good reason to believe there should be no fluctuations at all leaving at horizon exit during inflation. This is because the exponential growth of the scale factor, $a \propto e^{Ht}$, during inflation is absolutely ruthless in ironing out any spacetime wrinkles since momentum-dependent terms like $(k/a)^2$ in the field equations go to zero so quickly.

But the key words in the above are ``within the classical approximation". Quantum fluctuations are {\em not} ironed away during inflation, and persist at a level proportional to the Hubble scale. Because this Hubble scale is approximately constant the resulting fluctuations are largely scale-independent, providing a natural explanation for why primordial fluctuations seem to be close to the Zel'dovich spectrum. But $H$ during inflation also cannot be exactly constant since inflation must end eventually. In the explicit models examined earlier the time-dependence of $H$ arises at a level suppressed by the slow-roll parameters $\epsilon$ and $\eta$ and so deviations from scale invariance should arise at the few percent level. Because of this we shall find below that the prediction for $n_s$ in inflationary models is a bit smaller than unity, naturally agreeing with the observed value $n_s \simeq 0.97$.

\subsubsection{Linear evolution of metric-inflaton fluctuations}

The first task is to evolve fluctuations forward from the epoch of inflationary horizon exit until they re-enter during the later Hot Big Bang era. In particular our focus is on the perturbations of the metric, $\delta g_{\mu\nu}$, since these include perturbations of the Newtonian potential and so also the density fluctuations whose power spectrum is ultimately measured. The discussion here follows that of \cite{Mukhanov}.  

The symmetry of the FRW background allows the fluctuations of the metric to be classified by their rotational properties, with fluctuations of different spin not mixing at linear order in the field equations. Fluctuations of the metric come in three such kinds: {\sl scalar}, {\sl vector} and {\sl tensor}. Specializing to a spatially flat FRW background and transforming to conformal time, $\tau = \int \exd t/a$, the scalar perturbations may be written
\be
    \delta_S g_{\mu\nu} = a^2 \begin{pmatrix} 2 \phi &&
    \partial_j \cB \\ \partial_i \cB && 2 \psi \, \delta_{ij}
    + \partial_i\partial_j \cE \end{pmatrix} \,,
\ee
while the vector and tensor ones are
\be
    \delta_V g_{\mu\nu} = a^2 \begin{pmatrix} 0 && \cV_j \\
    \cV_i && \partial_i \cW_j + \partial_j \cW_i \end{pmatrix}
    \quad \hbox{and} \quad
    \delta_T g_{\mu\nu} = a^2 \begin{pmatrix} 0 && 0 \\
    0 && h_{ij} \end{pmatrix} \,.
\ee
Here all vectors are divergence-free, as is the tensor (which is also traceless). To these are added the fluctuations in the inflaton field, $\delta \varphi$.

There is great freedom to modify these functions by performing infinitesimal coordinate transformations, so it is useful to define the following combinations that are invariant at linearized order:
\bea \label{GaugeInvariantDefs}
    \Phi &=& \phi - \frac{1}{a} \Bigl[ a(\cB - \cE') \Bigr]'
    \,, \qquad
    \Psi = \psi + \frac{a'}{a} (\cB - \cE') \\
    \delta \chi &=& \delta \varphi
    - \varphi' ( \cB - \cE') \,, \quad
    V_i = \cV_i - \cW_i \quad \hbox{and} \quad
    h_{ij}  \,, \nn
\eea
in terms of which all physical inferences can be drawn. Here primes denote differentiation with respect to conformal time,
$\tau$. Notice that $\Phi$, $\Psi$ and $V_i$ reduce to $\phi$, $\psi$ and $\cV_i$ in the gauge choice where $\cB = \cE = \cW_i = 0$, and so $\Phi$ is the relativistic generalization of the Newtonian potential.


These functions are evolved forward in time by linearizing the relevant field equations:
\be
    \Box \varphi - V'(\varphi) = 0 \quad \hbox{and} \quad
    R_{\mu\nu} - \frac12 \,R g_{\mu\nu} = \frac{T_{\mu\nu}}{M_p^2}
    \,,
\ee
and provided we use the invariant stress-energy perturbations,
\bea
    \delta {\cT^0}_0 &=& \delta {T^0}_0 -
    \left[ {t^0}_0 \right]' (\cB - \cE') \,, \nn\\
    \delta {\cT^0}_i &=& \delta {T^0}_i -
    \left[ {t^0}_0 - \frac13 \, {t^k}_k \right]
    \partial_i (\cB - \cE') \,, \\
    \delta {\cT^i}_j &=& \delta {T^i}_j -
    \left[ {t^i}_j \right]' (\cB - \cE') \,, \nn
\eea
(where ${t^\mu}_\nu$ denotes the background stress-energy), the results can be expressed purely in terms of the gauge-invariant quantities, eqs.~\pref{GaugeInvariantDefs}.

The equations which result show that in the absence of vector stress-energy perturbations ({\em i.e.} if $\delta {\cT^0}_i$ is a pure gradient - as would be the case for perturbed inflaton), then vector perturbations, $V_i$, are not sourced, and decay very rapidly in an expanding universe, allowing them to be henceforth ignored. Similarly, in the absence of off-diagonal stress-energy perturbations ({\em i.e.} if $\delta {\cT^i}_j = \delta p \, \delta^i_j$) it is also generic that $\Psi = \Phi$.

Switching back to FRW time, the equations which govern the evolution of tensor modes then become (after Fourier transforming)
\be \label{TensorModeEqn}
    \ddot h_{ij} + 3 H \, \dot h_{ij}
    + \frac{k^2}{a^2} \, h_{ij} = 0
    \,,
\ee
showing that these evolve independent of all other fluctuations. Such primordial tensor fluctuations can be observable if they survive into the later universe, since the differential stretching of spacetime that they predict can contribute observably to the polarization of CMB photons that pass through them. The search for evidence for this type of primordial tensor fluctuations is active and ongoing, and (as is shown below) is expected in inflation to be characterized by a near scale-invariant tensor power spectrum, 
\be \label{tensorpower}
   P_h(k) = A_T \, k^{n_\ssT} \,,
\ee 
with $n_\ssT$ close to zero.

The equations evolving the scalar fluctuations are more complicated and similarly reduce to
\bea \label{ScalarModeEqn}
    &&\delta \ddot \chi + 3H \delta \dot \chi + \frac{k^2}{a^2}
    \delta \chi + V''(\varphi) \delta \chi - 4 \dot\varphi
    \,\dot\Phi + 2 V'(\varphi) \,\Phi = 0 \nn\\
    &&\qquad\qquad \hbox{and} \qquad
    \dot \Phi + H \,\Phi = \frac{\dot\varphi}{2M_p^2}\; \delta \chi
    \,.
\eea
The homogeneous background fields themselves satisfy the equations
\be
    \ddot \varphi + 3H \dot \varphi + V'(\varphi) = 0
    \quad\hbox{and}\quad
    3 M_p^2 H^2 = \frac12 \dot\varphi^2 + V(\varphi) \,.
\ee
These expressions show that although $\Phi$ and $\delta \chi$ would decouple from one another if expanded about a static background (for which $\dot \varphi = V' = 0$), they do not when the background is time-dependent. 

\subsubsection{Slow-roll evolution of scalar perturbations}

The character of the solutions of these equations depends strongly on the size of $k/a$ relative to $H$, since this dictates the extent to which the frictional terms can compete with the spatial derivatives. As usual the two independent solutions for $\delta \chi$ that apply when $k/a \gg H$ describe damped oscillations
\be \label{DampedOscillationEqn}
    \delta \chi_k \propto \frac{1}{a \sqrt k}
    \exp \left[ \pm i k \int^t
    \frac{\exd t'}{a(t')} \right] \,.
\ee
Our interest during inflation is in the limit $k/a \ll H$ in a slow-roll regime for  which $\delta\ddot\chi$, $\ddot \varphi$ and $\dot \Phi$ can all be neglected. In this limit the scalar evolution equations simplify to
\be \label{SlowRollDE}
    3H \delta\dot\chi + V''(\varphi) \delta \chi +
    2V'(\varphi)\Phi \simeq 0 \quad \hbox{and} \quad
    2M_p^2 H \,\Phi \simeq \dot\varphi \,\delta \chi \,,
\ee
and have approximate solutions (after Fourier transformation) of the form
\be \label{SlowRollPerturbationEqn}
    \delta \chi_k \simeq C_k \, \frac{V'(\varphi)}{V(\varphi)}
    \quad\hbox{and} \quad
    \Phi_k \simeq -\frac{C_k}{2} \, \left( \frac{
    V'(\varphi)}{V(\varphi)} \right)^2 \,.
\ee
where $C_k$ is a (potentially $k$-dependent) constant of integration. Since the background fields satisfy $M_p V'/V = \sqrt{2\epsilon}$ these equations show how the amplitude of $\delta \chi_k$ and $\Phi_k$ during inflation track the evolution of the slow-roll parameter, $\epsilon$, for super-Hubble modes, and therefore tend to grow in amplitude as inflation eventually draws to a close.

We have two remaining problems: ($i$) What is the origin of the initial fluctuations at horizon exit? ($ii$) How do we evolve fluctuations from the end of inflation through to the later epoch of horizon re-entry? The latter of these seems particularly vicious since it {\em a priori} might be expected to depend on the many details involved in getting the universe from its inflationary epoch to the later Hot Big Bang. 

\subsubsection{Post-Inflationary evolution}

For the case of single-field inflation discussed here, post-inflationary evolution of the fluctuation $\Phi$ actually turns out to be quite simple.  This is because it can be shown that when $k \ll aH$ the quantity
\be \label{ConservedCurvatureEqn}
    \zeta = \Phi + \frac23 \left( \frac{\Phi
    + \dot\Phi/H}{1 + w} \right) =
    \frac{1}{3(1+w)} \left[ (5 + 3w) \, \Phi +
    \frac{2\dot\Phi}{H} \right] \,,
\ee
is {\em conserved}, inasmuch as $\dot \zeta \simeq 0$ for $k \to 0$. 

This result follows schematically because the perturbed metric can be written as proportional to $e^\zeta g_{ij}$ and so spatially constant $\zeta$ is indistinguishable from the background scale factor, $a(t)$. Conservation has been proven under a wide variety of assumptions \cite{Mukhanov, ConstZeta}, but the form used here assumes that the background cosmology satisfies an equation of state $p = w \rho$, but $w$ is {\it not} assumed to be constant. The same result is known not to be true if there were more than a single scalar field evolving.

Conservation of $\zeta$ is a very powerful result because it can be used to evolve fluctuations using $\zeta(t_i) = \zeta(t_f)$, assuming only that they involve a single scalar field, and that the modes in question are well outside the horizon: $k/a\ll H$. Furthermore, although $\dot\Phi$ in general becomes nonzero at places where $w$ varies strongly with time, this time dependence quickly damps due to Hubble friction for modes outside the Hubble scale. 

We may therefore for most of the universe's history also neglect the dependence of $\zeta$ on $\dot\Phi$ provided we restrict $t_i$ and $t_f$ to epochs during which $w$ is roughly constant. This allows the expression $\zeta(t_i) = \zeta(t_f)$ to be simplified to
\be \label{UsefulConservationEqn}
    \Phi_f = \frac{1+w_f}{1+w_i}
    \left( \frac{5 + 3 w_i}{5+3w_f} \right) \Phi_i  \,,
\ee
where $w_i = w(t_i)$ and $w_f = w(t_f)$, implying in particular $\Phi_f = \Phi_i$ whenever $w_i = w_f$. Similarly, the values of $\Phi$ deep within radiation and matter dominated phases are related by $\Phi_{\rm mat} \simeq \frac{9}{10} \; \Phi_{\rm rad}$.

To infer the value of $\Phi$ in the later Hot Big Bang era we choose $t_i$ just after horizon exit (where  a simple calculation shows $w_i \simeq -1 + \frac23 \, \epsilon_{\rm he}$, with $\epsilon_{\rm he}$ the slow-roll parameter at horizon exit). $t_f$ is then chosen in the radiation dominated universe (where $w_f = \frac13$), either just before horizon re-entry for the mode of interest, or just before the transition to matter domination, whichever comes first. Eqs.~\pref{SlowRollPerturbationEqn} and \pref{UsefulConservationEqn} then imply
\be \label{MatchPhi}
    \Phi_f \simeq  \left( \frac{2 \Phi}{3\epsilon} \right)_{\rm he}
   \,.
\ee
It remains to grapple with what should be expected for the initial condition for $\Phi$ at horizon exit.

\subsubsection{Quantum origin of fluctuations}

The primordial fluctuation amplitude derived in this way depends on the integration constants $C_k$, which are themselves set by the initial conditions for the fluctuation at horizon exit, during inflation. But why should this amplitude be nonzero given that all previous evolution is strongly damped, as in eq.~\pref{DampedOscillationEqn}? The result remains nonzero (and largely independent of the details of earlier evolution) because quantum fluctuations in $\delta \chi$ continually replenish the perturbations long after any initial classical configurations have damped away.

The starting point for the calculation of the amplitude of scalar perturbations is the observation that the inflaton and metric fields whose dynamics we are following are quantum fields, not classical ones. For instance, for spatially-flat spacetimes the linearized inflaton field, $\delta\chi$, is described by the operator
\be
    \delta\chi(x) = \int \frac{\exd^3k}{(2\pi)^3} \Bigl[ \mathfrak{c}_k \,
    u_k(t) \, e^{i \bfk \cdot \bfr/a} + \mathfrak{c}_k^* \, u_k^*(t)
    \, e^{-i \bfk \cdot \bfr/a} \Bigr] \,,
\ee
where the expansion is in a basis of eigenmodes of the scalar field equation in the background metric, $u_k(t) \, e^{i \bfk \cdot \bfx}$, labelled by the co-moving momentum $\bfk$. For constant $H$ the time-dependent mode functions are
\be
    u_k(t) \propto \frac{H}{k^{3/2}} \left( i +
    \frac{k}{aH} \right) \exp \left( \frac{ik}{aH} \right) \,,
\ee
which reduces to the standard flat-space form, $u_k(t) \propto a^{-1} k^{-1/2} \, e^{-ik\int \exd t/a}$, when $k/a \gg H$. [This is perhaps easiest to see using conformal time, for which $\exp(ik/aH) = \exp(-ik\tau)$, or more directly by using $\exp\left( -ik \int \exd t/a \right) = \exp\left(ik/aH\right)$ when $a \propto e^{Ht}$.] The quantities $\mathfrak{c}_k$ and their adjoints $\mathfrak{c}_k^*$ are {\sl annihilation} and {\sl creation operators}, which define the adiabatic vacuum state, $|\Omega \rangle$, through the condition $\mathfrak{c}_k |\Omega \rangle = 0$ (for all $\bfk$).

The $\delta\chi$ auto-correlation function in this vacuum, $\langle \delta\chi(x) \delta \chi(x') \rangle$, describes the quantum fluctuations of the field amplitude in the quantum ground state, and the key assumption is that the quantum statistics of the mode leaving the horizon during inflation agrees with the classical fluctuations of the field $\delta \chi$ after evolving outside of the Hubble scale.  This assumes the quantum fluctuations to be decohered (for preliminary discussions see ref.~\cite{decoherence, decoherencewodecoherence}) into classical distribution for $\delta \chi$ sometime between horizon exit and horizon re-entry. 

It turns out that during inflation interactions with the bath of short-wavelength, sub-Hubble modes is extremely efficient at decohering  the quantum fluctuations of long-wavelength, super-Hubble modes \cite{BHT}. As is usual when a system is decohered through interactions with an environment, the resulting classical distribution is normally defined for the `pointer basis', that diagonalizes the interactions with the environment. It turns out that the freezing of super-Hubble modes has the effect of making them very classical (WKB-like), and so ensure the fields canonical momenta become functions of the fields themselves. This ensures that it is always the field basis that diagonalizes any local interactions, and so guarantees that quantum fluctuations become classical fluctuations for the fields (like $\delta \chi$) rather than (say) their canonical momenta. 

The upshot is that after several $e$-foldings even very weak interactions (like gravitational strength ones) eventually convert quantum fluctuations into classical statistical fluctuations for the classical field, $\varphi$, about its spatial mean. For practical purposes, this means in the above calculations we can simply use the initial condition $|\delta \chi_k| \sim [\langle \delta \chi_k \delta \chi_{-k} \rangle]^{1/2} \propto |u_k(t)|$. For observational purposes what matters is that the classical variance of these statistical fluctuations is well-described by the corresponding quantum auto-correlations -- a property that relies on the kinds of `squeezed' quantum states that arise during inflation \cite{squeezed,Mukhanov}.

Evaluating $\delta\chi_k \sim u_k$ at $t_{\rm he}$ (where $k = aH$) and equating the result to the fluctuation of eq.~\pref{SlowRollPerturbationEqn} allows the integration constant in this equation to be determined to be
\be
    C_k = u_k(t_{\rm he}) \left( \frac{V}{V'}
    \right)_{\varphi_{\rm he}} \,,
\ee
where both $t_{\rm he}$ and $\varphi_{\rm he} = \varphi(t_{\rm he})$ implicitly depend on $k$. Using this to compute $\Phi_k$ in eq.~\pref{SlowRollPerturbationEqn} then gives
\be \label{InflationaryPhiEvolution}
    \Phi_k(t) = - \frac12 u_k(t_{\rm he})
    \left( \frac{V}{V'}\right)_{\varphi_{\rm he}}
    \left( \frac{V'}{V} \right)^2_{\varphi(t)}
    = - \epsilon(t)
    \left( \frac{u_k}{\sqrt{2\epsilon} \, M_p}
    \right)_{t_{\rm he}} \,.
\ee
In particular, evaluating at $t = t_{\rm he}$ then gives
\be \label{InflationaryPhiEvolution2}
    \Phi_k(t_{\rm he}) =  - \left( \frac{u_k}{M_p}  \sqrt{\frac{\epsilon}{2}}
    \right)_{t_{\rm he}} \,.
\ee

\subsubsection{Predictions for the scalar power spectrum}

We are now in a situation to pull everything together and compute in more detail the inflationary prediction for the properties of the primordial fluctuation spectrum. Using \pref{InflationaryPhiEvolution2} in \pref{MatchPhi} gives
\be
    \Phi_k(t_f) \simeq  \left( \frac{2 \Phi}{3\epsilon} \right)_{\rm he} =  - 
    \left( \frac{2 u_k}{3\sqrt{2\epsilon} \, M_p}
    \right)_{t_{\rm he}}
   \,.
\ee
Using this in the definition of the dimensionless power spectrum for $\Phi$, $\Delta_\Phi^2 = k^3 P_\Phi/(2\pi^2)$, then leads to
\be
    \Delta^2_\Phi(k) = \frac{k^3 |\Phi_k(t_f)|^2}{2\pi^2} \propto
    \frac{k^{3} |u_k(t_{\rm he})|^2}{\pi^2 \epsilon(\varphi_{\rm he})
    M_p^2 }\,.
\ee
Once the order-unity factors are included one finds
\be \label{CurvaturePowerSpectrum}
    \Delta^2_\Phi(k) = \frac{k^3
    P_\Phi(k)}{2 \pi^2} = \left(
    \frac{H^2}{8 \pi^2 M_p^2 \, \epsilon} \right)_{\rm
    he} = \left( \frac{V}{24 \pi^2 M_p^4 \,
    \epsilon} \right)_{\rm
    he} \,,
\ee

It is the quantity $V/\epsilon$ evaluated at Hubble exit that controls the amplitude of density fluctuations, and so is to be compared with the observed power spectrum of scalar density fluctuations,
\be\label{PowerPivot}
    \Delta^2_\Phi({k}) =  \Delta^2_\Phi(\hat{k})  \left( \frac{k}{\hat k} \right)^{n_s} \,,
\ee
where $n_s = 0.968 \pm 0.006$ \cite{Ade:2015xua} and
\be
    \Delta^2_\Phi(\hat{k}) = 2.28 \times 10^{-9}  \,,
\ee
is the amplitude evaluated at the reference `pivot' point $k = \hat{k} \sim 7.5\, a_0 H_0$. In terms of $V$ this implies
\be
    \left( \frac{V}{\epsilon} \right)_{\rm pivot}^{1/4} = 6.6 \times 10^{16}
    \; \hbox{GeV} \,,
\ee
for the epoch when the pivot scale underwent Hubble exit. The smaller $\epsilon$ becomes, the smaller the required potential energy during inflation. For $\epsilon \sim 0.01$ we have $V \sim 2 \times 10^{16}$ GeV. This is titillatingly close to the scale where the couplings of the three known interactions would unify in Grand Unified models, which may indicate a connection between the physics of Grand Unification and inflation.\footnote{Of course, $V$ can be much smaller if $\epsilon$ is smaller as well, or if primordial fluctuations actually come from another source.}

Notice also that the size of $\Delta_\Phi^2(k)$ is set purely by $H$ and $\epsilon$ at horizon exit, and these only depend weakly on $k$ (through their weak dependence on time) during near-exponential inflation. This is what ensures the approximate scale-invariance of the primordial power spectrum which inflation predicts for the later universe. To pin down the value of $n_s$ more precisely notice that the power-law form of \pref{PowerPivot} implies  
\be
    n_s - 1 \equiv \left. \frac{\exd \ln \Delta^2_\Phi}{\exd
    \ln k} \right|_{\rm he}\,.
\ee

To evaluate this during slow-roll inflation use the condition $k = a H$ (and the approximate constancy of $H$ during inflation) to write $\exd \ln k = H \exd t$. Since the right-hand side of eq.~\pref{CurvaturePowerSpectrum} depends on $k$ and $t$ only through its dependence on $\varphi$, it is convenient to use the slow-roll equations, eq.~\pref{SRScalarFieldEqn} to further change variables from $t$ to $\varphi$: $\exd t = \exd\varphi/\dot\varphi \simeq - (3H/V') \, \exd \varphi$, and so
\be
    \frac{\exd}{\exd \ln k} = - M_p^2 \left( \frac{V'}{V} \right)
    \, \frac{\exd }{\exd \varphi}  =  \sqrt{2\epsilon} \, M_p \;\frac{\exd }{\exd \varphi}\,.
\ee

Performing the $\varphi$ derivative using \pref{CurvaturePowerSpectrum} finally gives the following relation between $n_s$ and the slow-roll parameters, $\epsilon$ and $\eta$
\be \label{nsFormulaEqn}
    n_s - 1 = - 6 \epsilon + 2 \eta \,,
\ee
where the right-hand side is evaluated at $\varphi = \varphi_{\rm he}$. For single-field models the right-hand side is negative and typically of order 0.01, agreeing well with the measured value $n_s \simeq 0.97$. 

\subsubsection{Tensor fluctuations}

A similar story goes through for the tensor fluctuations, though without the complications involving mixing between $\delta \chi$ and $\Phi$. Tensor modes are also directly generated by quantum fluctuations, in this case where the vacuum is the quantum state of the graviton part of the Hilbert space. Although tensor fluctuations have not yet been observed, they are potentially observable through the polarization effects they produce as CMB photons propagate through them to us from the surface of last scattering. 

Just like for scalar fluctuations, for each propagating mode the amplitude of fluctuations in the field $h_{ij}$ is set by $H/(2\pi)$, but because there is no longer a requirement to mix with any other field (unlike $\Phi$, which because it does not describe a propagating particle state has to mix with the fluctuating field $\delta \chi$), the power spectrum for tensor perturbations depends only on $H^2$ rather than on $H^2/\epsilon$. Repeating the above arguments leads to the following dimensionless tensor power spectrum
\be \label{DeltaSqTSlowRoll}
    \Delta^2_h(k) = \frac{8}{M_p^2} \left( \frac{H}{2 \pi} \right)^2
    = \frac{2V}{3\pi^2 M_p^4}
    \,.
\ee
This result is again understood to be evaluated at the epoch when observable modes leave the horizon during inflation, $\varphi = \varphi_{\rm he}$.

\begin{figure}[h]
\begin{center}
\includegraphics[width=120mm,height=80mm]{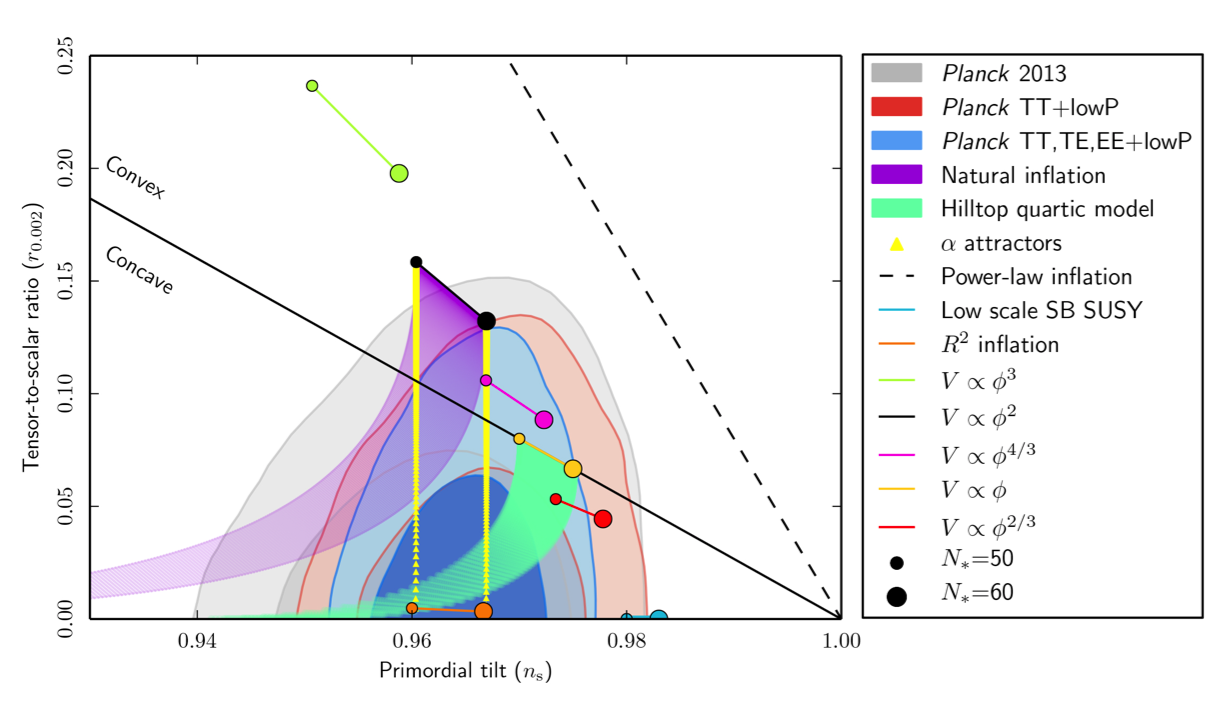} 
\caption{A comparison of a variety of inflationary models to a suite of cosmological measurements (taken from \cite{Ade:2015lrj}), with the ellipses showing the observationally preferred values for the scalar-to-tensor ratio, $r$, and primordial `tilt', $n_s$. Each model is portrayed as giving a range of values rather than a single point, with $N_e$ chosen for a variety of assumptions about the nature of reheating ({\em c.f.}~eq.~\pref{NIFromHorizonEqn}). The model labelled `Natural Inflation' corresponds to the `pseudo-Goldstone axion' model described in the text, while the ones called `$\alpha$-attractors' represent the text's `pseudo-Goldstone dilaton' models. } \label{Fig:rvsns} 
\end{center}
\end{figure}

Should both scalar and tensor modes be measured, a comparison of their amplitudes provides a direct measure of the slow-roll parameter $\epsilon$. This is conventionally quantified  in terms of a parameter $r$, defined as a ratio of the scalar and tensor power spectra
\be
    r := \frac{\Delta_h^2}{\Delta_\Phi^2}
    = 16 \, \epsilon \,.
\ee
The absence of evidence for these perturbations to date places an upper limit: $r \lsim 0.07$ \cite{Bicep} and so $\epsilon \lsim 0.004$. Because $\epsilon$ appears to be so small, the measured value for $n_s$ used with \pref{nsFormulaEqn} permits an inference of how large $\eta$ can be. Fitting to a global data set gives a less stringent bound on $r$ \cite{BicepPlanck, Ade:2015lrj}, leading to
\be
   \epsilon < 0.012 \; (\hbox{95\% CL}); \qquad \hbox{and} \qquad
   \eta = - 0.0080 \, {\scriptstyle{ +0.0080 \atop -0.0146}} \; (\hbox{68\% CL}) \,,
\ee
and it is this incipient evidence that $\epsilon \ne \eta$ that drives the tension with some of the model predictions described earlier. This information is given pictorially in Figure \ref{Fig:rvsns}.

The detection of tensor modes in principle also allows a measurement of the $k$ dependence of their power spectrum. This is usually quantified in terms of a tensor spectral index, $n_T$, defined by eq.~\pref{tensorpower} and so
\be \label{nTinSlowRoll}
    n_T \equiv \frac{\exd \ln \Delta^2_h}{\exd \ln k} = -2 \epsilon
    = -\frac{r}{8}\,,
\ee
where the second-last equality evaluates the derivative within inflation as before by changing variables from $k$ to $\varphi$. 

Ultimately single-field models have three parameters: $\epsilon$, $\eta$ and the Hubble scale during inflation, $H_I$. But the scalar and tensor fluctuation spectra provide four observables: $A_s$, $A_T$, $n_s$ and $n_T$. The ability to describe four observables using just three parameters implies a predicted relation amongst the observables: $n_T = - r/8$ (as seen from \pref{nTinSlowRoll}). This is a robust prediction shared by all single-field slow-roll inflationary models. 

\subsection{Flies in the ointment}

Although not really the main line of development of these lecture notes, it would be wrong to leave the impression that inflationary theories must be the last word in early universe cosmology. Indeed they have problems that motivate some to seek out better alternatives \cite{Alternatives}. Here are a few of the main complaints.

\medskip\noindent {\em Initial-condition problems}

\smallskip\noindent
A major motivation for inflation comes from trying to understand the peculiar initial conditions required for the success of the late-time cosmology of the $\Lambda$CDM model. But this cannot be regarded as being a success if inflation itself also requires contrived initial conditions. In particular there are concerns that inflation might not start unless the universe is initially prepared in a sufficiently homogeneous configuration over several Hubble scales. 

Although the fragility of the required initial conditions is in dispute \cite{InitialCondsOK}, it is true that there are not many explicit calculations done with more generic initial conditions. There are calculations involving more generic random potentials \cite{Random} that do indicate that inflation can be a rare occurrence but it is still being explored how much these calculations depend on the assumptions being made.

\medskip\noindent {\em Fine-tuning problems}

\smallskip\noindent
Slow-roll inflation requires relatively shallow potentials, and these are relatively difficult to obtain within the low-energy limit of explicit UV completions. Not all models are equally bad in this regard, with those based on pseudo-Goldstone bosons being able to arrange shallow potentials in more controlled ways. 

But even with these models inflationary predictions are notoriously sensitive to small effects. Because the inflationary effect being sought is gravitational it is Planck suppressed and so can be threatened even by other Planck-suppressed effective interactions that in any other circumstances would have been regarded as negligible.

\medskip\noindent {\em The multiverse and the landscape}

\smallskip\noindent
Once it gets going inflation can be hard to stop. And even if it ends in some parts of the universe, if it survives in others these inflating regions expand so quickly that they can come to dominate the volume of the universe. These kinds of effects are made even worse if even physical constants are really controlled by the expectation values of fields that can be different in different parts of the universe, as seems to be the case in theories like string theory. 

For such theories it becomes hard to see how to make definite predictions in a traditional way. A great variety of universes might arise within any given framework, and how does one falsify such a theory if the options available become too numerous? One way out is to use anthropic reasoning, but it is not yet clear what the proper rules should be for doing so. 

One point that is worth making is that problems with the multiverse (if problems they prove to be) are actually common to pretty much any cosmological framework in theories that admit a complicated landscape of solutions. That is because even if you think early-universe cosmology is described by something besides inflation (such as a bouncing cosmology) all the above predictivity problems in any case arise if inflation nevertheless should unintentionally get going in any remote corner of the landscape. 

\medskip\noindent {\em My two cents}

\smallskip\noindent
My own opinion is to accept that inflationary models are a work in progress, leaving many things to be desired. But even so they seem at this point to be in better shape than all of their alternatives, mostly because of the control they allow over all of the approximations being made during their use. This situation could change as alternatives get better explored (as they should certainly be), but the shortage of convincing alternatives shows that inflation already sets a fairly high bar for other theories to pass.

Although it may be premature to speculate about issues of the multiverse that are hard to compare with observations, inflationary models do seem to give a clean answer to the more limited practical question of what kind of extrapolation could be useful into our relatively immediate pre-Big-Bang past. Their predictions seem to be under good theoretical control and to agree well with the properties of the primordial fluctuations that have so far been revealed.

\section{EFT issues}
\label{Sec:EFT}

It may not yet be clear how EFT methods enter into the beautiful story presented above, but this section argues EFT methods are actually used throughout (as is also typically true essentially everywhere else in physics). Since these lectures are being delivered in a school entirely devoted to EFTs the logic of this section is not to explain what an EFT is (such as they arise in areas like chiral perturbation theory), but rather to sketch some of the issues that come up when they are applied to gravity- and cosmology-specific problems. 

In my opinion the lesson of these applications is twofold. First, there is no evidence (yet) for `gravitational exceptionalism:' the idea that there is nothing to learn about gravity from experience with other interactions because gravity is fundamentally different. The second lesson is that EFT applications to gravity can sometimes more resemble effective descriptions of particles moving through a medium than they do the traditional uses of EFTs in particle physics. As such they can be mind-broadening to those of us who approach the subject with a particle-physics training. 

Each of the subsections addresses different kinds of examples of this, in turn.

\subsection{Effective field theories and gravity}

The most important use of EFT methods in gravity-related problems is the one described in this subsection: the justification of the semiclassical approximation that underpins almost all theoretical studies of gravity, including cosmology. Although it is common to think of gravitational interactions as being classical, a question less often asked is why this is true (and, if so, what is the small parameter that suppresses quantum effects). 

The claim made here is that the issues for gravitational systems in may ways resemble those arising in nonlinear sigma-models, 
\be
  {\cal L} = - \frac{f^2}{2} \, \; G_{ij}(\phi) \, \partial_\mu \phi^i \, \partial^\mu \phi^j \,,
\ee
such as describe Goldstone (and pseudo-Goldstone) bosons (including those studied in chiral perturbation theory). This similarity arises because both are non-renormalizable, in that their interactions involve inverse powers of a mass scale ($f$ for the sigma-model and $M_p$ for gravity) and both are dominated at low energies by interactions involving only two derivatives but many powers of the interacting fields. 

Both of these properties lose their power to paralyze once it is recognized that the action should really also include all possible kinds of higher-derivative interactions, and it is recognized that predictive power is only possible for low-energy observables relative to $f$ (or $M_p$). For gravity this leads one to regard General Relativity (GR) as the leading part of what might be called (in analogy to the Standard Model Effective Field Theory -- or SMEFT) the General Relativity Effective Field Theory -- or GREFT. 

\subsubsection{GREFT}

To see how this works in detail for gravity we apply to GR the same steps seen in your other classes for sigma models. (For reviews on GR as an EFT see \cite{GREFT, GREFT2}.)

The low-energy degrees of freedom in this case are gravitons, whose field is the metric, $g_{\mu\nu}$, of spacetime itself. The low-energy symmetries that constrain the form of the action are general covariance and local Lorentz invariance. Invariance under these symmetries dictate the metric can appear in the action only through curvature invariants built from the Riemann tensor and its contractions and covariant derivatives. The Riemann curvature tensor is defined by
\bea
 {R^\mu}_{\nu\rho\lambda} &=& \partial_\lambda \Gamma^\mu_{\nu\rho} + \Gamma^\mu_{\lambda\alpha} \Gamma^\alpha_{\nu\rho} - (\lambda \leftrightarrow \rho) \nn\\
 \hbox{with} \quad
 \Gamma^\mu_{\nu\lambda} &=& \frac12 \, g^{\mu\beta} \Bigl( \partial_\nu g_{\beta\lambda} + \partial_\lambda g_{\beta\nu} - \partial_\beta g_{\nu\lambda} \Bigr) \,.\\
\eea 
and its only independent contractions are the Ricci curvature tensor $R_{\mu\nu} = {R^\alpha}_{\mu\alpha\nu}$ and its trace $R = g^{\mu\nu} R_{\mu\nu}$, where the inverse metric, $g^{\mu\nu}$, satisfies $g^{\mu\nu} g_{\nu\lambda} = \delta^\mu_\lambda$. What is important in what follows about these definitions is that, although complicated, the curvature tensors involve precisely two derivatives of the metric. 

GREFT is defined (as usual) by writing down a local action involving all possible powers of derivatives of the metric, which general covariance then requires must be built from powers of the curvature tensors and their derivatives. This leads to the following effective
lagrangian:
\bea
\label{gravaction}
 - \, {{\cal L}_{\rm GREFT} \over \sqrt{- g}} &=& \lambda
+ \frac{M_p^2}{2}  \, R \nn\\
  && \qquad + c_{41} \, R_{\mu\nu} \, R^{\mu\nu} + c_{42} \, R^2
+  c_{43} \, R_{\mu\nu\lambda\rho} R^{\mu\nu\lambda\rho} + c_{44} \, \Box R \\
 && \qquad \qquad\qquad + \frac{c_{61} }{ M^2}\; R^3 + \frac{c_{62}}{M^2} \partial_\mu R \, \partial^\mu R +  \cdots \,,\nn
\eea
where $\sqrt{-g} = \sqrt{- \det g_{\mu\nu}}$, as usual. The first line here includes all possible terms involving two or fewer derivatives, and is the Einstein-Hilbert action of General Relativity, with cosmological constant $\lambda$. The second line includes all possible terms involving precisely four derivatives, and (for brevity) the third line includes only the first two representative examples of the many possible terms involving six or more derivatives. 

The first, cosmological constant, term in eq.~(\ref{gravaction}) is the only one with no derivatives. Its appearance complicates power-counting arguments (in much the same was as does the appearance of a scalar potential when power-counting with a sigma-model --- more about this below). Such terms cause problems if their coefficients are too large (such as similar in size as for the two-derivative terms), and the good news is that if $\lambda$ is regarded as being the Dark Energy it is measured to be extremely small. The puzzle as to {\em why} this happens to be true is a well-known unsolved problem \cite{CCprob}. For simplicity of presentation the cosmological constant term is simply dropped in the power-counting argument that follows, though it returns once scalars and their potential are considered in later sections. Once this is done the leading term in the derivative expansion is the Einstein-Hilbert term of GR. Its coefficient defines Newton's constant (and so also the Planck mass, $M_p^{-2} = 8 \pi G$).

The constants $c_{dn}$ are dimensionless couplings, with the convention that $d$ counts the number of derivatives of the corresponding effective operator and $n = 1, \cdots, N_d$ runs over the number of such couplings. These couplings are dimensionless because the explicit mass scales, $M$ and $M_p$, are extracted to ensure this is so. Often one sees this action written with only the Planck scale appearing, {\em i.e.} with $M \sim M_p$. However, as is usual in an EFT, the scale $M$ is usually of order the lightest particle integrated out to produce this effective theory, leaving only the metric as the variable. Since it is the {\em smallest} such a mass that dominates, $M$ is generically expected to be much smaller than $M_p$. (For applications to the solar system $M$ might be the electron mass; for applications to post-nucleosynthesis Big-Bang cosmology $M$ might be of order the QCD scale, and so on.) Of course, contributions like $M^2 R$ or $R^3/M_p^2$ could also exist, but these are completely negligible compared to the terms displayed in eq.~(\ref{gravaction}). The central point of EFT methods is that the consequences of \pref{gravaction} should be explored as low-energy expansion in powers of $q/M$ and $q/M_p$, where $q$ is a typical energy/momentum or curvature scale characterizing the observables of interest.

\subsubsection*{Redundant interactions}

Just as is true in SMEFT, to save needless effort one should eliminate those redundant interactions that can be removed by integrating by parts or performing a field redefinition. The freedom to drop total derivatives allows us to set the coupling $c_{44}$ to zero, as well (in 4 dimensions) as $c_{43}$. (For $c_{44}$ this can be done because $\sqrt{-g} \, \Box R$ is a total derivative, and for $c_{43}$ the relevant observation is that the quantity
\begin{equation}
 \sqrt{-g} \; X = \sqrt{-g} \Bigl( R_{\mu\nu\lambda\rho} R^{\mu\nu\lambda\rho} -4 R_{\mu\nu}
 R^{\mu\nu} + R^2\Bigr)  \,,
\end{equation}
integrates to give a topological invariant in 4 dimensions, and so is locally also a total derivative. It is therefore always possible to replace, for example, $R_{\mu\nu\lambda\rho} R^{\mu\nu\lambda\rho}$ in the 4-derivative effective lagrangian with the linear combination $4 \, R_{\mu\nu}R^{\mu\nu} - R^2$, with no consequences for any observables, provided these observables are insensitive to the overall topology of spacetime (such as are the classical equations, or perturbative particle interactions). 

As discussed in your other lectures (see also \cite{GREFT}), the freedom to perform field redefinitions allows the dropping of any terms that vanish when evaluated at solutions to the lowest-order equations of motion. This freedom allows the removal of the other two 4-derivative terms because (in the absence of other, matter, fields) the lowest order equations of motion are $R_{\mu\nu} = 0$, and the remaining terms vanish when this is imposed. For pure gravity (without a cosmological constant) the first nontrivial effective interaction involves more than 4 derivatives, such as the term proportional to the cube of the Riemann tensor. This irrelevance of all of the 4-derivative terms must be re-examined once matter fields are included, however, since once these are included $R_{\mu\nu}$ need no longer vanish. 

\subsubsection{Power counting (gravity only)}

In any EFT the central question asks which interactions are relevant when computing observables at a specific order in the low-energy expansion in powers of $q/M$ and $q/M_p$. Because of the similarity in the structure of derivatives appearing in sigma models and General Relativity, power-counting for the two types of theories is very similar. This section briefly recaps the result without repeating the details (see however \cite{GREFT}), highlighting those features that differ.  

To this end start by considering the interactions of gravitons propagating in flat space (returning to curved space below). In this case we expand\footnote{A factor proportional to $1/M_p$ would appear with $h_{\mu\nu}$ if fluctuations were to be canonical normalized, but this normalization is not required in what follows.} $g_{\mu\nu} = \eta_{\mu\nu} + h_{\mu\nu}$ and identify propagators and interactions for perturbative calculations in the usual way. For the purposes of this power counting all we need to know about the curvatures is that they each involve all possible powers of  $h_{\mu\nu}$, but with only precisely two derivatives. Consider an arbitrary graph that contributes at $L$ loops to the amputated\footnote{Amputation means that the graphs have no external lines, such as might be encountered when computing the size of coefficients in a low-energy effective action.} $E$-point graviton-scattering amplitude, ${\cal A}_E(q)$, performed with energy $q$. Suppose also the graph contains $V_{id}$ vertices involving $d$ derivatives and the emission or absorption of $i$ gravitons. Using arguments identical to those used for sigma models in your other lectures leads to the following dependence\footnote{Technical point: as is usually the case this power counting result is computed in dimensional regularization, since not including a spurious cutoff scale makes arguments based on dimensional analysis particularly simple.} of ${\cal A}_E(q)$ on the scales $q$, $M$ and $M_p$:  
\begin{equation}
\label{GRcount1a}
 {\cal A}_E(q) \sim q^2 M_p^2 \left( {1
 \over M_p} \right)^{E}
 \left( {q \over 4 \pi M_p} \right)^{2
 L} {\prod_{i} \prod_{d>2}} \left[{q^2 \over M_p^2}
 \left( {q \over M} \right)^{(d-4)}  \right]^{V_{id}} \,.
\end{equation}
Notice that since $d$ is even for all of the interactions, the condition $d > 2$ in the product implies there are no negative powers of $q$ in this expression.

Eq.~(\ref{GRcount1a}) shows that the weakness of a graviton's coupling (much like the weak couplings of a Goldstone boson) comes purely from the low-energy approximations, $q \ll M_p$ and $q \ll M$. It is also clear that even though the ratio $q/M$ could be much larger than $q/M_p$, it only arises in ${\cal A}_{E}$ together with a factor of $q^2/M_p^2$, and only when including interactions with $d \ge 6$ (coming from curvature-cubed interactions or higher).

Furthermore \pref{GRcount1a} shows that the dominant contributions to low-energy graviton scattering amplitudes correspond to graphs with $L = 0$ and $V_{id} = 0$ for all $d > 2$. That is to say, graphs built using ony tree graphs constructed purely from the Einstein-Hilbert ($d=2$) action: it is classical General Relativity that governs the low-energy dynamics of gravitational waves. 

But EFTs excel when computing next-to-leading contributions. In this case these come in one of the following two ways. Either:
\begin{itemize}
\item $L = 1$ and $V_{id} = 0$ for any $d\ne 2$ but $V_{i2}$ is arbitrary, or
\item $L = 0$, $\sum_i V_{i4} = 1$, $V_{i2}$ is arbitrary, and all other $V_{id}$ vanish.
\end{itemize}
That is, the next to leading contribution is found using one-loop graphs using only the interactions of General Relativity, or by working to tree level and including precisely one insertion of a curvature-squared interaction in addition to any number of interactions from GR. Both of these are suppressed compared to the leading term by a factor of $(q/M_p)^2$. The next-to-leading tree graphs provide precisely the counter-terms required to absorb the UV divergences in the one-loop graphs. And so on.

What this shows is that the small parameter that controls the loop expansion ({\em i.e.} the semi-classical expansion) for graviton scattering is the ratio $q^2/(4\pi M_p)^2$; the semiclassical approximation {\em is} the low-energy approximation. 

But the above argument was made specifically for gravitons propagating in flat space. How reliable should these power-counting arguments be for drawing conclusions for more general curved environments? Related to this, how important is it to be able to work in momentum space, as is usually done in sigma-model type arguments (and those adapted from them to gravity)?

The issue of momentum space can be put aside, because the arguments for sigma models can equally well be made in position space.  The key estimate made to arrive at \pref{GRcount1a} is based on dimensional analysis: all of the factors of $M$ and $M_p$ are tracked by counting how they appear as factors in propagators and vertices, and the remaining dimensions are all filled in as the common low-energy scale $q$. The analogous argument works also in position space, provided there is also only one scale $q$ that characterizes the observables of interest in the low-energy theory.\footnote{General EFT arguments still apply when there is more than one scale, but are more complicated. Indeed much of the complications encountered in other lectures when non-relativistic particles are present can be traced to their having more than a single scale, and the same is true for non-relativistic particles interacting with gravity \cite{GRNR}.}

Physically, the equivalence of the short-distance position-space and high-energy momentum-space estimates happens because the high-energy contributions arise due to the propagation of modes having very small wavelength, $\lambda$. Provided this wavelength is very small compared with the local radius of curvature, $r_c$, particle propagation behaves just as if it had taken place in flat space. One expects the most singular behaviour to be just as for flat space, with curvature effects appearing in subdominant corrections as powers of $\lambda/r_c$. This expectation is borne out explicitly in curved-space calculations using heat-kernel methods \cite{HeatKernel, HeatKernelReview}.

It is often true that the low-energy gravitational system is characterized by a single scale. For cosmological models this scale is often the Hubble scale $q \sim H$. (For black holes it is instead $q \sim r_s^{-1}$ where $r_s = 2G\cM = \cM/(4\pi M_p^2)$ is the Schwarzschild radius of a black hole with mass $\cM$.) In this case the above power-counting arguments imply the semiclassical expansion arises as powers of $H^2/(4\pi M_p)^2$ [or $(4\pi M_p r_s)^{-2} \sim (M_p/ \cM)^2$ in the case of black holes]. We require $H/M_p \ll 1$ (or $\cM \gg M_p$ for black holes) in order to believe inferences about their properties using semiclassical methods. 

\subsubsection{Power counting (scalar-tensor theories)}
\label{sec:PCinf}

So far so good, but for inflationary applications these power-counting rules must be extended to include both the metric and inflaton, as is now done following \cite{InflationPC}. A little more detail is given because the introduction of the scalar potential changes the reasoning somewhat relative to the case of pure gravity.

To this end add $N$ dimensionless scalar fields, $\theta^i$, expanding the effective lagrangian to the form
\bea \label{Leffdef}
 - \frac{ \cL_{\rm eff}}{\sqrt{-g}} &=& v^4 V(\theta) + \frac{M_p^2}{2}
 \, g^{\mu\nu} \Bigl[  W(\theta) \, R_{\mu\nu}
 + G_{ij}(\theta) \, \partial_\mu \theta^i
 \partial_\nu \theta^j \Bigr] \\
 && \quad + A(\theta) (\partial \theta)^4 + B(\theta)
 \, R^2 + C(\theta) \, R \, (\partial \theta)^2
 + \frac{E(\theta)}{M^2} \, (\partial \theta)^6
 + \frac{F(\theta)}{M^2} \, R^3 + \cdots \,,\nn
\eea
with terms involving up to two derivatives written explicitly and the rest written schematically, inasmuch as $R^3$ collectively represents all possible independent curvature invariants involving six derivatives, and so on. The explicit mass scales $M_p$ and $M$ are explicitly written, as before, so that the functions $W(\theta)$, $A(\theta)$, $B(\theta)$ {\it etc}, are dimensionless. As before it is natural to assume $M \ll M_p$, where $M$ is the lowest scale integrated out to obtain $\cL_{\rm eff}$. A new scale, $v$, is also added so that the scalar potential $V(\theta)$ is also dimensionless. The kinetic term for the scalars is chosen to be normalized with the Planck mass as its coefficient, and this has the effect of generically suppressing the couplings of canonically normalized scalars by powers of $M_p$. With inflationary applications in mind take $H^2 M_p^2 \sim V \sim v^4 \ll M^4 \ll M_p^4$ when $\theta \simeq \cO(1)$. 

\subsubsection*{Semiclassical expansion}

Expanding about a classical solution using fields that have canonical dimension, $\theta^i(x) = \vartheta^i(x) + {\phi^i(x)}/{M_p}$ and $g_{\mu\nu} (x) = \hat g_{\mu\nu} (x) + {h_{\mu\nu}(x)}/{M_p}$ allows this lagrangian  to be written  
\be \label{Leffphih}
 \cL_{\rm eff} = \hat \cL_{\rm eff} + M^2 M_p^2 \sum_{n}
 \frac{c_{n}}{M^{d_{n}}} \; \cO_{n} \left(
 \frac{\phi}{M_p} , \frac{ h_{\mu\nu}}{M_p} \right)
\ee
where $\hat\cL_{\rm eff} = \cL_{\rm eff}(\vartheta,\hat g_{\mu\nu})$ and the interactions, $\cO_{n}$, involve $N_n = N^{(\phi)}_n + N^{(h)}_n \ge 2$ powers of the fields $\phi^i$ and $h_{\mu\nu}$. Using a parameter $d_{n}$ to count the number of derivatives appearing in $\cO_n$, the coefficients $c_n$ are dimensionless and the prefactor, $M^2 M_p^2$, ensures the kinetic terms (and so also the propagators) are independent of $M$ and $M_p$. 

The lagrangians of \pref{Leffphih} and \pref{Leffdef} make equivalent predictions for physical observables provided an appropriate dependence on $M$, $M_p$ and $v$ is assigned to the coefficients $c_n$. In particular, reproducing the $m$-dependence of the coefficients of the curvature-cubed and higher terms in \pref{Leffdef} implies
\be \label{cndgt2h}
 c_n = \left( \frac{ M^2}{ M_p^2} \right) g_n
 \qquad \hbox{(if $d_n > 2$)} \,,
\ee
where $g_n$ is at most order-unity and independent (up to logarithms) of $M$ and $M_p$. For terms with no derivatives --- {\it i.e.}~those coming from the scalar potential, $V(\theta)$ --- one instead finds
\be \label{cndeq0}
 c_n = \left( \frac{v^4}{M^2 M_p^2} \right) \lambda_n
 \qquad \hbox{(if $d_n = 0$)} \,,
\ee
where the dimensionless couplings $\lambda_n$ are also independent of $M_p$ and $M$.  In terms of the $\lambda_n$'s the above assumptions mean that the scalar potential has the schematic form
\be \label{assumedVform}
 V(\phi) = v^4 \left[ \lambda_0 + \lambda_2 \left(
 \frac{\phi}{M_p} \right)^2 + \lambda_4 \left(
 \frac{\phi}{M_p} \right)^4 + \cdots \right] \,,
\ee
which shows that $V$ ranges through values of order $v^4$ as $\phi^i$ range through values of order $M_p$. These choices capture qualitative features of many explicit inflationary models.

For cosmological applications it also proves useful to normalize amplitudes differently than is done in earlier sections, which treated $\cA_\ssE$ as appropriate for the sum of amputated Feynman graphs (as would be useful when computing an effective action, say). For cosmology it is more useful to track correlation functions of fields, since our interest is in tracking how quantities like $\langle \phi^2 \rangle$ depend on the scales $M$, $M_p$, $v$ and $q \sim H$.  To obtain correlation functions from amputated graphs a propagator is attached to to each external line followed by an integration over the space-time where the external line is attached to the rest of the graph. Using $q \sim H$ for the common low-energy scale, the Feynman amplitude for a correlation function with $E$ external lines scales as $\cB_\ssE(H) \simeq \cA_\ssE(H)  H^{2E-4}$.

Another complication for cosmology is that one separately tracks dependence on {\em two} low-energy scales and not just one, since correlators are required as functions of both $H$ and mode momentum $k/a$. This need not alter the power-counting dimensional analysis argument, however, if these are the same size (as they are during the epoch of most interest: Hubble exit).

Combining these observations and repeating the usual power-counting steps leads to the result
\bea \label{PCresult2}
 \cB_\ssE (H) &\simeq& \frac{M_p^2}{H^2} \left( \frac{H^2}{M_p} \right)^E
 \left( \frac{H}{4 \pi \, M_p}
 \right)^{2L}  \prod_{d_n = 0} \left[ \lambda_n \left( \frac{v^4}{H^2 M_p^2}
 \right) \right]^{V_n} \\
 && \qquad \qquad \qquad \times \left[ \prod_{d_n = 2}  c_n^{V_n} \right] 
 \prod_{d_n \ge 4} \left[ g_n \left( \frac{H}{M_p}
 \right)^2 \left( \frac{H}{M}
 \right)^{d_n-4} \right]^{V_n}  \,, \nn
\eea
which shows that the presence of vertices with no derivatives ($d_n = 0$) introduces factors where the low-energy scale $q \sim H$ appears in the denominator rather than the numerator. Such terms are potentially dangerous because their repeated insertion threatens to undermine the entire low-energy expansion. 

These dangerous scalar-potential terms do not pose a problem for the class of inflationary potentials, eq.~\pref{assumedVform}, of present interest however \cite{HIPCCrit}. That is because once the relationship, $H \simeq v^2/M_p$, connecting the size of $H$ to the scale in the potential is used, the potentially dangerous $d_n = 0$ term becomes
\be
 \prod_{d_n = 0} \left[ \lambda_n \left( \frac{v^4}{H^2 M_p^2}
 \right) \right]^{V_n} \simeq \prod_{d_n=0} \lambda_n^{V_n} \,,
\ee
leading to the final result
\bea \label{PCresult}
 \cB_\ssE (H) &\simeq& \frac{M_p^2}{H^2} \left( \frac{H^2}{M_p} \right)^E
 \left( \frac{H}{4 \pi \, M_p}
 \right)^{2L} \left[ \prod_{d_n = 2}  c_n^{V_n} \right] \\
 && \qquad \qquad \qquad \times
 \left[ \prod_{d_n = 0}  \lambda_n^{V_n} \right]
 \prod_{d_n \ge 4} \left[ g_n \left( \frac{H}{M_p}
 \right)^2 \left( \frac{H}{M}
 \right)^{d_n-4} \right]^{V_n}  \,, \nn
\eea

Eq.~\pref{PCresult} shows why scalar fields do not undermine the validity of the low-energy approximation underlying gravitational semiclassical expansion. The basic loop-counting parameter is small provided $H \ll 4 \pi M_p$, and the leading contribution describes classical ($L=0$) physics, as is assumed in standard treatments of cosmology.   

Eq.~\pref{PCresult} also shows why trans-Planckian field values need not be a threat to the validity of the low-energy expansion that underpins EFT methods, even though this expansion demands low energies compared with mass scales that are Planck size or lower. The point is that what would be bad is high energies and this does not necessarily follow from large fields. This fact is baked into the above power-counting analysis by the assumption that the scalar potential $V = v^4 U(\theta)$ remained $\cO(v^4)$ for generic $\theta = \phi/M_p$ order unity. 

This emphasizes the important conceptual difference between expanding in powers of fields, like $\phi$, and expanding in powers of derivatives of fields. The derivative expansion is part and parcel of low-energy methods, while the field expansion is only relevant to the exploration of a particular neighbourhood of field space. It can happen that a potential has a known and bounded asymptotic form at large fields rather than small fields --- such as perhaps\footnote{This is not a purely hypothetical example, since the energy of an extra-dimensional modulus (such as the radius, $r$, of an extra-dimensional sphere, say) typically arises as a curvature expansion and so as a series in powers of $1/r$. When regarded as a field in the low-energy 4D EFT this gives lagrangians of the form $\cL \propto (\partial r)^2/r^2 + V(r)$ with $V(r) = V_0 + V_1/r + \cdots$, with the kinetic term coming from the dimensionally reduced Einstein action, generically leading to the exponential form so attractive for inflation \cite{ExpRadius}.} $V(\varphi) \sim V_0 -V_1 \, e^{-\varphi/f} + \cdots$ --- while not knowing its behaviour for small $\varphi$.

\subsubsection*{Slow-roll suppression}

With additional assumptions the previous arguments can be refined to track suppression by slow-roll parameters as well as powers of $H/M_p$. This is done by recognizing that derivatives of background fields and derivatives of fluctuation fields may be very different sizes. In inflation, derivatives of background fields are suppressed by slow-roll parameters while fluctuations need not be, and suppressed background derivatives can make the effective couplings like $g_n$ or $\lambda_n$ parametrically small. 

There are two ways that slow-roll parameters can enter into these effective couplings. 
First, they can do so because of the assumed flatness of the inflationary potential. For instance, if all slow-roll parameters are similar in size then the $s$th derivative of the scalar potential to be written $(\exd^sV/\exd \varphi^s) \simeq \epsilon^{s/2} \, V/M_p^s \simeq \epsilon^{s/2} \; v^4/M_p^s$ and so 
\be
   \lambda_n \simeq \epsilon^{N_n/2} \hat \lambda_n \,,
\ee
where now it is $\hat \lambda_n$ that is order unity. Here $N_n$ counts the number of scalar lines that meet at the vertex in question, and having all slow-roll parameters of a similar size means there is a factor of $\sqrt\epsilon$ arising for each scalar line that meets in the vertex.

The other way slow-roll parameters enter into eq.\ \pref{PCresult} is through derivatives of the background scalar field, which we assume satisfies  
\be
  \frac{1}{M_p} \, \frac{\exd^n \varphi}{ \exd t^n} \simeq \epsilon^{n/2} H^n  \,,
\ee
and so in particular satisfies the slow-roll relation $H M_p \,\dot \varphi \simeq V'$. It is straightforward then to track how powers of $\epsilon$ appear in any particular graph contributing to $\cB_\ssE(H)$, the details of which are in \cite{InflationPC}.

For single-field slow-roll models these power-counting rules imply the standard estimates for the size of the leading few $n$-point functions. For instance, the leading contributions to metric and inflaton two-point functions correspond to using $E = 2$ and $L = 0$ and taking vertices only from the 2-derivative interactions. The diagonal terms then arise unsuppressed by powers of $H/M_p$ or $\epsilon$, while the leading off-diagonal terms are down by at least one power of $\sqrt\epsilon$. Mixed terms are suppressed because they arise at leading order by expanding the scalar kinetic term $\sqrt{-g}\; (\partial\phi)^2 \simeq h \dot \varphi \dot \phi+ \cdots$ with $\dot\varphi \propto \sqrt\epsilon$. 

This leads to the estimates 
\be
  \langle hh \rangle \sim \langle \phi\phi\rangle \sim H^2 \,, \quad \hbox{while} \quad \langle \phi h\rangle \sim \sqrt\epsilon \; H^2 \,.
\ee  
Keeping in mind that the standard gauge-invariant variables $\zeta \sim \Phi$ are related to the basic inflaton and metric by
\be \label{zetadef}
   \zeta \sim \frac{ \phi}{\dot \varphi/H} \sim \frac{ \phi}{\sqrt\epsilon \; M_p} \qquad \hbox{and} \qquad
   t_{\mu\nu} \sim \frac{h_{\mu\nu}}{M_p} \,.
\ee 
then leads to the usual estimates
\be
   \langle\zeta\zeta\rangle \sim \frac{H^2}{\epsilon\, M_p^2}  \qquad \hbox{and} \qquad
    \langle tt \rangle \sim \frac{H^2}{M_p^2} \,.
\ee  
The first of these agrees with the more explicit arguments given earlier for $\Delta_\Phi^2$ at Hubble crossing.

The leading powers of $H/M_p$ in the 3-point functions (called `bispectra' by cosmologists) are similarly obtained by choosing $E=3$ and $L=0$ and no vertices used except those with $d_n =2$.  For the quantities $\langle hhh \rangle$ and $\langle h\phi \phi \rangle$ the leading contributions then are
\be\label{eqn:bispecest1}
  \langle hhh\rangle\sim \langle h\phi\phi\rangle \sim \frac{H^4}{M_p} \,, 
\ee  
since unsuppressed cubic vertices comes from either the Einstein-Hilbert action or the inflaton kinetic term. By contrast, the correlators $\langle hh\phi \rangle$ or $\langle \phi \phi \phi \rangle$ all come suppressed by at least one power of $\sqrt\epsilon$, with the leading contribution obtained by inserting a single $h$-$\phi$ kinetic mixing into $\langle hhh \rangle$ or $\langle hh\phi \rangle$. This leads to the estimates 
\be\label{eqn:bispecest2}
  \langle hh\phi\rangle \sim \langle \phi\phi\phi\rangle \sim \frac{\sqrt\epsilon \; H^4}{M_p} \,.
\ee  

Again converting to dimensionless strain and curvature fluctuation using eq.\  \pref{zetadef} then leads to the usual results \cite{Maldacena}
\be
 \langle ttt \rangle \sim \langle tt \zeta\rangle \sim \frac{H^4}{M_p^4}  \qquad \hbox{and} \qquad
   \langle t\zeta\zeta\rangle \sim \langle\zeta\zeta\zeta\rangle \sim  \frac{H^4}{\epsilon \, M_p^4} \,,
\ee  
and so on. The last of these, $\langle\zeta\zeta\zeta\rangle$, is a measure of the amount of non-gaussianity associated with the primordial distribution of density fluctuations, and because CMB measurements tell us $\langle\zeta\zeta\rangle \sim \Delta_\Phi^2 \sim H^2/(\epsilon M_p^2) \sim 10^{-10}$ a simple estimate of the size of this non-gaussianity in single-field slow-roll models is
\be
 \langle \zeta\zeta\zeta\rangle \sim  \frac{H^4}{\epsilon \, M_p^4} \sim \epsilon \left( \frac{H^2}{\epsilon M_p^2} \right)^2 
\ee
which is too small to be detected at present. This suppression need not be present for more complicated models, making searches for non-gaussianity a useful benchmark when testing the single-field hypothesis.

\subsection{Conceptual issues for EFTs with time-dependent backgrounds}

Besides issues specific to gravity, use of EFTs in cosmology can also involve other complications that are often not seen in particle physics (but do arise in other areas of physics where time-dependent background fields are encountered).

\subsubsection{Importance of the adiabatic approximation}

An issue specific to cosmology arises due to the appearance there of time-dependent backgrounds. The issue asks: if EFTs are defined by dividing systems into low- and high-energy states how can they be defined in time-dependent problems where energy is not conserved? The short version of this section is that time-dependence (in gravity and elsewhere) imposes additional restrictions on the domain of validity of EFTs, the most important of which is the requirement that the background time-dependence should be adiabatic. (That is, $\dot\varphi/\varphi$ should be smaller than the UV scales of interest, for every time-dependent background field $\varphi$ in the problem.)

Adiabatic motion is important because in essence EFTs organize states according to their energy, and energy is generically not conserved (and so is not useful) in the presence of time-dependent backgrounds. In the special case of adiabatic motion, however, an approximately conserved hamiltonian, $\mathfrak{H}(t)$, can exist, even though it may drift slowly with time as the background evolves. This allows both the definition of an approximate ground state and an energy in terms of which the low-energy/high-energy split can be defined. 

Once the system is partitioned in this way into low-energy and high-energy states, one can ask whether a purely low-energy description of time evolution is possible using only a low-energy, local effective lagrangian. The main danger is that the time evolution of the system need not keep low-energy states at low energies, or high-energy states at high energies. For instance, this could happen if the background's time-dependence is rapid enough to allow particle-production of what were regarded as high-energy states, making the initial ground state unstable. Or it could happen that the initial gap between high and low energies decreases with time, and so the approximation of expanding in the ratio of these energies becomes a poor approximation. Level-crossing is an extreme example of the evolution of gap size, in which the gap eventually vanishes and high- and low-energy states  nominally cross one another as time evolves. 

A related issue can arise if there is a transfer of states from high-energy to low-energy as the dividing line between them, $\Lambda(t)$, evolves.  For example, this could happen for a charged particle in a decreasing magnetic field if the effective theory is set up so that the dividing energy, $\Lambda(t)$, between low- and high-energies is not similarly time dependent. In this case then Landau levels continuously enter the low-energy theory as the magnetic field strength wanes. Such a migration of states can also happen in cosmology, such as during an inflationary phase (the so-called trans-Planckian `problem'). This usually is only a problem for the effective-theory formulation if the states which enter in this way are not in their adiabatic ground state when they do so. If they are in their adiabatic ground state they do not affect low-energy observables, but if they are not they can since then new physical excitations appear at low energies. 

What emerges from this is that EFTs can make sense despite the presence of time-dependent backgrounds, provided one can focus on the evolution of low-energy states, ($q < \Lambda(t)$), without worrying about losing probability into high-energy states ($q > \Lambda(t)$). This can often be ensured if the background time evolution is sufficiently adiabatic.

\subsubsection{Predicting background evolution with EFTs}

There is another issue at stake when using EFTs in cosmology (or other time-dependent settings). Up to now the evolution of the background field is regarded as being given, and the EFT issues of the previous section are to do with understanding how to split the system into low and high energy states relative to an adiabatic energy defined in the presence of this time-dependent background. 

But it is often also of interest to know how the background itself responds to events within time-dependent systems. For instance the background might back-react in response to changes in the state of fluctuations with which it interacts. This can also be amenable to EFT analysis, often by solving self-consistently for the background using the field equations of the low-energy theory. Central to this approach is the assumption that solutions to field equations within an EFT actually capture the behaviour of solutions to field equations within the full theory. 

Need this always be true? This section --- following \cite{GhostBusters} --- argues in general the answer is `no', although it usually is for adiabatic motion. 

To see why EFTs and UV completions can agree on their solutions to the equations of motion one must hark back to the definitions of the EFT itself. (The EFT formulation used here follows the review \cite{EFTmine}.) Consider therefore a theory with high-energy and low-energy fields $h$ and $\ell$, with action $S(h,\ell)$. We wish to integrate out $h$ to obtain the effective action, $S_{\rm eff}(\ell)$, to examine its equations of motion. For simplicity we do so at the classical level, in which case integrating out $h$ is equivalent to solving its classical field equations as a function of the light field, $h_c(\ell)$ and plugging the result back into the original action: 
\be \label{EFTclass}
 S_{\rm eff}[\ell] = S[h_c(\ell), \ell] \,, \qquad \hbox{where} \quad
 \left( \frac{\delta S}{\delta h} \right)_{h = h_c(\ell)} = 0 \,.
\ee
(Exercise: verify this statement explicitly by showing that it is equivalent to integrating out $h$ using only tree graphs.)

An immediate consequence of the above derivation seems to be that any solution to the low-energy EFT 
\be
 \left( \frac{\delta S_\EFT}{\delta \ell} \right)_{\ell = \ell_c} = 0 \,,
\ee
must also be extrema of the full theory, by virtue of the choice $h = h_c(\ell)$. How can this argument ever fail? 

The key step in deriving any EFT, glossed over in the previous paragraphs, is the necessity of expanding to some finite order in powers of the heavy mass scale, $1/M$. It is only after this expansion that an effective action like \pref{EFTclass} is given by a local lagrangian density. Because of this we should only trust the equations of motion of any local EFT up to the same order in powers of $1/M$. Solutions of the full theory can differ from those of the effective theory if they are not captured by such a $1/M$ expansion. 


It is actually a good thing that the solutions to an EFT are not completely equivalent to solutions to the full theory from which the EFT is derived. One upside is that EFTs often involve higher time derivatives, and so naively should generically have unstable runaway solutions \cite{Ostrogradsky}, even if the underlying theory has none. 

To see why instabilities might arise within the EFT  consider the following toy effective lagrangian:
\be
 \frac{L}{v^2} = \frac12 \, \dot \theta^2 + \frac{1}{2M^2} \, \ddot \theta^2 \,,
\ee
whose variation $\delta L = 0$ gives the linear equation of motion
\be
 - \ddot \theta  + \frac{1}{M^2} \,  \ddddot \theta  = 0 \,.
\ee
The general solution to this equation is
\be
 \theta = A + B t + C e^{Mt} + D e^{-Mt} \,,
\ee
where $A$, $B$, $C$ and $D$ are integration constants. 

Now comes the main point. Only the solutions with $C = D = 0$ go over to the solutions to the lowest-order field equation, obtained from the $M \to \infty$ lagrangian, $L_0 = \frac12 \dot \theta^2$. The others make no sense at any finite order of $1/M$ because for them the $\dot\theta^2$ and $\ddot\theta^2$ terms are always comparably large. Since a local EFT is only meant to capture the full theory order-by-order in $1/M$ these exponential solutions should not be expected to be reproducing the low-energy approximation of the full theory.

\subsubsection{EFT for inflationary fluctuations}

The previous sections touch on a general issues associated with time-dependent backgrounds. For inflation it is particularly interesting if the time evolution describes a slow roll near de Sitter spacetime, since this is the maximally symmetric spacetime obtained when solving Einstein's equations with a positive cosmological constant, with equation of state $w = -1$ (corresponding to the case of exponential expansion $a \propto e^{Ht}$ that arises during a potential-dominated epoch).

Slow-roll evolution of a scalar field near a maximally symmetric space like de Sitter spacetime lends itself to symmetry arguments. This is because maximal symmetry in 4 dimensions implies de Sitter space enjoys a 10-parameter group of isometries (transformations that preserve the form of the metric). For de Sitter space the group is $O(4,1)$, which has the same number of generators as the Poincar\'e symmetry group of flat space (though with different commutation relations).  

This abundance of symmetries gives the discussion of fluctuations about a near-de Sitter slow roll some fairly universal features, that allow a more general parameterization than might otherwise be possible. These features are described by yet another kind of effective field theory within the inflationary literature, one that has come to be known as `the' Effective Theory of Inflation \cite{Cheung:2007st} (for a review see \cite{Senatore:2016aui}). This section provides a telegraphic summary of this specific theory, in order to put it into its context within the broader EFT pantheon.

The EFT of Inflation is aimed at single-field inflationary models including, but not restricted to, the simple models considered above. The starting observation of this theory is that homogeneous roll of the single inflaton field, $\varphi(t)$, provides the clock that breaks the symmetries the de Sitter spacetime otherwise would have had. There are many equivalent ways to set up spatial slices in maximally symmetric spaces, but the level-surfaces of an evolving scalar field pick out a specific preferred frame within which to build spatial slices. 

The background $\varphi$ (and the evolving metric to which it also gives rise) spontaneously breaks the symmetries of de Sitter space and acts as the Goldstone boson for this breaking (and is ultimately eaten by the metric, in the same way that would-be Goldstone bosons for local internal symmetries get eaten by the corresponding gauge bosons. Because of this symmetry representation it is possible to use the nonlinearly realized symmetries to classify the kinds of allowed low-energy interactions in a model-independent way, providing a more robust description of the kinds of observables that could potentially arise at low energies for cosmologies based any single-field slow-roll system near de Sitter space.

Concretely, because the slowly rolling field, $\phi$, is assumed to be a scalar, by definition under arbitrary spacetime motions, $\delta x^\mu = V^\mu(x)$, it transforms as $\delta \phi = V^\mu \partial_\mu \phi$. But the {\em fluctuation} in such a scalar field about a homogeneous background $\varphi(t)$ transforms differently than this, since if $\phi = \varphi(t) + \hat\phi(x,t)$, then although $\delta \hat \phi(x,t) = V^i \partial_i \hat \phi(x,t)$ remains a scalar under purely spatial motions, $\delta x^i = V^i(x, t)$, at fixed $t$, it transforms nonlinearly under motions $\delta t = V^0(x,t)$, since
\be \label{inhomoV0}
  \delta \hat \phi(x,t) = V^0(x,t) \dot \varphi + V^0 \partial_0 \hat \phi \,.
\ee
This kind of inhomogeneous transformation indicates a would-be Goldstone mode.

There are two natural ways to describe the interactions of a field that transforms like \pref{inhomoV0}. One way --- called `unitary gauge' --- uses this transformation to completely remove the field $\hat \phi$ from the problem. A second way trades $\hat\phi$ for a `Stueckelberg field' $\hat \pi(x,t) := - V^0(x,t)$, though for brevity's sake this summary sticks with the unitary gauge formulation.\footnote{The formulation involving $\hat\pi(x,t)$ has the advantage that for scales of order $H$ many (but not all) quantities can be computed fairly easily using only relatively simple self-interactions of $\hat \pi$, without needing the full complications of expanding the metric dependence of the Einstein action. } 

In unitary gauge the physics of $\hat \phi$ gets transferred into any other fields that are present and transform under the symmetry. In the present case the only other field is the metric, and in this sense $\hat \phi$ gets eaten by the metric. In this representation the remaining symmetries are spatial motions, $\delta x^k = V^k(x)$ and 
\bea
   \delta g_{00} &=& V^k \partial_k g_{00} \nn\\
   \delta g_{0i} &=& V^k \partial_k g_{0i} + \partial_iV^k g_{0k} \\
   \hbox{and} \quad 
   \delta g_{ij} &=& V^k \partial_k g_{ij} + \partial_i V^k g_{jk} + \partial_jV^k g_{ik} \,,
\eea 
and so respectively transform as a scalar, vector or tensor. The most general invariant action is built using the spatial scalar $g_{00}$ and the spatial tensors built from the intrinsic and extrinsic curvatures, ${R^i}_{jkl}$ and $K_{ij}$, of the spatial slices (which -- from the Gauss-Codazzi relations -- together encode all of the information of the full 4D curvature ${R^\mu}_{\nu\lambda\rho}$). Whereas ${R^i}_{jkl}$ involves two derivatives of the metric, $K_{ij}$ only involves one. 

The most general lagrangian density describing these degrees of freedom is
\be \label{cheungetalL}
  \frac{\cL}{\sqrt{-g}} = \frac{M_p^2}2 \, R - \alpha(t)  - \beta(t) g^{00} + \frac12 \, M_2^4(t) (g^{00} + 1)^2 + \frac{1}{3!} \, M_3^4(t) \, (g^{00}+1)^3 + \cdots \,,
\ee
where the ellipses include terms with more powers of $g^{00} +1$ and its derivatives, as well as the variation of the extrinsic curvatures, $\delta K_{ij}$, computed relative to the extrinsic curvatures evaluated in the background geometry, and so on. The coefficients $\alpha$, $\beta$, $M_i$ and so on are in principle arbitrary functions of $t$, though in a near-de Sitter framework this $t$-dependence should be suppressed by slow-roll parameters.

An important property of the lagrangian \pref{cheungetalL} is that the only terms linear in a metric fluctuation are the two terms with coefficients $\alpha$ and $\beta$. This is why in all other terms $g^{00}$ is chosen always to appear in the combination $g^{00}+1$. Although not obvious, this claim relies on the observation that the other linear terms are redundant, in that they are total derivatives or can be removed by a field redefinition. Because of this only $\alpha$ and $\beta$ play a role in the evolution of the background metric, $H(t)$. Since the background has the symmetries of an FRW geometry the stress energy computed from $\cL$ evaluated at the background is characterized by an energy density and a pressure, $\rho$ and $p$, and these are related to $\alpha$ and $\beta$ by $\rho = \alpha + \beta$ and $p = \beta - \alpha$.  The background Einstein equations then imply
\be
 H^2 = \frac{\alpha + \beta}{3M_p^2} \qquad \hbox{and} \qquad
 \frac{\ddot a}{a} = \dot H + H^2 = - \frac{(2\beta - \alpha)}{3M_p^2} \,,
\ee
and so \pref{cheungetalL} can be rewritten as
\bea \label{cheungetalL2}
  \frac{\cL}{\sqrt{-g}} &=& \frac{M_p^2}2 \, R -M_p^2(3H^2 + \dot H) + M_p^2 \dot H g^{00} \\
  && \qquad\qquad\qquad  + \frac12 \, M_2^4(t) (g^{00} + 1)^2 + \frac{1}{3!} \, M_3^4(t) \, (g^{00}+1)^3 + \cdots \,.\nn
\eea

The lagrangian obtained by setting all coefficients except for $\alpha$ and $\beta$ to zero corresponds to the simple single-field slow-roll models considered in the rest of these notes. This can be seen by taking the scalar lagrangian, $\cL_\phi/\sqrt{-g} := - \frac12 (\partial \phi)^2 - V(\phi)$ and going to unitary gauge, for which $\phi = \varphi(t)$ and so $\cL_\phi/\sqrt{-g} = -\frac12 \, \dot \varphi^2 g^{00} - V(\varphi)$. Comparing with \pref{cheungetalL} shows $\alpha = V(\varphi)$ and $\beta = \frac12 \, \dot \varphi^2$. In this case all of the nonlinear interactions amongst the fluctuations are contained within the Einstein-Hilbert part of the action. 

The remaining coefficients $M_i(t)$ and so on describe deviations from the simplest scalar models. These could correspond to supplementing the basic scalar lagrangian with higher derivative interactions, like higher powers of the kinetic term --- such as $(\partial \phi)^4$ and so on --- or through more exotic kinds of choices \cite{MoreExotic}. Part of the utility of \pref{cheungetalL} is that one does not need to know what these choices might have been in order to use $\cL$ to compute how observables can depend on the coefficients $M_i$. This allows a relatively model-independent survey of what kind of observables are possible at low energies, without having to go through all possible microscopic models beforehand.

\subsubsection{Open systems}

EFTs applied to gravitational systems can surprise in other ways as well. In particular, during inflation we have seen that the main observational consequences are tied up with super-Hubble modes, for which $k/a \ll H$. Since these are the longest-wavelength modes in the system the effective action that describes them has long been sought as the most efficient way to capture inflationary predictions in as model-independent way as possible.  But no such an effective action was ever found. 

This doesn't mean that an EFT description for these modes does not exist, it just turns out it need not necessarily be usefully described by a traditional Wilsonian effective action \cite{OpenEFT1}. This unusual situation arises because during inflation the long-wavelength modes are described by the EFT for an {\em open} system \cite{OpenEFT1, OpenEFT2, OpenEFT3, OpenEFT4}, in that modes are continually moving from sub-Hubble to super-Hubble throughout the inflationary epoch. This should be contrasted with the usual situation with a Wilsonian effective theory, for which high- and low-energy states are forbidden from transitioning into one another by energy conservation. 

Because of this mode migration the long- and short-distance sectors can interact in more complicated ways than are normally entertained, such as by entangling and/or decohering with one another. The appropriate language for describing long-wavelength modes in this kind of situation is to use the reduced density matrix, $\varrho_L = \hbox{Tr}_S\, \rho$, in which the full system's density matrix is traced over the unwatched (in this case, short-wavelength) sector. It turns out that $\varrho_L$ evolves in time according to a Lindblad-type equation, which need not be writable as a Liouville equation for some choice of effective Hamiltonian. 

Using these kinds of arguments it is possible to show that the evolution of the leading effective description of the diagonal parts of the reduced density matrix, $P[\varphi,t] := \langle \varphi | \rho_L | \varphi \rangle$, describing fluctuations amongst super-Hubble modes during inflation, is given by a Fokker-Planck equation. The description of the properties of this equation is called `stochastic inflation' \cite{StochasticInf}, and the stochastic description of the quantum system is possible because quantum states become very classical, in that they are well-described by the WKB approximation, when outside the Hubble scale. Although a full description goes beyond the scope of these lectures, the evidence now is that this open-system evolution (starting with stochastic inflation) does a good job capturing the late-time evolution of super-Hubble modes. In particular, it resolves problems to do with infrared divergences and the secular breakdown of perturbation theory at late times that are encountered in all but the simplest inflationary models.

A bonus with this picture comes when evolving the off-diagonal components of the reduced density matrix,  $\langle \varphi | \rho_L | \psi \rangle$, since this evolution tends towards a diagonal matrix in a basis that diagonalizes the field operators $\phi(x)$. This provides the explanation of why perturbations that leave the Hubble scale as quantum fluctuations eventually re-enter the sub-Hubble regime well-described as a statistical distribution of classical fields \cite{OpenEFT1}.

\subsubsection*{The Bottom Line}

These notes trace four beautiful threads woven deeply into the great cosmic tapestry. 

First, a very beautiful and pragmatic picture is emerging wherein the special initial conditions required of late-time cosmology can be understood in terms of quantum fluctuations during a much earlier accelerating epoch. Although there is debate as to whether the state being accelerated was initially growing --- inflation --- or shrinking --- a bounce --- everyone sings from the same quantum scoresheet. If true, quantum effects in gravity are literally written across the entire sky.

Second, as in all other areas of physics, EFT methods provide extremely valuable ways to handle systems with more than one scale. This is especially true for gravitational physics in particular, whose nonrenormalizability is a red flag telling us we are dealing with the low-energy limit of something more fundamental. Indeed, EFT ideas provide the bedrock on which all standard semiclassical reasoning is founded. With our present understanding of physics, any departure from an EFT framework when working with gravity always brings loss of control over theoretical error. Cosmologists cannot afford to do so now that cosmology has become a precision science.

Third, at this point there is no uncontroversial evidence in favour of gravitational exceptionalism, inasmuch as the issues encountered applying EFTs to gravitational problems also seem to arise in other areas of physics (for which many powerful tools have been developed). But there are also unsolved puzzles associated with quantum fields interacting gravitationally, and it is always prudent to have one eye out for surprises.  

Finally, although the EFTs used in gravity seem to have counterparts elsewhere in physics, these other areas are often not particle physics and so gravity {\em can} provide surprises to those coming from a particle-physics training. These surprises include secular evolution and a generic breakdown of perturbation theory at late times (that arise because the gravitational field never goes away, and so even small secular effects eventually can accumulate to become large). They include stochastic effects where Wilsonian actions need not be the useful way to formulate the problem, because the physics of interest is an open system. For these the effective interactions of particles moving within a larger medium can be better models than intuition based on traditional low-energy Wilsonian EFTs.

But this is a relatively young field and yours is the generation likely to be reaping the rewards of (or discovering) new directions. In situations like this the field is likely to belong to those who bring diverse tools and an open mind: be broad and good luck!

\end{document}